\documentclass[trackchanges]{aastex701}

\usepackage{amsmath}
\usepackage{caption}
\usepackage{subcaption}
\usepackage{multirow}
\begin{document}

\title{Sun-to-Earth Coronal Mass Ejection Simulations From a Vector Magnetogram}

\author[orcid=0009-0006-0187-3964]{Yifu An}
\affiliation{University of Michigan}
\email{anyifu@umich.edu}

\author[orcid=0000-0001-5074-2246]{G\'abor T\'oth}
\affiliation{University of Michigan}
\email{gtoth@umich.edu}  

\author[orcid=0000-0001-5447-5456]{Beatrice Popescu Braileanu}
\affiliation{Universitetet i Bergen}
\email{beatrice.braileanu@uib.no}  
% \author{River Europe}
% \affiliation{University of Heidelberg}
% \email{fakeemail4@google.com}

% \author{River Europe}
% \affiliation{University of Heidelberg}
% \email{fakeemail4@google.com}

% \author[0000-0000-0000-0003,sname=Asia,gname=Mountain]{Asia Mountain}
% \altaffiliation{Astrosat Post-Doctoral Fellow}
% \affiliation{Tata Institute of Fundamental Research, Department of Astronomy}
% \email{fakeemail5@google.com}

% \author[0000-0000-0000-0004]{Coral Australia}
% \affiliation{James Cook University, Department of Physics}
% \email{fakeemail6@google.com}

% \author[gname=IceSheet]{Penguin Antarctica}
% \affiliation{Amundsen–Scott South Pole Station}
% \email{fakeemail7@google.com}

% \collaboration{all}{The Terra Mater collaboration}

\begin{abstract}

We implement a novel approach to performing Sun-to-Earth coronal mass ejection (CME) simulations and test it on three geo-effective space weather events.
Using a vector magnetogram observed prior to the CME as the boundary condition, we reconstruct non-linear force free field (NLFFF) solutions in solar active regions with an established magneto-frictional method.
We find a pre-eruption solar corona containing the NLFFF in the AWSoM model, which then spontaneously erupts.
We apply STITCH, a photospheric driving method, when needed, to increase the strength of the CME shock.
The eruptions successfully produce magnetic flux ropes (MFRs) that propagate to 1\,au in the full MHD simulation.
The synthetic white light images of the simulated CMEs share a striking resemblance in shape to observations.
The interplanetary MFRs (IMFRs) arrive at 1\,au with a 1.5- to 9-hour error.
A comparison of simulated solar wind plasma with in-situ measurements shows that IMFR crossing can reproduce a southward $B_z$ and often its magnitudes, which determine the geo-effectiveness of the event.

\end{abstract}



\section{Introduction}
Coronal mass ejections (CMEs) are one of the most violent yet spectacular events in our solar system.
During a CME, a massive amount of solar plasma and magnetic energy is released rapidly. 
When interplanetary CMEs (ICMEs) travel in the Earth's direction, they can cause geomagnetic storms in the magnetosphere and damage space assets. %, or CMEs that affect the interplanetary space
Due to CMEs' profound impacts on human activities, understanding the initiation of CMEs and predicting their impact at the location of Earth have always been an active research topic.

% ADAPT, SHARP and GONG. Fitting potential field.
In recent decades, numerical methods have become dominant in CMEs research \citep{Manchester2017}.
Numerical modeling of CMEs, in particular those aimed at understanding magnetic processes, starts with the observed magnetic field in the photosphere.
The Helioseismic and Magnetic Imager (HMI) is an instrument onboard NASA’s Solar Dynamics Observatory (SDO) \citep{deanpesnell2011solar, schou2011design}.
Among its core measurements are the line-of-sight magnetograms and vector magnetograms, which are maps of all three components of a magnetic field.
Both are available at high spatial and temporal resolutions \citep{Scherrer:2012}.
\citet{Bobra:2014} implemented the Spaceweather HMI Active Region Patch (SHARP) pipeline to identify and track active regions (ARs) in HMI magnetograms.
The pipeline derives vector magnetograms by inverting HMI Stokes profiles within each SHARP.
At the Joint Science Operation Center (JSOC), SHARP vector magnetograms are available in cylindrical equal-area (CEA) coordinates and in the form of $B_r$, $B_\theta$ and $B_\phi$, where $r$, $\theta$ and $\phi$ are the radial coordinate, colatitude and longitude, respectively.
Although SHARP data account for visible active regions, magnetograms for the entire solar surface are not directly available because only the half-disk of the Sun facing the Earth is visible.
One solution is to stitch together magnetograms observed at various times during the Sun's rotation, which results in a synoptic magnetogram.
The Global Oscillations Network Group (GONG) revolutionizes this process by providing seamless and rapidly updated synoptic magnetograms using observations from six solar observatories across the globe \citep{gong2026}.
Another approach is to estimate the magnetogram on the far side of the Sun using physics-based modeling.
Air Force Data Assimilative Photospheric Flux Transport (ADAPT) magnetograms provide nowcast of the solar surface by modeling the far side with flux transport, which accounts for known solar forces such as differential rotation, meridional flow and supergranular diffusion \citep{Arge2010ADAPT,Arge2013ADAPT}.
ADAPT runs an ensemble of 12 simulations for each time instant, each with varied model parameters, to address uncertainties in unpredictable solar activities.
The synoptic or ADAPT magnetogram is then used in the Potential Field Source Surface (PFSS) model to construct a potential magnetic field for the solar corona as the background.
The potential field can be obtained by finding the spherical harmonics that fit the magnetogram or by applying a finite difference method such as the Finite Difference Iterative Potential Field Solver (FDIPS) \citep{Toth:2011}.
%The spherical harmonics that fit the magnetogram can be found or solved for the potential field using a finite difference method \citep{Toth:2011}.

%%modeling with MFR
It is widely accepted that CMEs involve magnetic flux ropes (MFRs).
A popular way to initialize numerical simulations for CMEs is to insert a MFR into the potential background field.
The pioneering work by \citet{Gibson1998} produced the 3-part structure of CMEs as in observation: ``a surrounding high-density region, an internal low-density cavity, and a high-density core''. 
The Gibson-Low (GL) MFR is a self-similar time-dependent solution found by radially stretching a spheromak configuration.
As the GL MFR is not force-balanced, it immediately accelerates and expands after its insertion, resulting in a spontaneous eruption \citep{Manchester:2004,Jin:2017a,Singh_2018,Manchester_2025}.
\citet{Titov1999} proposes a twisted MFR (TD MFR) generated by a uniform ring current embedded in a dipole field, which models the magnetic field of an active region.
A potential arcade acts as the strapping field for the MFR and maintains a delicate force balance.
Later, \citet{Titov2014} improve their work by allowing for more realistic current distributions, and more general shape of the active region dipole, known as the modified TD (TDm) MFR.
The TDm MFR is not initially in force balance as a sacrifice for physical versatility.
Depending on the purpose of the investigation, a TDm MFR can undergo a magneto-hydrodynamics (MHD) relaxation step to settle in the background field \citep{Torok_2018,Liu2024}, or be deliberately initialized in an unstable state \citep{Regnault2023,Linan2023}.
\citet{sokolov2023titovdemoulin} propose an analytical solution for TD-type MDR with finite plasma $\beta$, marking the latest development of MFR models.

%% photospheric driving, STITCH
Photospheric driving is another approach to initiate CMEs simulations.
By directly modifying the velocity or magnetic and electric fields at the bottom of the computational domain, a photospheric driving method adds free energy to a potential field or drives a stable pre-eruption configuration out of equilibrium.
One can modify the boundary condition using time-dependent observational data \citep{Weinzierl_2016,Hayashi_2018,Guo2019} or by introducing a prescribed flow near the polarity inversion line (PIL), such as shearing flow \citep{Linker:1995,Soenen2009,cai2025}, flux emergence \citep{Chen2000,Manchester:2004b,Torok2024} and convergent flow that often drives flux cancellation \cite{Linker2003,Zhao2017}.
Rotational flow can also drive a stable solar corona into eruption.
\citet{Antiochos2013} argues that helicity, injected by photospheric motions, cascades, and condenses near PILs.
On top of this theory, \citet{Dahlin2022} proposes STatistical InjecTion of Condensed Helicity (STITCH), a reduced-order representation of helicity injection. 
The microscopic injection of free energy by vortical cells and their reconnection are abstracted into an ensemble term in the induction equation and Ohm's law, which directly injects tangential magnetic flux into the low corona.
More recently, \citet{vanderHolst2025} shows that STITCH can add free magnetic energy to a potential field, lead to the formation of a filament channel over the PIL and produce an eruption.

%% NLFFF modeling
Although the tangential magnetic field is available from HMI observations, it is often ignored in model-based CME simulations such as the insertion of a MFR.
%The tangential magnetic field at the foot points of a MFR is simply a superposition of the tangential components of the MFR field and the background field.
Non-linear force free field (NLFFF) modeling methods are capable of reconstructing the magnetic structures in an active region using only observational data and a few assumptions.
Under the magnetostatic assumption (zero velocity) and taking the zero plasma beta limit ($\beta=2\mu_0P/B^2\ll1$), the momentum equation of the MHD equations reduces to 
\begin{equation}
    \label{eqn_force_free_1}
    \mathbf{J}\times\mathbf{B} = 0,
\end{equation}
where $\mathbf J=\nabla\times\mathbf B/\mu_0$ is the current density. This implies
\begin{equation}
\label{eqn_force_free_2}
\nabla\times\mathbf{B} = \alpha \mathbf{B}.
\end{equation}
$\alpha=0$ is the trivial solution that $B$ is a potential field; constant $\alpha$ suggests that $B$ is a linear force free field (LFFF), which is mathematically unfit for modeling the solar corona \citep{Wheatland2000}.
When $\alpha$ is a function of spatial coordinates, equation \ref{eqn_force_free_2} describes a NLFFF, which usually has to be solved numerically.
A thorough review of numerical methods for NLFFF modeling can be found in \citet{Wiegelmann2021}.
MHD relaxation is one such method that evolves a form of the MHD equations until equation \ref{eqn_force_free_1} is satisfied \citep{chodura1981,Craig1986,Inoue2011}, and the magneto-frictional method is its simplified and efficient variant.
A fictitious velocity is determined algebraically by a simplified momentum equation such that it is parallel to the Lorentz force, e.g.:
\begin{equation}
\label{eqn_magnetofriction0}
\mathbf{v}=\nu^{-1}(\mathbf{J}\times \mathbf{B}),
\end{equation}
where $\nu$ is a friction coefficient  \citep{Klimchk1992}.
The magnetic field can then be evolved using the induction equation.
\citet{Roumeliotis1996} proposes the ``stress-and-relax'' strategy of iteratively alternating a stressing stage and a relaxing stage.
The stressing stage changes the photospheric boundary condition closer to the observed vector magnetogram, while the relaxing stage reduces the Lorentz force in the domain by solving the time evolution based on equation \ref{eqn_magnetofriction0} and the induction equation.
\citet{Valori2005} and \citet{Valori2007} implement a form of equation \ref{eqn_magnetofriction0} and test their method on idealized paradigms such as the force-free field given by \citet{Low:1990}.
They use a diffusion based $\nabla\cdot \mathbf{B}$ cleaning, which results in an additional term in the induction equation:
\begin{equation}
    \frac{\partial \mathbf{B}}{\partial t} = \nabla \times (\mathbf{v}\times \mathbf{B}) + \delta\nabla( \nabla \cdot \mathbf{B}), \label{eqn_induction_guo}
\end{equation}
where $\delta$ is a parameter to control the dissipation of the error in $\nabla\cdot\mathbf B$.
\citet{Guo:2016I} implement this method in the Message Passing Interface Adaptive Mesh Refinement Versatile Advection Code (MPI-AMRVAC) \citep{Xia2018,Keppens2023}, which is a versatile simulation code featuring support for a wide range of discretization schemes, block-based adaptive mesh refinement (AMR), parallel computing capabilities, etc.
They validate the implementation with various paradigms of NLFFF, using a combination of boundary conditions, in both Cartesian and spherical coordinates, and further obtain good-quality NLFFF solutions from observational data in real active regions \citep{Guo:2016II}.
See section \ref{section_methods} for more details on the implementation of the \cite{Guo:2016II} magneto-frictional method.
It is worth noting that there are other implementations of the magneto-frictional method such as \citet{Jiang2012global}, which reconstructs a full-sphere NLFFF solution using a special grid; time-dependent magneto-frictional model (TMFM) that takes time-dependent boundary conditions \citep{Daei2023}; and a fully-implicit method that uses a Newtonian, viscous formulation for the magneto-frictional velocity \citep{Liu2025NLFFF}.
More recently, magneto-frictional methods have been employed to study the formation and initiation of CMEs in detail \citep{Guo2024}, relax force-imbalanced MFRs for a pre-eruption state \cite{Guo2023}, or compare simulation results from a TMFM with in-situ observations at 1\,au using a simplified propagation model \citep{Kilpua2021}.

%% SWMF
Versatile software frameworks facilitate ICME simulations, such as the Virtual Space Weather Modeling Center (VSWMC) \citep{Poedts2020} and the Space Weather Modeling Framework (SWMF) \cite{Toth:2005swmf,gombosi:2021}.
%Some only focus on the solar corona (SC) and inner heliosphere (IH), such as CORona-HELiosphere (CORHEL) \citep{Riley2012} and the European Heliospheric Forecasting Information Asset (EUHFORIA) \citep{Pomoell2018}; 
The SWMF is our in-house software framework that has played an extensive role in simulating space weather events \citep{Toth:2007,Manchester:2014b,Jin:2017b}.
The SWMF covers a dozen physical or spatial domains, from the solar corona (SC) to the outer heliosphere (OH).
Each domain is simulated by its own model, and in a simulation involving multiple domains, some domains may be coupled together.
ICME simulations require the SC and the inner heliosphere (IH) components, which are coupled at around 20\,$R_\odot$.
A fundamental piece of SC and IH is the Block Adaptive-Tree Solar-wind Roe-type Upwind Scheme (BATS-R-US), an efficient extended MHD solver with support for AMR \citep{Powell:1999,toth2012}.
The versatility of BATS-R-US is mainly achieved by switching its \textit{equation module} and \textit{user module} before compilation.
The \textit{equation module} determines the form of the governing equations, and the \textit{user module} can introduce custom models, corrections to governing equations, boundary conditions, etc.
The SC and IH use the Alfv\'{e}n Wave Solar-atmosphere Model (AWSoM) \citep{van_der_Holst2010,vanderHolst:2014awsom}, which is a state-of-the-art solar wind model that extends the MHD equations to analytically account for Alfv\'{e}n wave heating.
The inclusion of the low corona in our model enables the production of synthesized extreme ultraviolet (EUV) images and Thomson-scattered white light images of the corona, which can then be compared to observation \citep{Lemen2012,howard2008}.

Compared with model-based methods involving MFRs, NLFFF modeling methods are a powerful tool for studying the initial stages of a CME while respecting the tangential components of the observed vector magnetograms.
However, few have adopted NLFFF modeling methods in Sun to Earth CME simulations. 
We thus endeavor to fill this gap with a novel approach to simulating Earth-directed CMEs. 
This approach reconstructs NLFFF solutions from observed vector magnetograms, which then produces CMEs that propagate to Earth's location in interplanetary space.
The remainder of this paper is organized as follows.
Section \ref{section_methods} details the workflow of our approach.
Section \ref{section_results} presents simulation results for 3 Earth-directed CMEs, compares them to various in-situ observations, and discusses potential improvements
Section \ref{section_conclusion} summarizes our findings.

\section{Methods}
\label{section_methods}

\subsection{Selection of Events}
We select three geo-effective CMEs as the events to model.
CME \#\,1 originates from AR 12158 on 10 September 2014 during Carrington rotation (CR) 2154.
It first entered the C2 FOV at 18:00\,UT and traveled at a linear speed of 1267\,km/s \citep{Kilpua2021}.
The interplanetary shock arrived at the WIND spacecraft, which is located at the first Sun–Earth Lagrange point (L1), at 15:17\,UT on 12 September 2014.
CME \#\,2 originates from AR 12297 on 15 March 2015 during CR 2161.
It is a partial CME that causes the largest geomagnetic storm of solar cycle 24, known as the St. Patrick's Day event \citep{wang2015}.
It first appeared in the FoV of LASCO C2 camera at around 01:36\,UT and traveled through the FoVs at 817\,km/s.
The interplanetary shock arrived at 04:00\,UT on 17 March 2015.
CME \#\,3 originates from active region (AR) 11692 on 15 March 2013 during CR 2134.
The halo CME first appeared in the field of view (FOV) of the LASCO C2 camera at 07:12\,UT on 15 March and expanded through the C2 and C3 FOV at around 980\,km/s \citep{wu2016}.
The interplanetary shock arrived at around 05:00 UT on 17 March 2013.
Table \ref{tab:Events_selection} summarizes the time, origin, and speed for the selected events.
We will then refer to the events with the years in which they took place.

\begin{deluxetable}{
cccclcc
}[htb!]
\tablecaption{Summary of selected events. AR coordinates are in a Carrington coordinate system.\label{tab:Events_selection}}
\tablehead{
  \colhead{CME \#} & \colhead{Year} & \colhead{Time} & \colhead{CR} & \colhead{AR (Longitude, Latitude)} & \colhead{Observed speed [km/s]} & \colhead{L1 Arrival Time (Transit Time [h])}
}
\startdata
 1 & 2014 & 18:00 UT, 10 Sept & 2154 & 12158 $(87.27^\circ,~14.97^\circ)$ & 1267 & 15:17 UT, 12 Sept (45.3)\\
 2 & 2015 & 01:36 UT, 15 Mar & 2161 & 12297 $(190.8^\circ,-17.86^\circ)$ & 817 & 04:00 UT, 17 Mar (50.4)\\
 3 & 2013 & 07:12 UT, 15 Mar & 2134 & 11692 $(71.95^\circ,~13.51^\circ)$ & 980 & 05:00 UT, 17 Mar (45.8)\\
\enddata
\end{deluxetable}

\subsection{Data Preprocessing} \label{sec:preprocessing}

\begin{figure}[htb!]
\centering
\includegraphics[width=\textwidth]{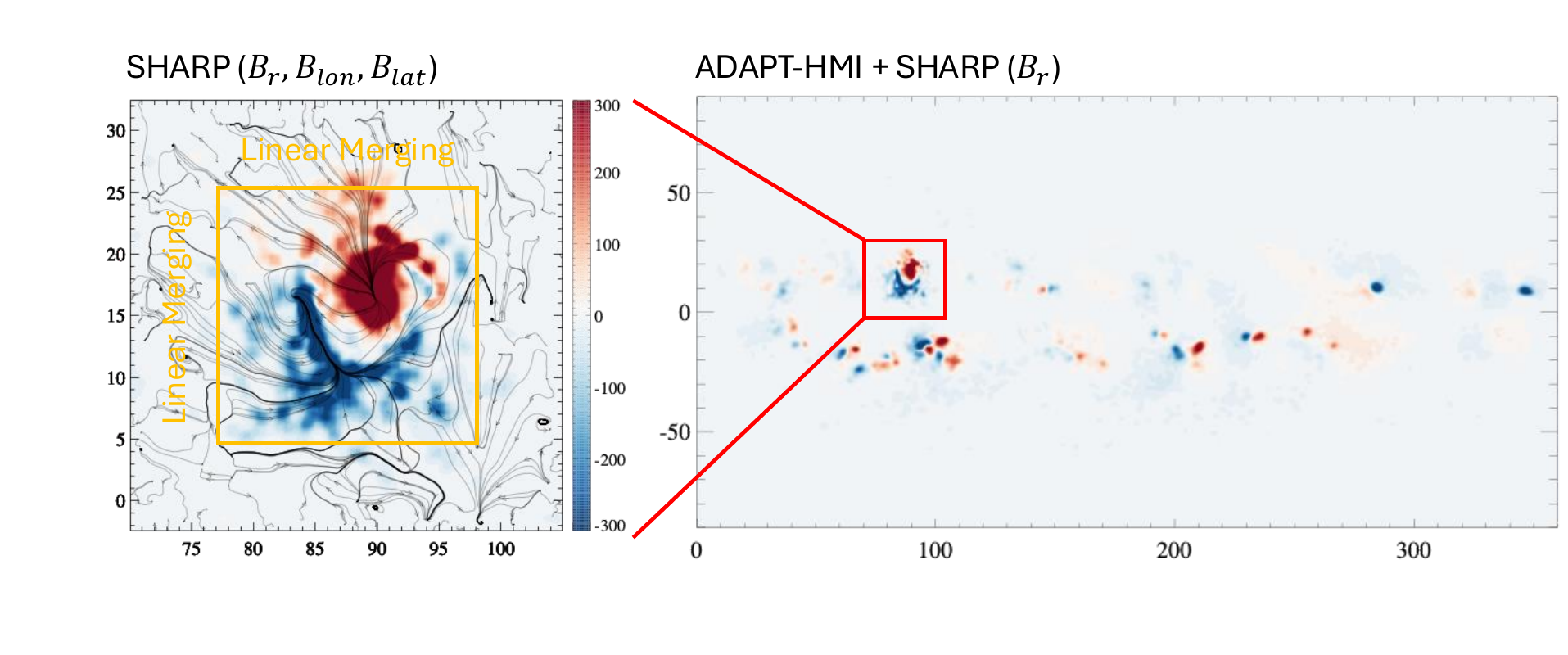}
\caption{Illustration of merging the enlarged SHARP data with the ADAPT-HMI magnetogram. In the outermost 20\% of the SHARP region in each direction, we apply linear merging of $B_r$. The boundary of the twice-enlarged SHARP defines the longitude and latitude ranges for the NLFFF solution. Contours of $B_r$ are saturated at $\pm 300$ G.}
\label{fig:methods_preprocessing}
\end{figure}

For each event, we first obtain the ADAPT-HMI map with the closest timestamp to the time of eruption and the SHARP patch for the active region of interest observed just prior to the time of eruption. 
ADAPT-HMI maps are available on the website of the National Solar Observatory (NSO); SHARP patches are available from the JSOC.
The ADAPT-HMI map contains the radial component, $B_r$, of the full disk.
There are 12 realizations for an ADAPT-HMI map, but as they often do not result in drastically different background solar wind conditions, we always use the first realization.
The SHARP patch contains all three components of the magnetic field, $B_r$, $B_{lon}$ and $B_{lat}$.
For $B_r$, the SHARP patch is merged with the ADAPT-HMI map using a linear transition on all four sides of the patch.
The merging process is illustrated by Figure \ref{fig:methods_preprocessing}.
By fitting spherical harmonics with a maximum degree of 180 using the merged $B_r$, which combines values from the SHARP patch in the AR and those from the ADAPT-HMI map elsewhere, we find a potential field $\mathbf{B_p}$ to be used as the background.
%We crop out the active region from this $B_r$.
The vector magnetogram for the AR is then composed of $B_{lon}$ and $B_{lat}$ from the SHARP patch and the merged $B_r$.
%Together with the longitudinal and latitudinal components of the SHARP patch, we obtain the vector magnetogram for the active region.
To reduce noise, we apply a boxcar smoothing of $B_r$, $B_{lon}$ and $B_{lat}$ with $1^\circ$, $0.5^\circ$ and $0.5^\circ$ widths respectively.

Past studies of active regions or CMEs using the AWSoM have reached a maximum resolution of $0.35^\circ$. 
This is not enough to resolve the NLFFF solution given the small scale of an active region in SHARP data, where the field strength can change by a few thousand Gauss over less than $5^\circ$ across the PIL.
On the other hand, despite BATS-R-US' support for AMR, we cannot afford to arbitrarily refine the active region due to the computational cost.
We use a maximum resolution of $0.175^\circ$ and choose to enlarge the active region by a factor of 2.
This gives an equivalent maximum resolution of $0.088^\circ$ for the active region.
To compensate for the enlarged strong field region, we divide the field strength by a factor $c_B$.
$c_B$ serves as a free parameter to control the free energy in the NLFFF solution, which then controls the energy and speed of the CME.
A range of $c_B\in [1.0,1.5]$ gives a reasonable bulk speed for eruptions in our simulation.
%In two modeled events (AR 11692 and AR 12158), the enlargement's impact on the background solar wind turns out to be minimal; in one other (AR 12297), the background slightly worsens.

We acknowledge that the preprocessing method by \citet{Wiegelmann:2006} may improve the quality of the NLFFF solution.
We do not apply such preprocessing to the vector magnetogram, mainly because the method minimizes a discretized version of $\int \mathbf J\times \mathbf B$, rather than $\int |\mathbf J\times \mathbf B|$. 
This makes it a necessary condition for force-freeness in the computational volume rather than a sufficient one.
It also requires considerable work to efficiently implement the method in the SWMF.
On the other hand, the work of \citet{Guo:2016II} shows the promise of this preprocessing method applied to real ARs.
%\citet{Guo:2016II} evaluates the quality of the NLFFF solution using a metric $\sigma_J$, i.e. the averaged sine of the angle between $\mathbf J$ and $\mathbf B$, weighted by $\mathbf J$.
%For various preprocessed vector magnetograms, $\sigma_J$ for their NLFFF solutions ranges from 0.2 to 0.52, which shows promise for this preprocessing method.
%0.20 to 0.41 (equivalent to an average angle of $11.5^\circ$ to $24.2^\circ$), but values as high as 0.52 ($31.3^\circ$) is also reported for local active regions.
Therefore, we hope to include this preprocessing method as one of the first steps to improve our work in the future.

\subsection{NLFFF Modeling}
To solve for a NLFFF solution from vector magnetograms, we integrate the magneto-frictional method developed and validated by \citet{Guo:2016I} into BATS-R-US.
We use this NLFFF modeling technique to find a local solution in a curvilinear box enclosing the active region of interest.
The radial extent of the domain ranges from 1\, $R_\odot$ to 1.75\, $R_\odot$, while the longitudinal and latitudinal extents range from $35^\circ$ to $45^\circ$.
The domain is resolved with a maximum angular resolution of $0.175^\circ$.
Rather than the full MHD equations, the magneto-frictional method evolves only the magnetic field, $\mathbf B$, and the velocity, $\mathbf v$.
%The implementation of \citet{Guo:2016I} is given by
%\begin{align}
%    \mathbf v &= \frac{c_c c_y}{\max(|\hat{\mathbf v}|)}\frac{\Delta x}{\Delta t}\hat{\mathbf v} f_w(x), %\label{eqn_mf1}\\
%    \hat{\mathbf v} &= \Delta x \frac{\mathbf J\times \mathbf B}{|\mathbf B|^2}, \label{eqn_mf2}\\
%        \frac{\partial \mathbf B}{\partial t} &= \nabla \times (\mathbf v\times \mathbf B)
%     + \delta \nabla (\nabla\cdot\mathbf B). \label{eqn_induction_full}
%\end{align}
%where $c_c$ and $c_y$ are free parameters that control the magnitude of $\mathbf v$.
%$f_w(x)\in [0,1]$ is a weighing function depending on the position within the magneto-frictional domain.
%The factor $\frac{\Delta x}{|\mathbf B|^2}$ in equation \ref{eqn_mf2} makes $\hat{\mathbf v}$ dimensionless and speeds up relaxation in weak-field regions.
%$\mathbf v$ is then used in the induction equation with a diffusive $\nabla\cdot\mathbf B$ cleaning term (Eqn. \ref{eqn_induction_full}) to update $\mathbf B$.

Our implementation of this method is based on \citet{Guo:2016I}, but differs in a few aspects.
We have simplified the formulation, improved the stability condition by setting the time step, and use a different method to control the error in $\nabla\cdot\mathbf B$. The magneto-friction equations are implemented as
\begin{align}
\label{eqn_velocity_ours}
\mathbf v &= \left\{ 
\begin{aligned}
& \hat{\mathbf v} & \quad \text{if } |\hat{\mathbf v}|\leq u_\mathrm{max};\\
& u_\mathrm{max}\frac{\hat{\mathbf v}}{|\hat{\mathbf v}|} & \quad \text{if } |\hat{\mathbf v}|> u_\mathrm{max}.
 \end{aligned}
 \right.\\
\hat{\mathbf v} &= c_{mf}\frac{\mu_0\mathbf J \times \mathbf B}{|\mathbf B|^2}. \label{eqn_velocity_ours_raw} \\
\frac{\partial \mathbf B}{\partial t} & = \nabla \times (\mathbf v\times \mathbf B) - \mathbf v\nabla\cdot\mathbf B - \nabla \psi 
\label{eqn_induction_ours}\\
\frac{\partial \psi}{\partial t} &= -c_h^2\nabla\cdot\mathbf{B} \label{eqn_hyp_cleaning}.
\end{align}
Equations \ref{eqn_velocity_ours}, \ref{eqn_velocity_ours_raw} for the magneto-frictional velocity $\mathbf v$ are complemented by the induction equation with the 8-wave term $-\mathbf v\nabla\cdot\mathbf B$ \citep{Powell1999} and hyperbolic $\nabla \cdot \mathbf B$ cleaning (Eqn. \ref{eqn_induction_ours}, \ref{eqn_hyp_cleaning}) \citep{Toth:2000,Dedner:2001}.
$u_\mathrm{max}$ is an artificially imposed maximum speed, $c_{mf}$ is a constant coefficient, $\psi$ is a scalar function introduced to clean $\nabla \cdot \mathbf{B}$ and $c_h$ is the hyperbolic cleaning speed.
The values are $c_{mf}=10^{11}$\,$\mathrm{m^2/s}$, $u_\mathrm{max}=30$\,km/s and $c_h=30$\,km/s in our runs.
We implement equations \ref{eqn_velocity_ours} through \ref{eqn_velocity_ours_raw} in a \textit{user module} in BATS-R-US.
Equations \ref{eqn_induction_ours}, \ref{eqn_hyp_cleaning} utilize BATS-R-US's generic state update module, which implements the 8-wave scheme \citep{Powell:1999}, after BATS-R-US is compiled with the \textit{equation module} containing the MHD variables and the scalar $\psi$.
Equations \ref{eqn_velocity_ours} through \ref{eqn_hyp_cleaning} evolve until we find a steady-state.
Note that the $\nabla \cdot \mathbf B$ cleaning technique is not necessary in the 8-wave scheme normally, as $\nabla\cdot \mathbf B$ is designed to be advected away with the flow velocity.
Here, in contrast, $\nabla \cdot \mathbf B$ accumulates because a steady state NLFFF should have zero velocity.

In our implementation, time stepping is controlled by
\begin{equation}
  \Delta t \le C \frac{\Delta x}{|\mathbf v| + c_{mf}/\Delta x}
\end{equation}
where the CFL number is $C=0.8$ and $\Delta x$ is the characteristic size of the grid cell.
The imposed speed limit, $u_\mathrm{max}$, improves the robustness of the method. 
\citet{Guo:2016I} gradually reduce the velocity to 0 near all boundaries except the bottom, which is intended for a smooth transition in $\mathbf B$ near the boundaries.
However, we find that reduced $\mathbf v$ at the boundaries prevents those regions from converging to a force-free state; therefore, we do not use this approach.
We instead improve the merging at the boundaries by selecting a sufficiently large domain.
In practice, numerical diffusion also prevents discontinuities at the boundary.

To improve the quality of the NLFFF solution, we separate the runs into two stages.
The first stage is a steady-state run, or the ``stressing stage''; the second stage that restarts at the end of the stressing stage is a time-accurate, or the ``relaxing stage''.
The two stages differ only in time stepping and boundary conditions.
In the stressing stage, BATS-R-US advances in steady state mode, where each grid cell takes its maximum allowable time step.
The bottom boundary at $r=R_\odot$ sets the three components of $\mathbf B$ to the vector magnetogram and uses a float velocity boundary condition.
On the other five surfaces of the curvilinear box, we set $\mathbf B$ to the potential field and $\mathbf v$ to zero.
In the relaxing stage, BATS-R-US advances in time-accurate mode, where all grid cells take the same time step.
The boundary conditions are the same except for the $r=R_\odot$ surface.
At the bottom boundary, we only enforce $B_r$ to the vector magnetogram but set the other two components to float.
We set $\mathbf v$ such that the foot points of the field lines in the NLFFF solution remain stationary, but the field lines can twist or bend in order to reduce the residual Lorentz force.
This means that the field-aligned component of $\mathbf v$ is float while the perpendicular component is reflected.
The stopping time of the relaxing stage is determined empirically: we stop when the NLFFF has small residual forces, but loses only a small fraction of free energy built up in the stressing stage.

%For better robustness, we limit $|v|\leq u_\mathrm{max}$: if $|v|> u_\mathrm{max}$, let $v'=vu_\mathrm{max}/|v|$ and set $v\longleftarrow v'$ instead.
%Thus, $|v_\mathrm{max}|=u_\mathrm{max}$ must exist and our simplification, equation \ref{eqn_magnetofriction_ours} holds.
%To improve the quality of the NLFFF equation at the end of the steady-state run, we evolves the same equations for some time with more relaxed boundary conditions.

We use the same metrics as \cite{Guo:2016I} for the quality of the NLFFF solution:
\begin{align}
\langle|f_i|\rangle &= \frac{\sum_i |f_i| \Delta V_i}{\sum_i\Delta V_i},\quad & f_i &=\frac{(\nabla\cdot \mathbf B)_i \Delta V_i}{|\mathbf B_i| A_i} \label{eqn_metric_div} \\
\sigma_J &= \frac{\sum_i \sin \theta_i |\mathbf J_i| \Delta V_i}{\sum_i |\mathbf J_i| \Delta V_i},\quad & \sin \theta_i &= \frac{|\mathbf J_i\times \mathbf B_i|}{|\mathbf J_i||\mathbf B_i|} \label{eqn_metric_sine}. 
\end{align}
$\langle|f_i|\rangle$ is a measure of the average $\nabla\cdot \mathbf B$ error in the volume and $\sigma_J$ is the average sine of the angle between $\mathbf J$ and $ \mathbf B$, weighted by current and volume.
We also introduce two useful metrics in our magneto-frictional runs:
\begin{align}
R_f &= \frac{\sum_i |\mathbf J_i\times \mathbf B_i|\Delta V_i}{\max_t(\sum_i |\mathbf J_i\times \mathbf B_i|\Delta V_i)}, \label{eqn_metric_jxb} \\
\epsilon_{1} &= \frac{\sum_i (|\mathbf B_i|^2-|\mathbf B_{P,i}|^2) \Delta V_i}{\sum_i |\mathbf B_{P,i}|^2 \Delta V_i}. \label{eqn_metric_emagnp}
\end{align}
$R_f$ measures the volume integral of the residual Lorentz force, normalized by its maximum during the magneto-frictional run, while $\epsilon_1$ measures the free energy in the NLFFF solution normalized by the energy of the potential field.
The summations in equations \ref{eqn_metric_div}-\ref{eqn_metric_emagnp} are all carried out in the active region, which is the computational domain where the magneto-frictional equations evolve.

\subsection{Pre-eruption Quasi-steady State}
We denote the potential field obtained from the spherical harmonics by $\mathbf B_P$ and the NLFFF solution obtained from the previous step by $\mathbf B_N$.
Note that the NLFFF solution only involves the magnetic field, so we run MHD simulations using the AWSoM model \citep{vanderHolst:2014awsom} in the SWMF to obtain a pre-eruption solar corona.
The pre-eruption solution contains a SC component and an IH component.
The SC component extends from the chromospheric surface at 1\,$r_\odot$ to 24\,$r_\odot$; the IH component extends from 18\,$R_\odot$ to slightly beyond 1\,au, or about 215\,$R_\odot$.
The coupling of the SC and IH components takes place in a buffer zone where the two grids overlap.
The background magnetic field in the SC component is $\mathbf B_0= \mathbf B_P$ except in the curvilinear box where we have obtained a NLFFF solution -- there, we use $\mathbf B_0= \mathbf B_N$ as the background.
The residual Lorentz force in the NLFFF solution is small but still large enough to cause large accelerations in the low corona, so we temporarily neglect the current in $\mathbf B_0$ to obtain a quasi-steady state.
Of course, the current in $\mathbf B_N$ is not zero, but this trick allows us to obtain a reasonable initial MHD condition based on the NLFFF solution.

We first run the SC component for 80,000 steps in steady state mode and then briefly couple the SC component with the IH component, which sets the inner boundary of the IH component using the outer boundary of the SC component.
After that, IH runs in steady state mode for another 5,000 steps. This is sufficient to reach an initial steady state, because the radial flow in IH is super-magnetosonic.

The pre-eruption state gives rise to background solar wind at 1\,au, which can affect subsequent simulations for CMEs.
We only adjust the Poynting flux parameter in the AWSoM model to tune the pre-eruption state \citep{Sachdeva_2021}.
The background solar wind observed at Earth for the entire Carrington rotation is compared with OMNI data \citep{king2005}.
It is known that the AWSoM can perform poorly for background runs near solar maximum, so we prioritize matching the solar wind around the L1 arrival time of the CMEs.
For the Carrington rotation each event took place, we run a few simulations and select the one with the best agreement in the number density and/or speed.
%The quality of the best matching background solar wind varies for each event.
We refer the reader to \citet{Jivani_2023} for a detailed parametric study of the AWSoM model.
%We have good agreements for two events (CR 2134, CR 2154), but for another (CR 2161) we end up with a lower speed and a lower density.
Furthermore, by comparing simulations with and without an enlargement of the AR, we find that the enlargement only has a minimal impact on the background solar wind at 1\,au.

We extensively use AMR in the grid to reduce computational cost.
The SC grid is spherical.
The central part of the AR, which is roughly 0.25\,$r_\odot\times$ 20$^\circ\times$ 20$^\circ$ in size, is refined to an angular resolution of $0.175^\circ$.
The most interesting parts of the domain, including the rest of the curvilinear box selected for the magneto-frictional run, a cone pointing in Earth's direction and a cone pointing upwards from the AR, are refined to a $0.35^\circ$ resolution.
Current sheets are refined to a $0.7^\circ$ resolution with dynamic AMR that follows the evolution of current sheets.
The radial extent of the grid cells in SC is determined by the local AMR level and the radial stretching of the grid that is used in AWSoM to resolve the transition region.
The minimum, found near $r=$\,$R_\odot$ in the active region, is only $1\times10^{-4}$\,$R_\odot$;
the maximum, found around the outer boundary, is several $R_\odot$ in length.
The IH grid is Cartesian, with higher resolution in a cone pointing from the Sun to Earth and lower resolution elsewhere.

\subsection{Time Accurate CME Simulation}
Time-accurate CME simulations restart from the quasi-steady states obtained previously with just one crucial change in their setup: we return to using $\mathbf B_0 = \mathbf B_P$ as the background in the entire corona so that the current in $\mathbf B_N$ is now fully included.
This essentially turns on the free energy in the approximate NLFFF solution and puts the corona into force imbalance due to finite Lorentz force $\mathbf J\times\mathbf B$.
The force imbalance then leads to spontaneous eruptions that propagate to 1\,au.
We compare the synthetic white light images of the corona to LASCO C2/C3 observations \citep{Yashiro2004} and the plasma states at the location of Earth to the Merged Interplanetary Data from L1 (MIDL) dataset \citep{MIDL2026}.

To speed up the simulations, we switch on partial local time stepping at later times.
Grid cells below a certain height take their maximum allowable time steps, while grid cells above this height still use a global time step.
This has minimal impact on the solution if we only switch on this feature when the CME has traveled far enough.

The residual Lorentz force is not guaranteed to trigger a rapid release of most free energy.
In some cases, the initial eruption is not strong enough to break free from the background field, and the remaining NLFFF solution slowly defuses away.
We apply STITCH for a short time to produce a successful eruption.
By injecting helicity into all grid cells below 1.01\,$R_\odot$, the magnetic field can acquire enough energy in a very short time (less than 30 seconds of simulation time).
The shock produced by STITCH coalesces with the initial shock generated by force imbalance very close to the active region.
The strength of the combined shock can be tuned by the duration of STITCH or the amount of energy injected.
Therefore, the division factor $c_B$ used in preprocessing (see subsection~\ref{sec:preprocessing}), together with the energy injected by STITCH, controls the strength of the CME shock.

\subsection{Procedure Summary}
Our method consists of the following steps:
\begin{enumerate}
\item Merge enlarged SHARP data with an ADAPT-HMI map and apply smoothing.
\item Obtain a potential field $\mathbf{B_P}$ based on the combined photospheric radial magnetic field $B_r$.
\item In a stressing stage and then a relaxing stage, solve the magneto-frictional equations for the approximate NLFFF solution $\mathbf{B}_N$ using the preprocessed $B_r$, $B_{lon}$, $B_{lat}$ as the bottom boundary condition in the AR.
%The stressing stage strictly applies the boundary condition, and the relaxing stage applies a relaxed one.
The NLFFF solution contains only the magnetic field.
\item Merge $\mathbf{B_P}$ and $\mathbf{B_N}$ into the background magnetic field $\mathbf{B}_0$. 
Neglect the current in $\mathbf{B}_0$ and obtain a pre-eruption AWSoM steady state.
The AWSoM steady state contains full MHD states.
\item Restart from the pre-eruption steady state and use $\mathbf{B_P}$ as $\mathbf{B}_0$.
The current in $\mathbf{B_N}$ is switched on and leads to a spontaneous eruption.
\item Apply STITCH to produce a stronger eruption if needed.
\end{enumerate}

\section{Results}
\label{section_results}

\subsection{NLFFF Solutions}

\begin{figure}[htb!]
     \centering
     \begin{subfigure}[b]{0.33\textwidth}
         \centering
         \includegraphics[width=\textwidth]{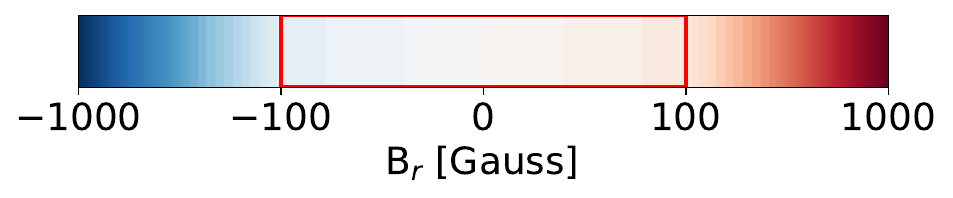}
         %\caption{$y=x$}
         %\label{fig:NLFFF_12297}
     \end{subfigure}%
     \begin{subfigure}[b]{0.33\textwidth}
         \centering
         \includegraphics[width=\textwidth]{plots/NLFFF/VM_colorbar.pdf}
         %\caption{$y=3\sin x$}
         %\label{fig:NLFFF_11692}
     \end{subfigure}%
     \begin{subfigure}[b]{0.33\textwidth}
         \centering
         \includegraphics[width=\textwidth]{plots/NLFFF/VM_colorbar.pdf}
         %\caption{$y=5/x$}
         %\label{fig:NLFFF_12158}
     \end{subfigure}
\vspace{-\lineskip}
     \begin{subfigure}[b]{0.33\textwidth}
         \centering
         \includegraphics[width=\textwidth]{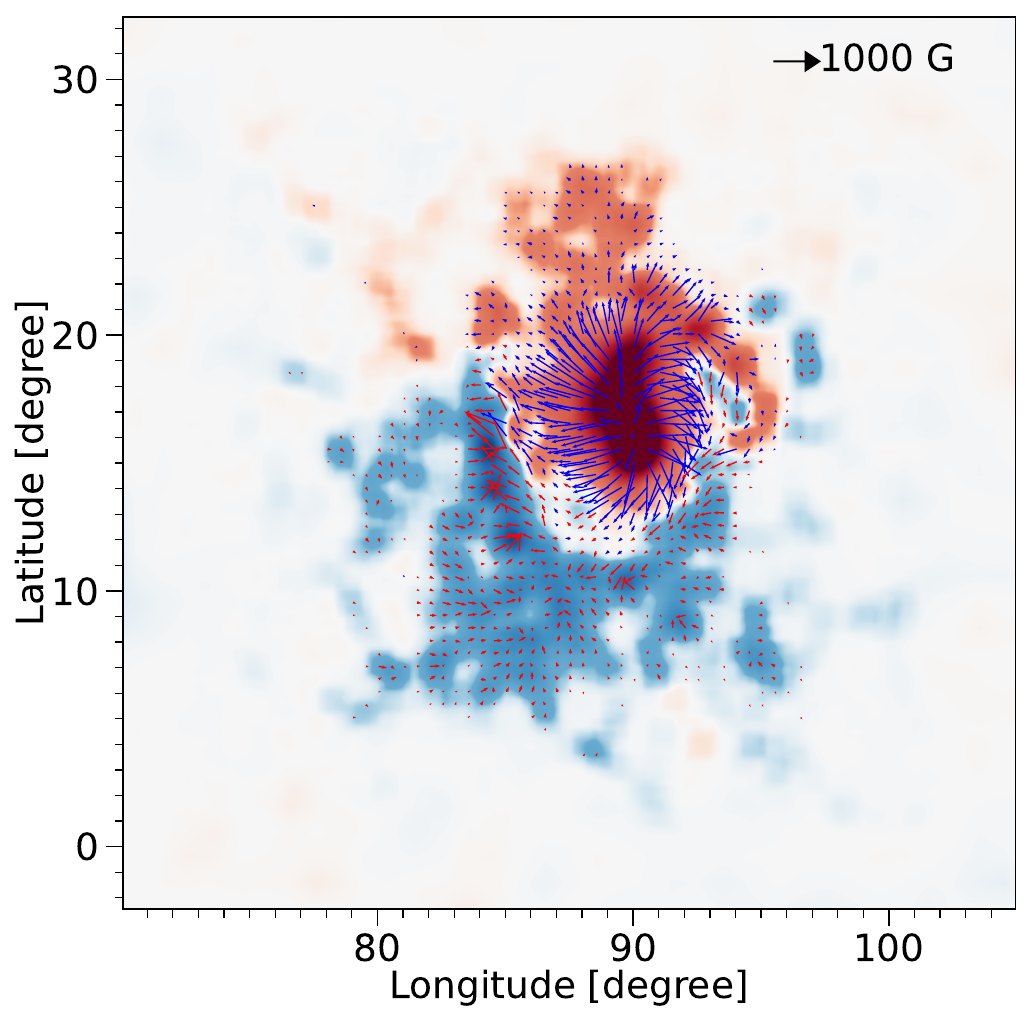}
         \caption{AR 12158 (2014 event).}
         \label{fig:VM_12158}
     \end{subfigure}%
     \begin{subfigure}[b]{0.33\textwidth}
         \centering
         \includegraphics[width=\textwidth]{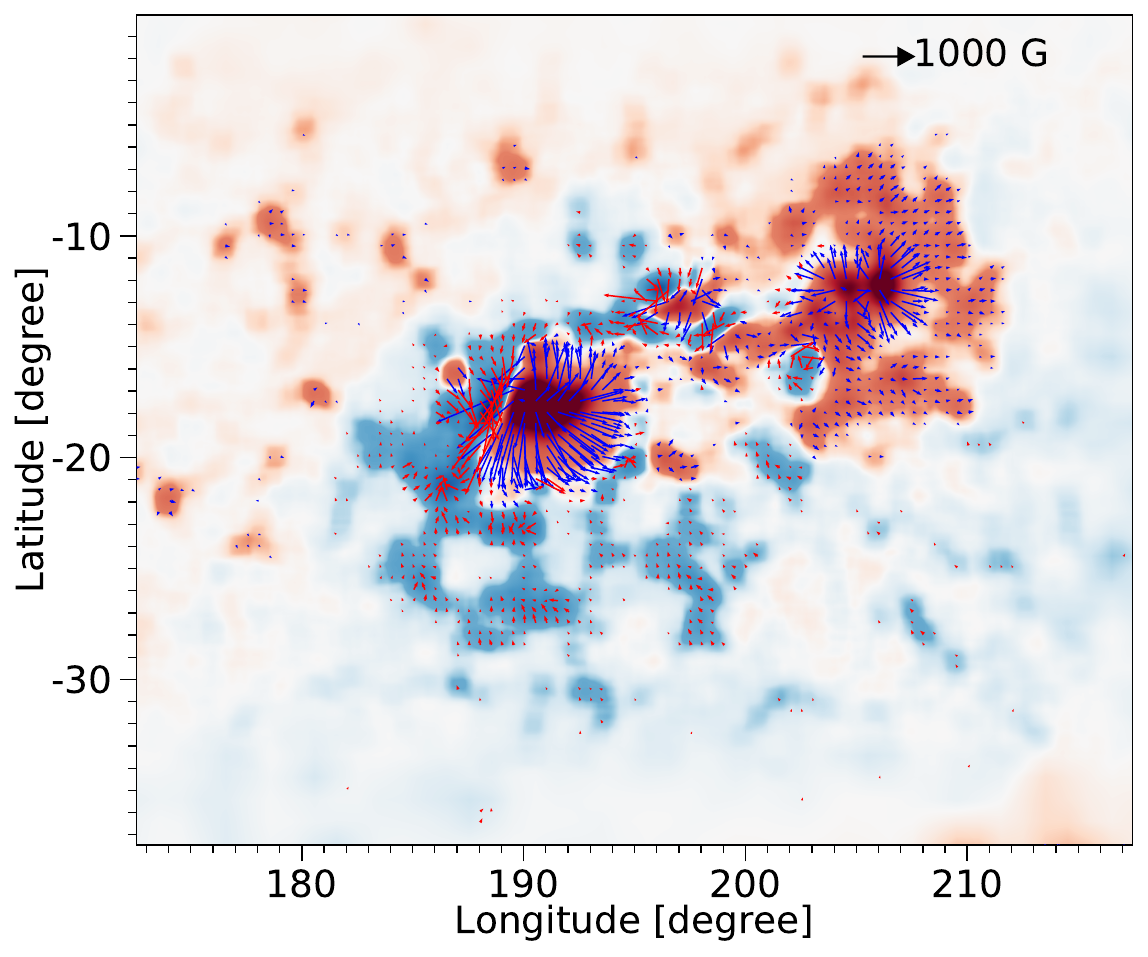}
         \caption{AR 12297 (2015 event).}
         \label{fig:VM_12297}
     \end{subfigure}%
     \begin{subfigure}[b]{0.32\textwidth}
         \centering
         \includegraphics[width=\textwidth]{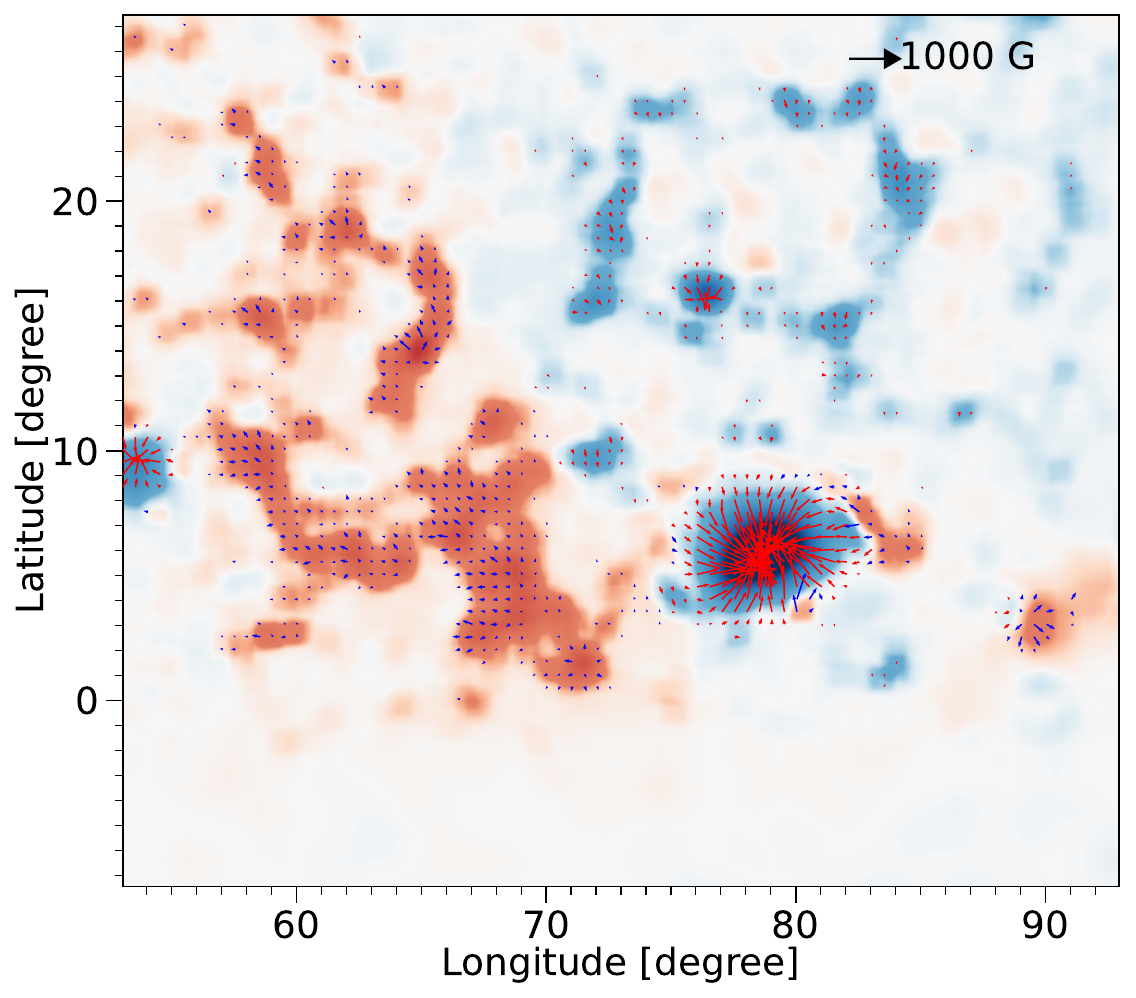}
         \caption{AR 11692 (2013 event).}
         \label{fig:VM_11692}
     \end{subfigure}
        \caption{Preprocessed vector magnetograms to be used as the boundary conditions for AR 12158, AR 12297 and AR 11692. The contours show the radial component ($B_r$) while the vectors show the tangential component ($B_{lon},B_{lat}$). The black arrows in the top right corners indicate the length corresponding to 1000\,G. The color mapping is capped at $\pm1000$\,G. To better show weak field regions, the color mapping is piecewise uniform in $[-1000,-100]$\,G, $[-100,100]$\,G and $[100,1000]$\;G, as is highlighted by the box in the colorbar. The vectors are colored red/blue for better contrast.}
        \label{fig:NLFFF_VM}
\end{figure}
%hmi.sharp_cea_720s.5298.20150315_010000_TAI.Br
%centerlon: 190.83, centerlat: -17.86
%hmi.sharp_cea_720s.2546.20130315_060000_TAI.Br
%center lon = 71.95, centerlat: 13.51
%hmi.sharp_cea_720s.4536.20140910_170000_TAI.Br
%center lon = 87.27, centerlat: 14.97
%Carrington coordinate system
For each event, we reconstruct a NLFFF solution from a SHARP vector magnetogram observed prior to its commencement and a concurrent ADAPT-HMI magnetogram.
The first realization of the ADAPT-HMI magnetograms is selected for all three events at our discretion.
The JSOC data series for the SHARP data is \texttt{hmi.sharp\_cea\_720s}.
The SHARP data observed around an hour before the eruption time are used.
For AR 12158, the patch is observed at 17:00 UT on September 10 2014.
For AR 12297, we use the patch observed at 01:00 UT on March 15, 2015.
% , which is centered at $190.83^\circ$ lon, $-17.86^\circ$ lat in Carrington coordinate system.
For AR 11692, the patch is observed at 06:00 UT on March 15, 2013.
%nd centered at $71.95^\circ$ lon, $13.51^\circ$ lat.
%and centered at $87.27^\circ$ lon, $14.97^\circ$ lat.
%As the build-up of energy for a CME can last days to weeks in real life, we acknowledge that this discretionary choice could be improved in future. %%% CONCLUSION
The SHARP data are spatially enlarged by a factor of two while their strength is reduced, merged with the ADAPT-HMI map, and smoothed as described in section \ref{section_methods}.
In the results presented in the following, the reduction factor, $c_B$, is 1.25 for all events. 
Additionally, we model the 2013 event with $c_B=1.0$ to test a stronger shock.
Figure \ref{fig:NLFFF_VM} shows the three vector magnetograms after preprocessing.

\begin{figure}[htb!]
     \centering
     \begin{subfigure}[b]{0.33\textwidth}
         \centering
         \includegraphics[width=\textwidth]{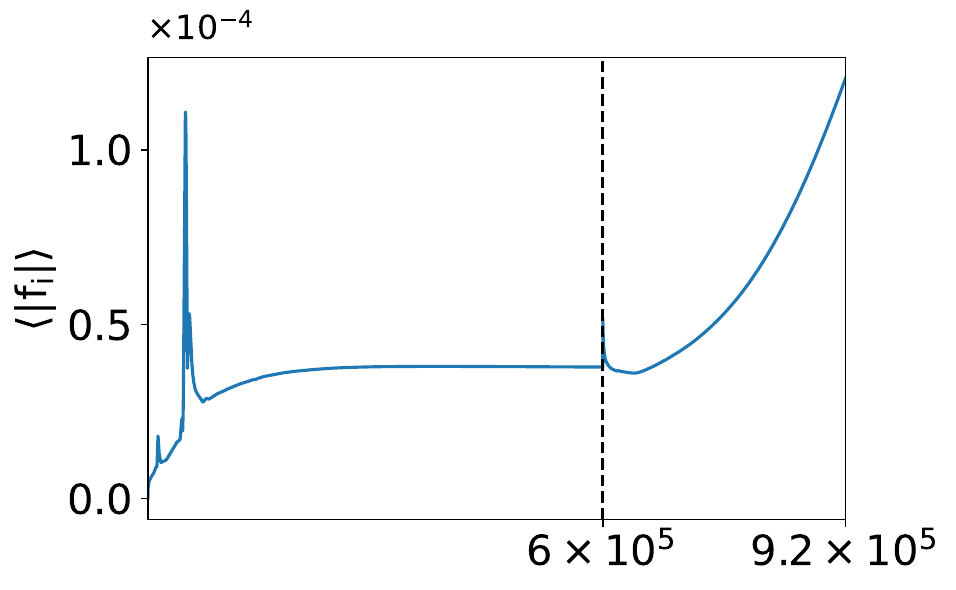}
         %\caption{$y=x$}
         %\label{fig:NLFFF_12297}
     \end{subfigure}%
     \begin{subfigure}[b]{0.33\textwidth}
         \centering
         \includegraphics[width=\textwidth]{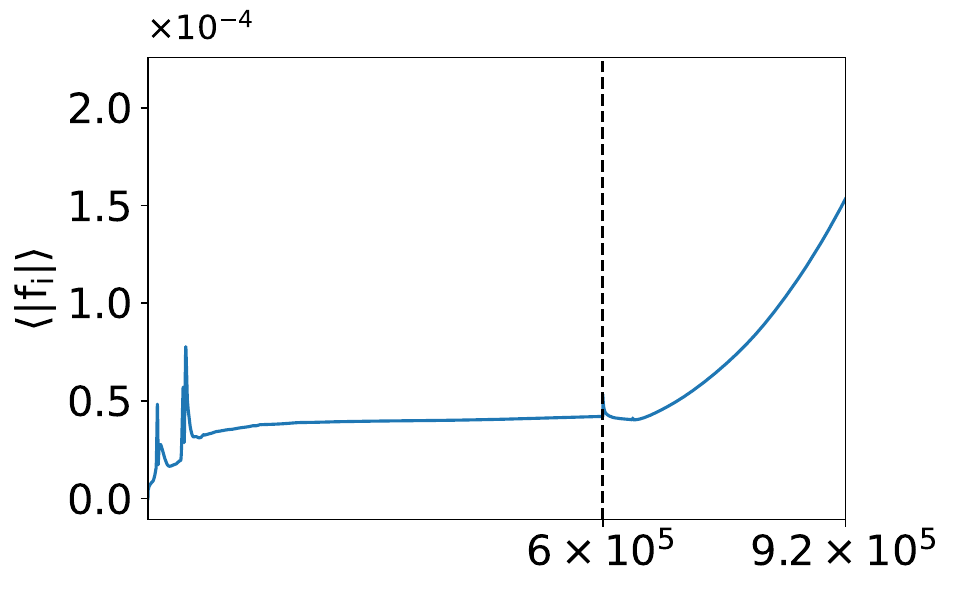}
         %\caption{$y=3\sin x$}
         %\label{fig:NLFFF_11692}
     \end{subfigure}%
     \begin{subfigure}[b]{0.32\textwidth}
         \centering
         \includegraphics[width=\textwidth]{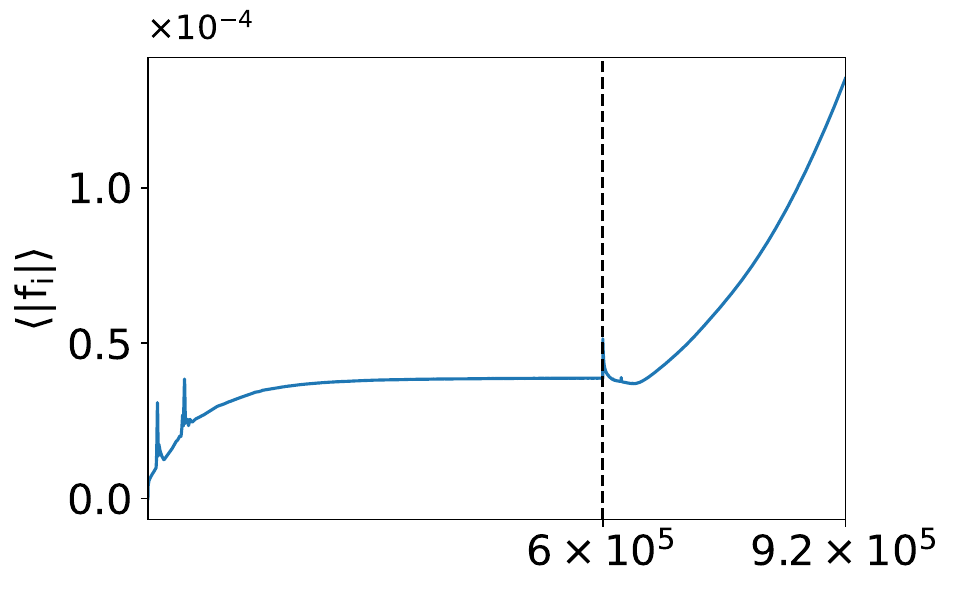}
         %\caption{$y=5/x$}
         %\label{fig:NLFFF_12158}
     \end{subfigure}
\vspace{-\lineskip}
    \begin{subfigure}[b]{0.33\textwidth}
     \centering
     \includegraphics[width=\textwidth]{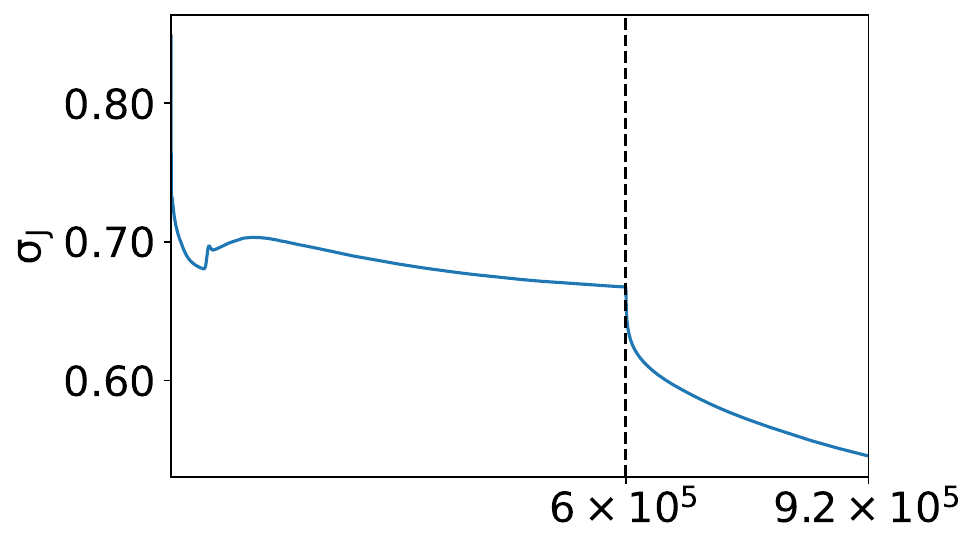}
     %\caption{$y=x$}
     %\label{fig:NLFFF_12297}
     \end{subfigure}%
     \begin{subfigure}[b]{0.33\textwidth}
         \centering
         \includegraphics[width=\textwidth]{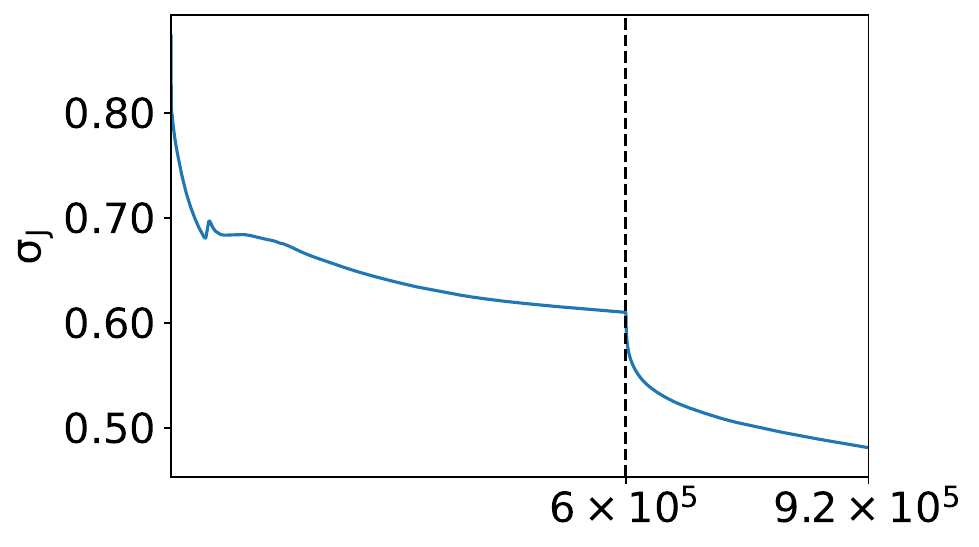}
         %\caption{$y=3\sin x$}
         %\label{fig:NLFFF_11692}
     \end{subfigure}%
     \begin{subfigure}[b]{0.33\textwidth}
         \centering
         \includegraphics[width=\textwidth]{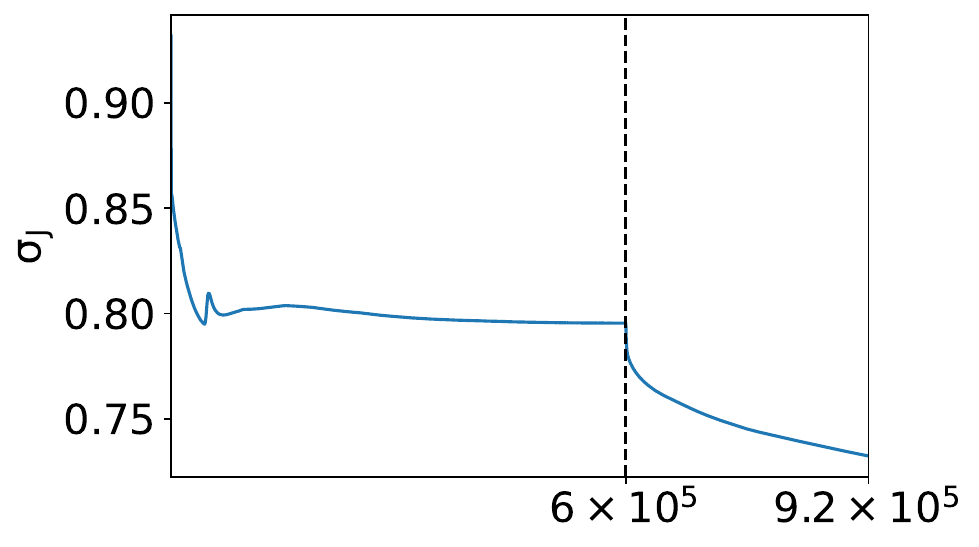}
         %\caption{$y=5/x$}
         %\label{fig:NLFFF_12158}
     \end{subfigure}
\vspace{-\lineskip}
    \begin{subfigure}[b]{0.33\textwidth}
     \centering
     \includegraphics[width=\textwidth]{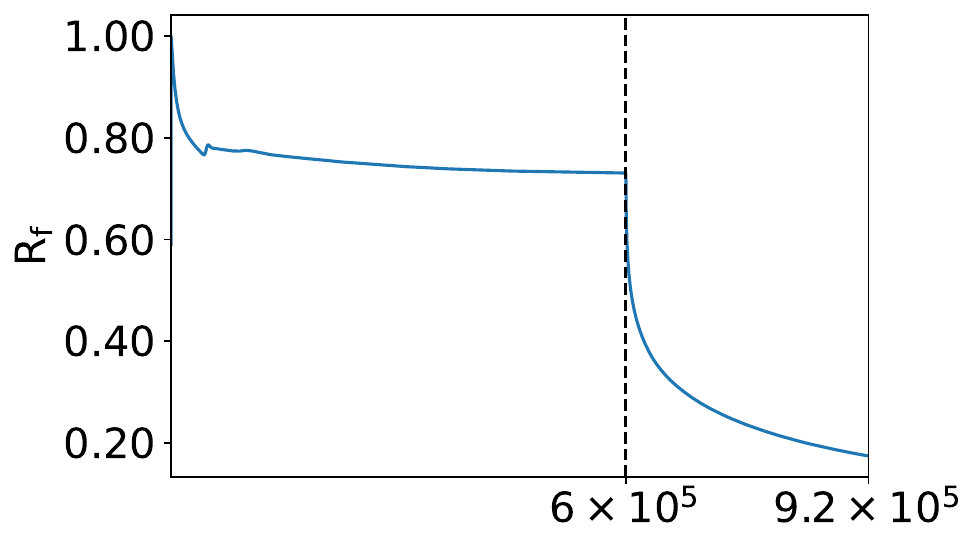}
     %\caption{$y=x$}
     %\label{fig:NLFFF_12297}
     \end{subfigure}%
     \begin{subfigure}[b]{0.33\textwidth}
         \centering
         \includegraphics[width=\textwidth]{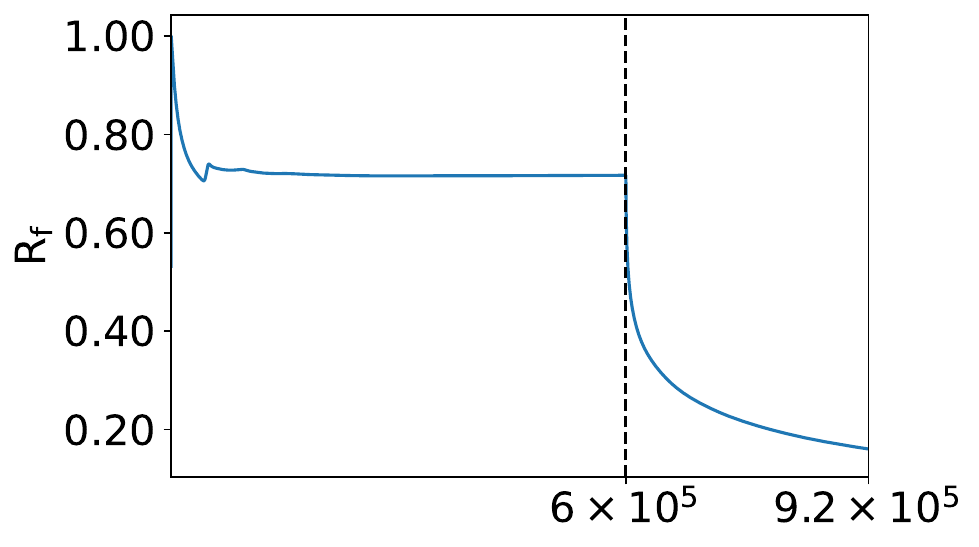}
         %\caption{$y=3\sin x$}
         %\label{fig:NLFFF_11692}
     \end{subfigure}%
     \begin{subfigure}[b]{0.33\textwidth}
         \centering
         \includegraphics[width=\textwidth]{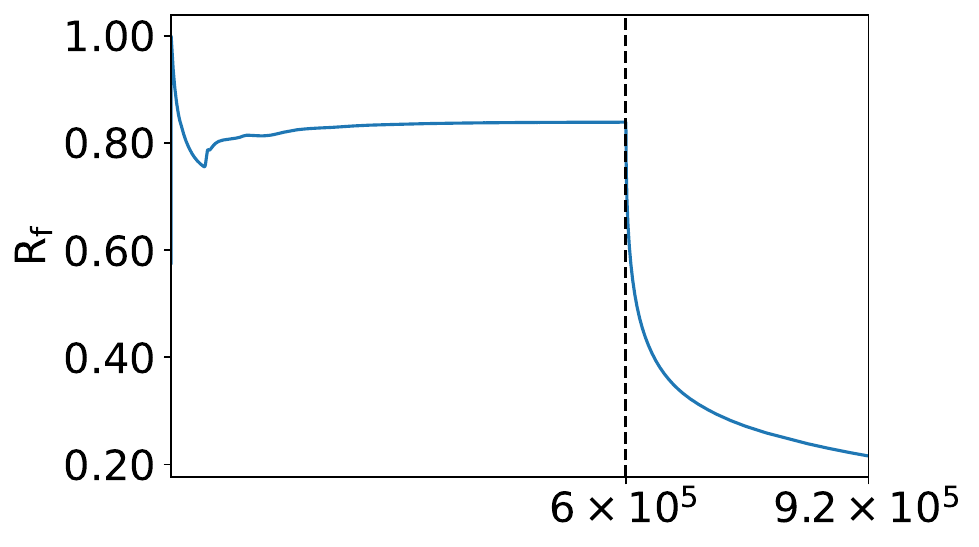}
         %\caption{$y=5/x$}
         %\label{fig:NLFFF_12158}
     \end{subfigure}
\vspace{-\lineskip}
    \begin{subfigure}[b]{0.33\textwidth}
     \centering
     \includegraphics[width=\textwidth]{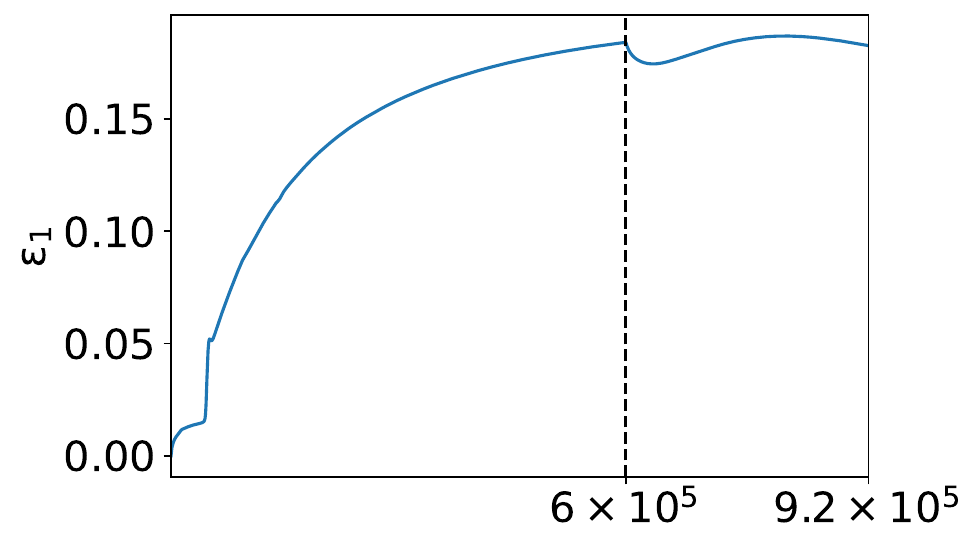}
     \caption{AR 12158 (2014 event).}
     \label{fig:NLFFF_12158_mfrun}
     \end{subfigure}%
     \begin{subfigure}[b]{0.33\textwidth}
         \centering
         \includegraphics[width=\textwidth]{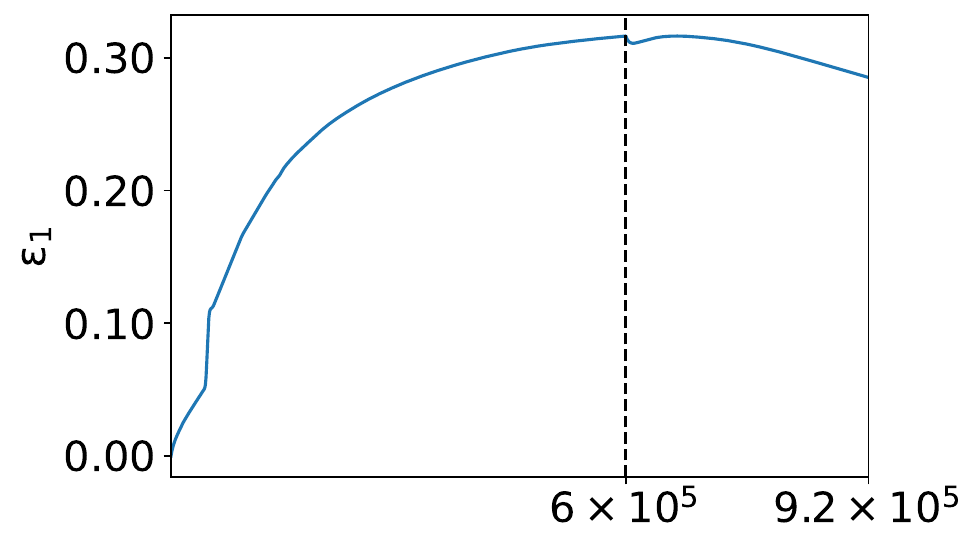}
         \caption{AR 12297 (2015 event).}
         \label{fig:NLFFF_12297_mfrun}
     \end{subfigure}%
     \begin{subfigure}[b]{0.33\textwidth}
         \centering
         \includegraphics[width=\textwidth]{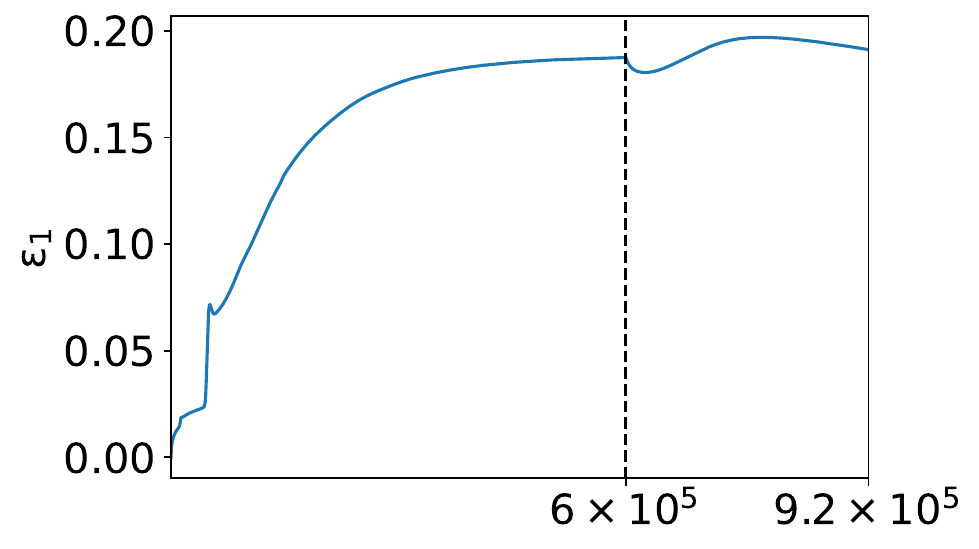}
         \caption{AR 11692 (2013 event).}
         \label{fig:NLFFF_11692_mfrun}
     \end{subfigure}
        \caption{Divergence-free metric $\langle|f_i|\rangle$, averaged sine of angle $\sigma_J$, normalized residual force $R_f$ and free energy $\epsilon_1$ as a function of iteration numbers for (a) AR 12158; (b) AR 12297 and (c) AR 11692. The dashed vertical line at $6\times 10^5$\,iterations separates the stressing stage from the relaxing stage.}
        \label{fig:NLFFF_mfrun}
\end{figure}

\begin{figure}[htb!]
     \centering
     \begin{subfigure}[b]{0.33\textwidth}
         \centering
         \includegraphics[width=\textwidth]{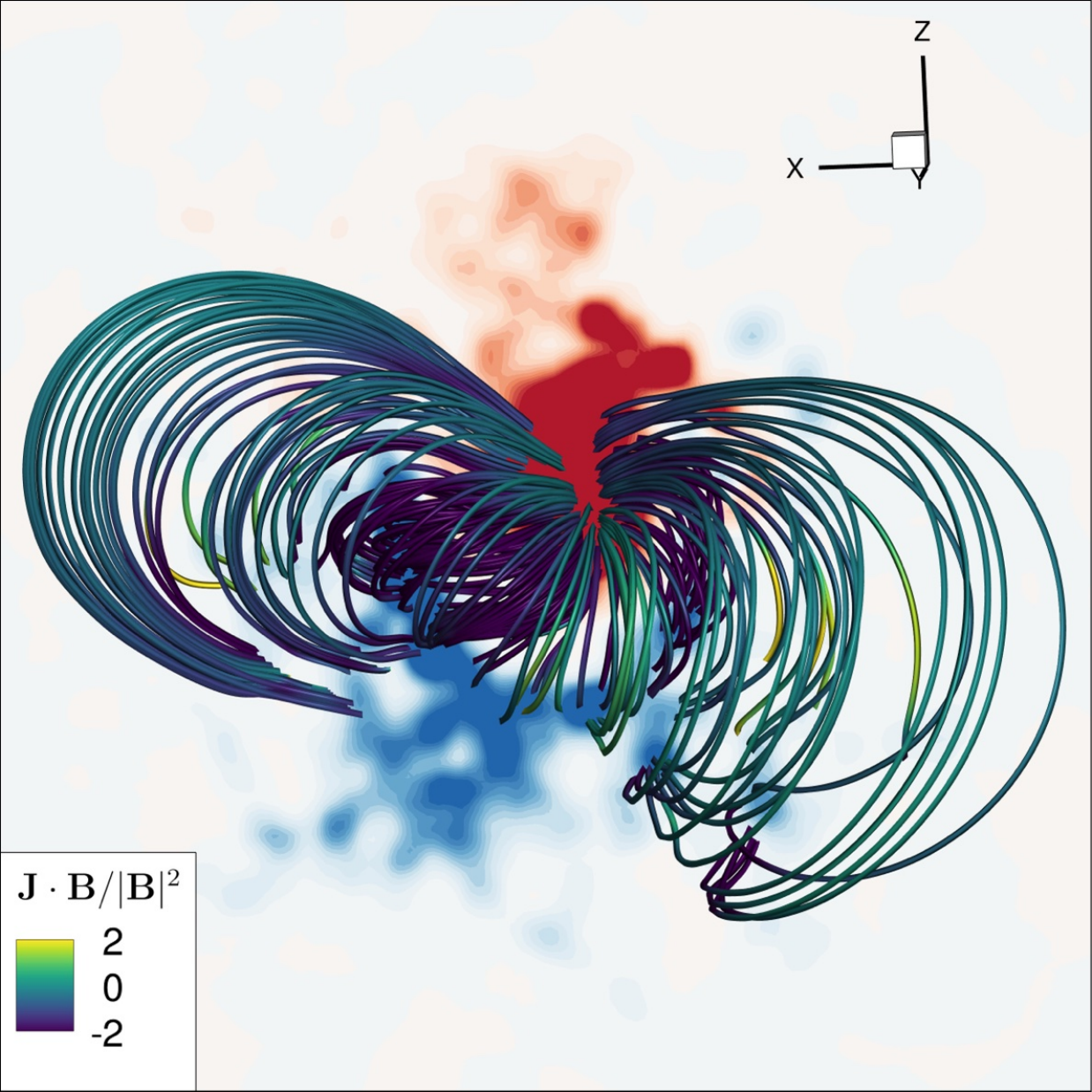}
         %\caption{$y=x$}
         %\label{fig:NLFFF_12297}
     \end{subfigure}%
     \begin{subfigure}[b]{0.33\textwidth}
         \centering
         \includegraphics[width=\textwidth]{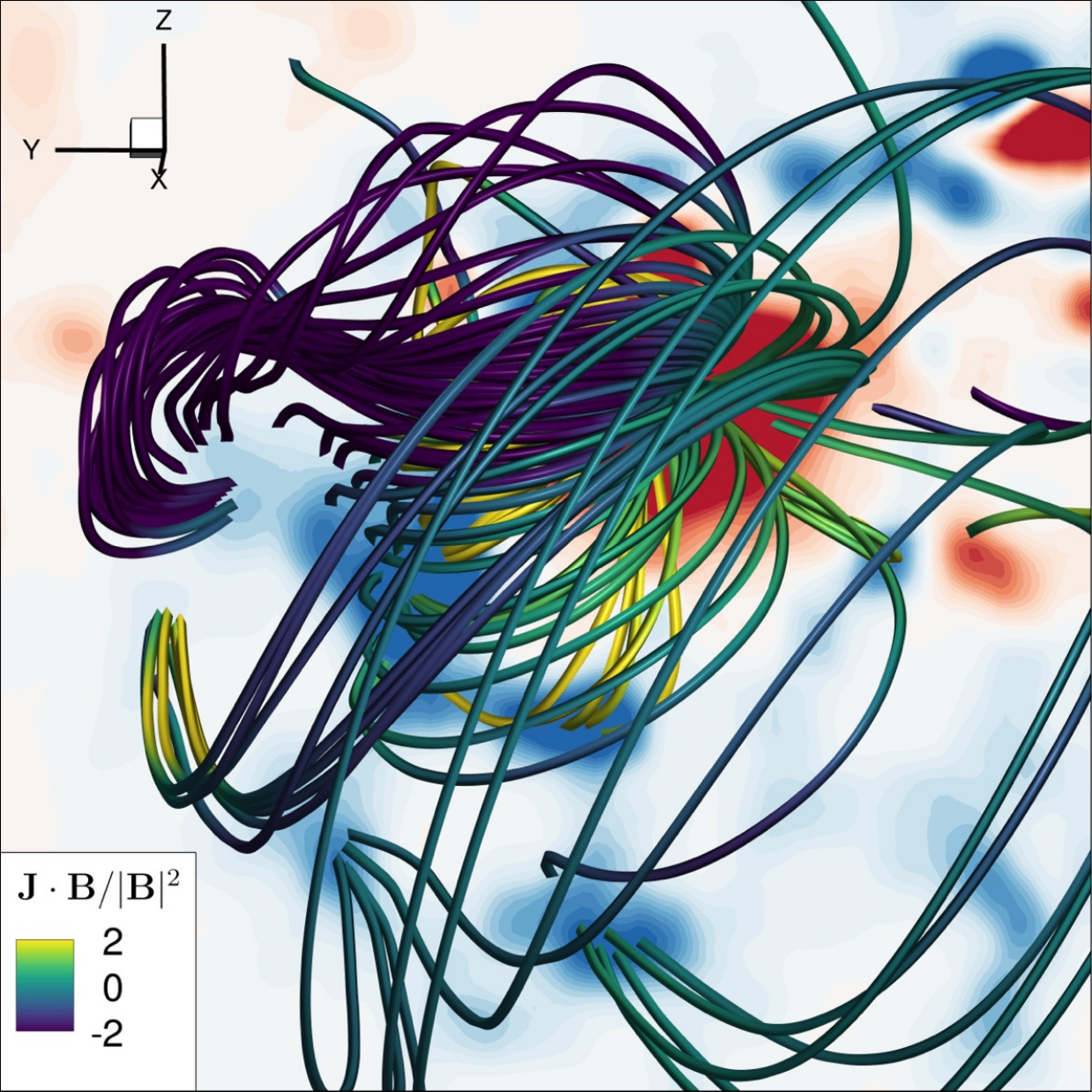}
         %\caption{$y=3\sin x$}
         %\label{fig:NLFFF_11692}
     \end{subfigure}%
     \begin{subfigure}[b]{0.33\textwidth}
         \centering
         \includegraphics[width=\textwidth]{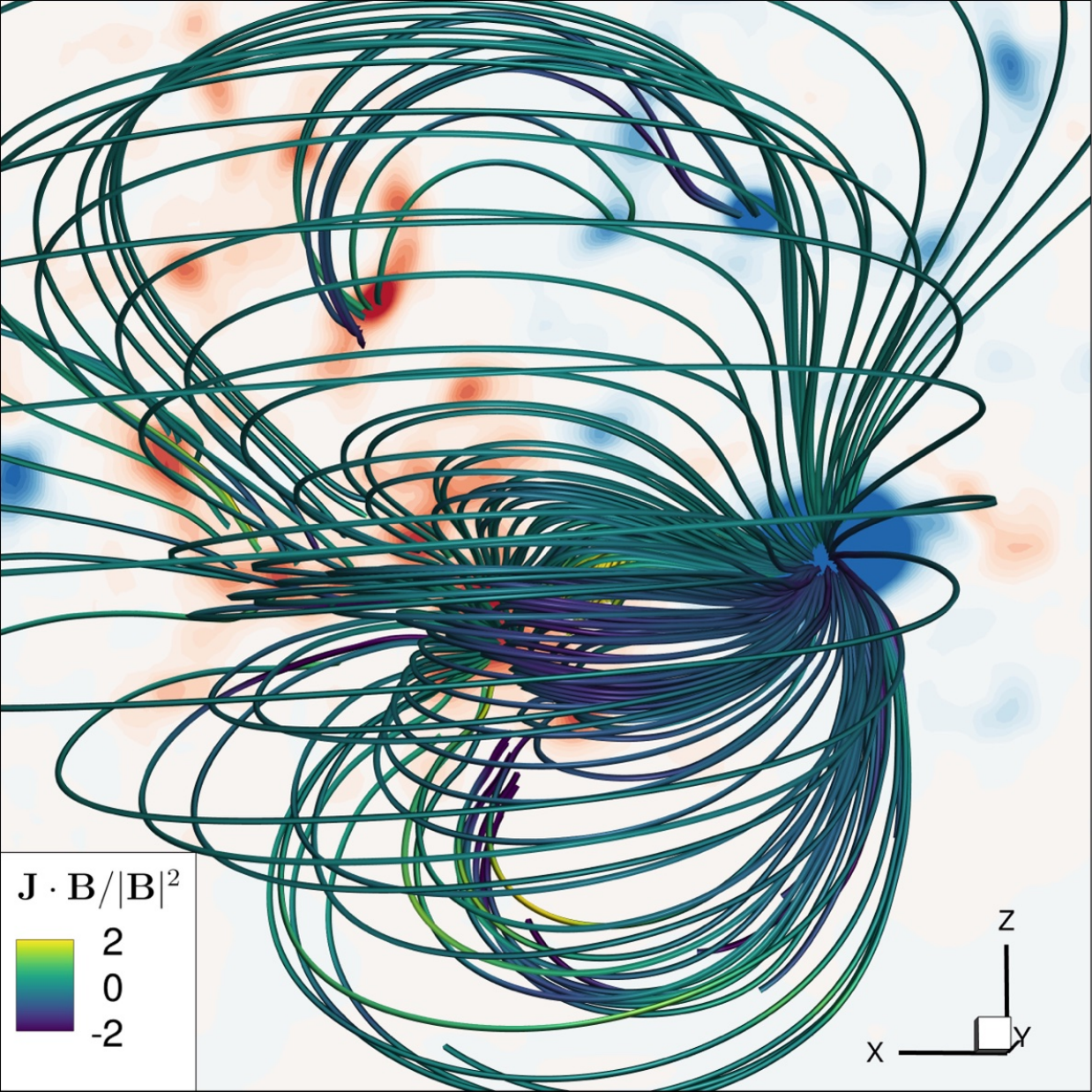}
         %\caption{$y=5/x$}
         %\label{fig:NLFFF_12158}
     \end{subfigure}
\vspace{-\lineskip}
     \begin{subfigure}[b]{0.33\textwidth}
         \centering
         \includegraphics[width=\textwidth]{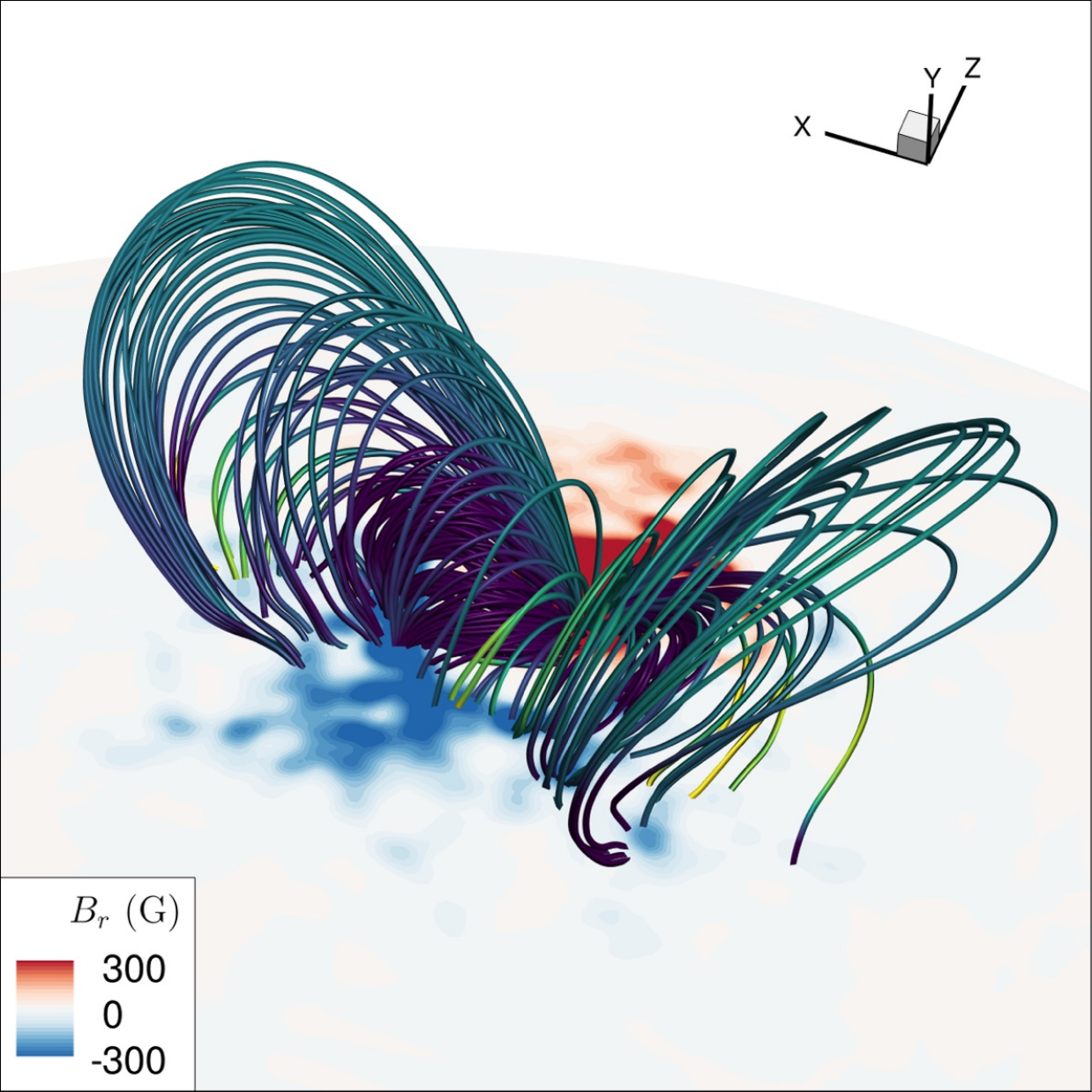}
         \caption{AR 12158 (2014 event).}
         \label{fig:NLFFF_12297}
     \end{subfigure}%
     \begin{subfigure}[b]{0.33\textwidth}
         \centering
         \includegraphics[width=\textwidth]{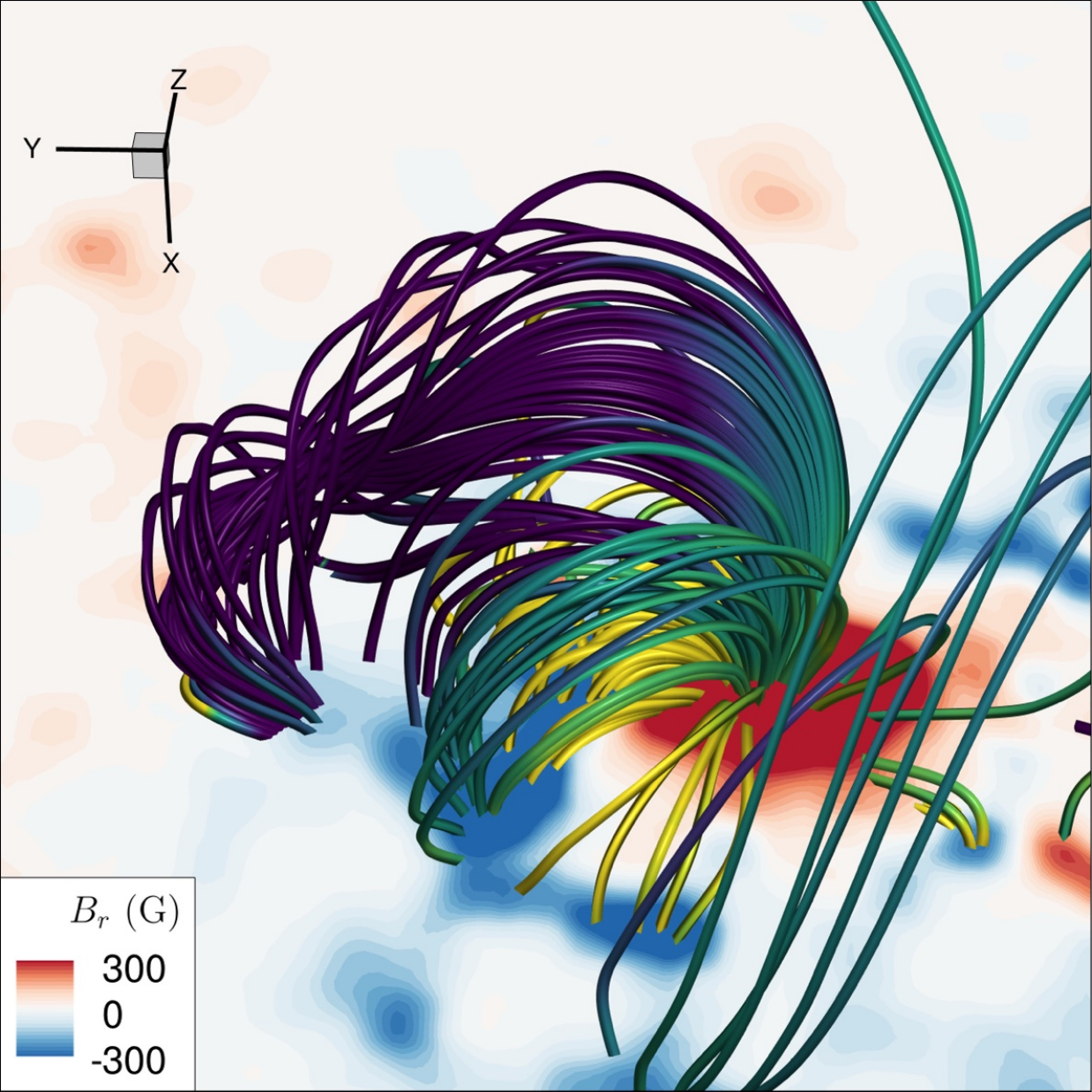}
         \caption{AR 12297 (2015 event).}
         \label{fig:NLFFF_11692}
     \end{subfigure}%
     \begin{subfigure}[b]{0.33\textwidth}
         \centering
         \includegraphics[width=\textwidth]{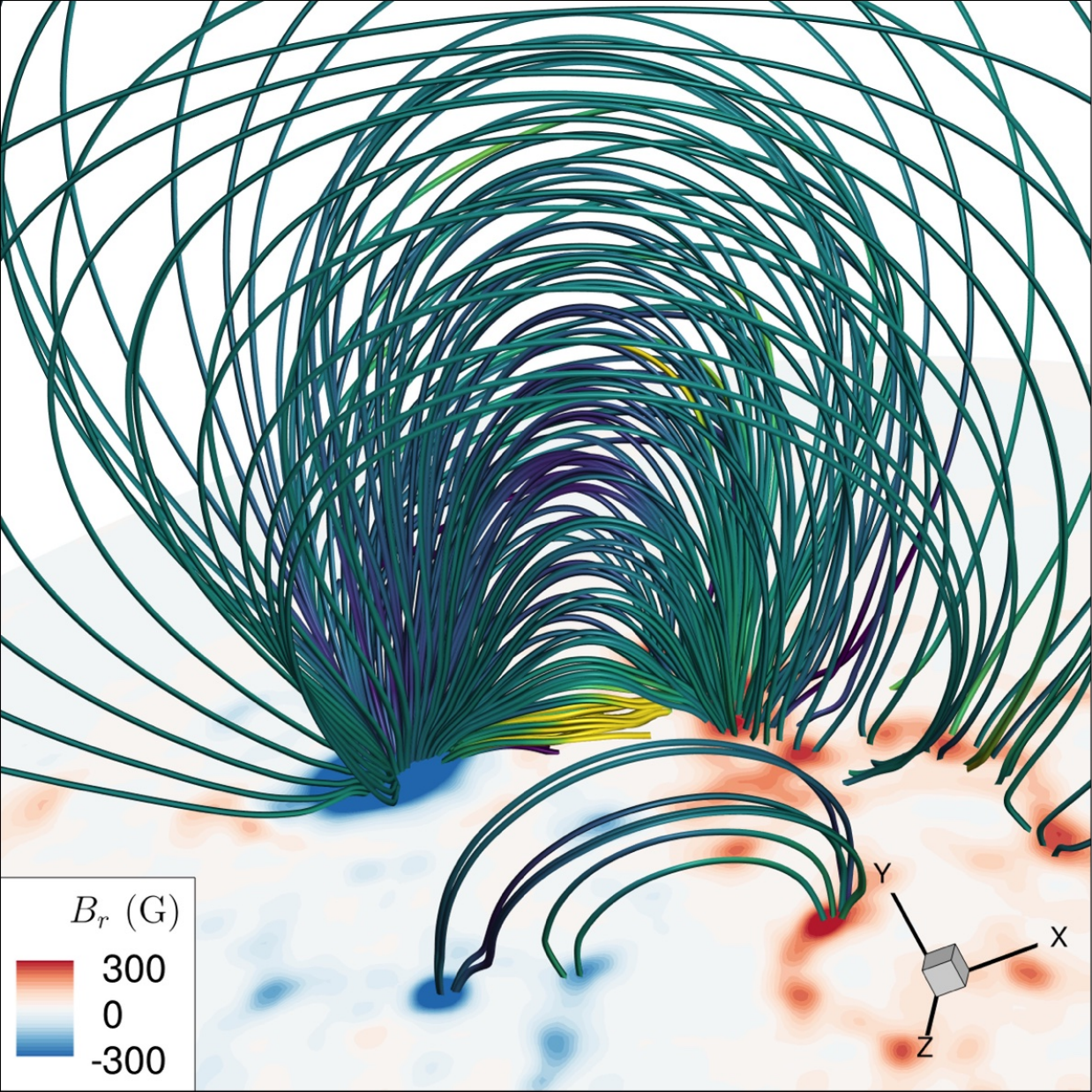}
         \caption{AR 11692 (2013 event).}
         \label{fig:NLFFF_12158}
     \end{subfigure}
        \caption{3-D representations of NLFFF solutions for (a) AR 12158, (b) AR 12297 and (c) AR 11692. The top row and the bottom row shows two different angles. The field lines are colored with the local twist number (up to a constant factor), $\mathbf J\cdot \mathbf B/|\mathbf B|^2$. The bottom boundary shows colors of $B_r$ saturated at $\pm 300$\,G. The windows are roughly $30^\circ$ in extent.}
        \label{fig:NLFFF_3D}
\end{figure}

\begin{deluxetable}{
ccccccc
}[htb!]
\tablecaption{$\langle|f_i|\rangle$, $\sigma_J$, $R_f$ and $\epsilon_1$ values at the end of the stressing and relaxing stages.\label{tab:NLFFF}}
\tablehead{
  \colhead{Event} & \colhead{AR} & \colhead{Stage} & \colhead{$<|f_i|>$} & \colhead{$\sigma_J$} & \colhead{$R_f$} & \colhead{$\epsilon_1$}
}
\startdata
\multirow{2}{*}{2014}
& \multirow{2}{*}{12158} & stressing & $3.77\times10^{-5}$ & 0.67 & 0.73 & 0.18 \\
&                        & relaxing  & $1.21\times10^{-4}$ & 0.55 & 0.17 & 0.18 \\
\hline
\multirow{2}{*}{2015}
  &\multirow{2}{*}{12297}& stressing & $4.20\times10^{-5}$ & 0.61 & 0.72 & 0.32 \\
  &                      & relaxing  & $1.53\times10^{-4}$ & 0.48 & 0.16 & 0.29 \\
  \hline
\multirow{2}{*}{2013}
& \multirow{2}{*}{11692}  & stressing & $3.87\times10^{-5}$ & 0.80 & 0.84 & 0.19 \\
&                         & relaxing  & $1.35\times10^{-4}$ & 0.73 & 0.22 & 0.19 \\
\enddata
\end{deluxetable}

\begin{figure}[htb!]
     \centering
     \begin{subfigure}[b]{0.35\textwidth}
         \centering
         \includegraphics[width=\textwidth]{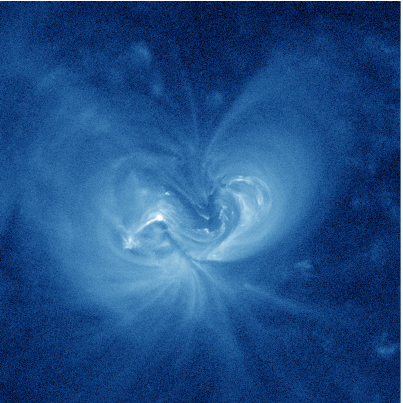}
         \caption{AIA 335\AA~observation of AR 12158.}
         \label{fig:NLFFF_AIA_12158}
     \end{subfigure}%
     \begin{subfigure}[b]{0.35\textwidth}
         \centering
         \includegraphics[width=\textwidth]{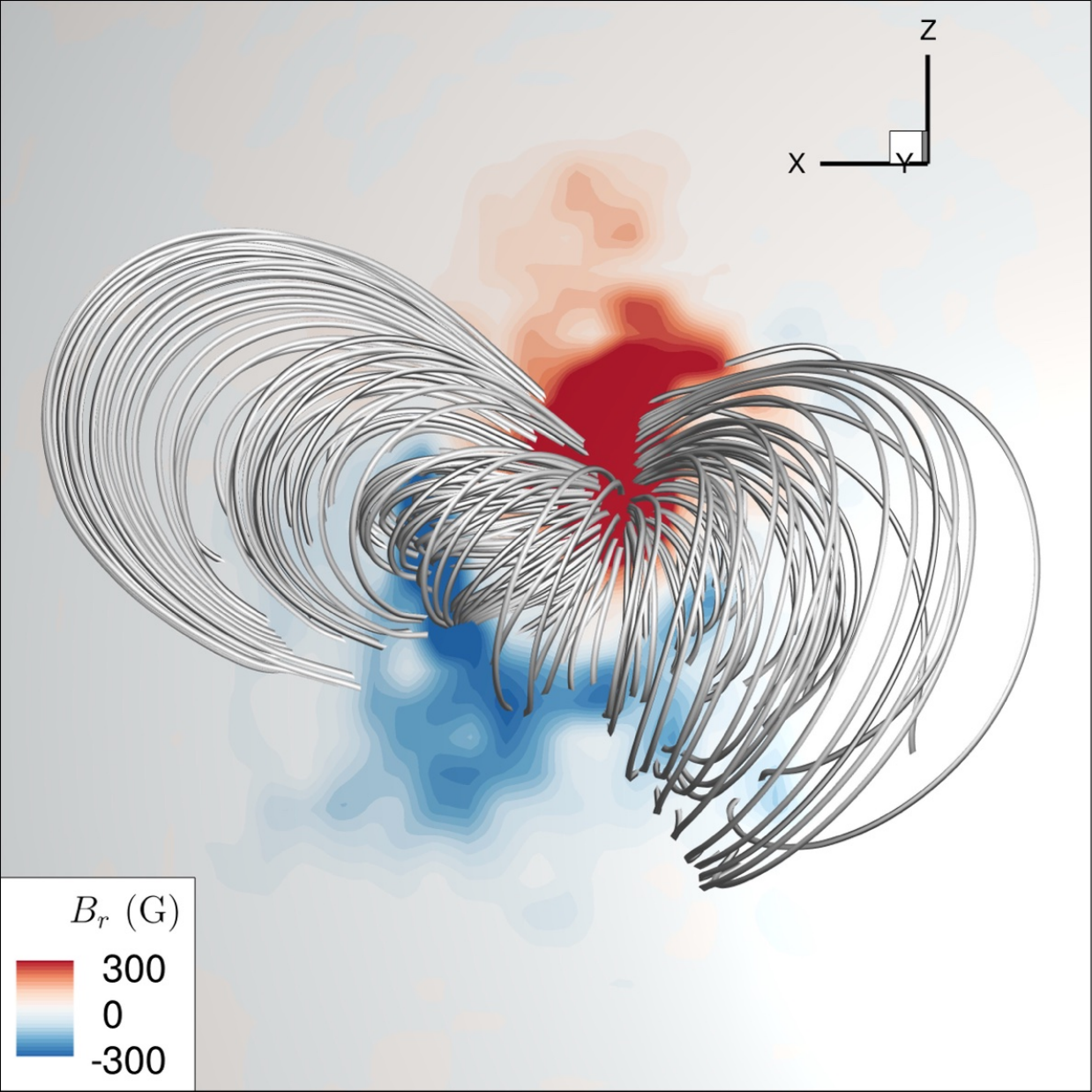}
         \caption{NLFFF field lines in AR 12158.}
         \label{fig:NLFFF_SIM_12158}
     \end{subfigure}
\vspace{-\lineskip}
          \begin{subfigure}[b]{0.35\textwidth}
         \centering
         \includegraphics[width=\textwidth]{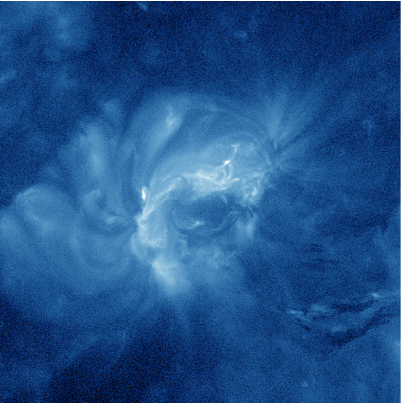}
         \caption{AIA 335\AA~observation of AR 12297.}
         \label{fig:NLFFF_AIA_12297}
     \end{subfigure}%
     \begin{subfigure}[b]{0.35\textwidth}
         \centering
         \includegraphics[width=\textwidth]{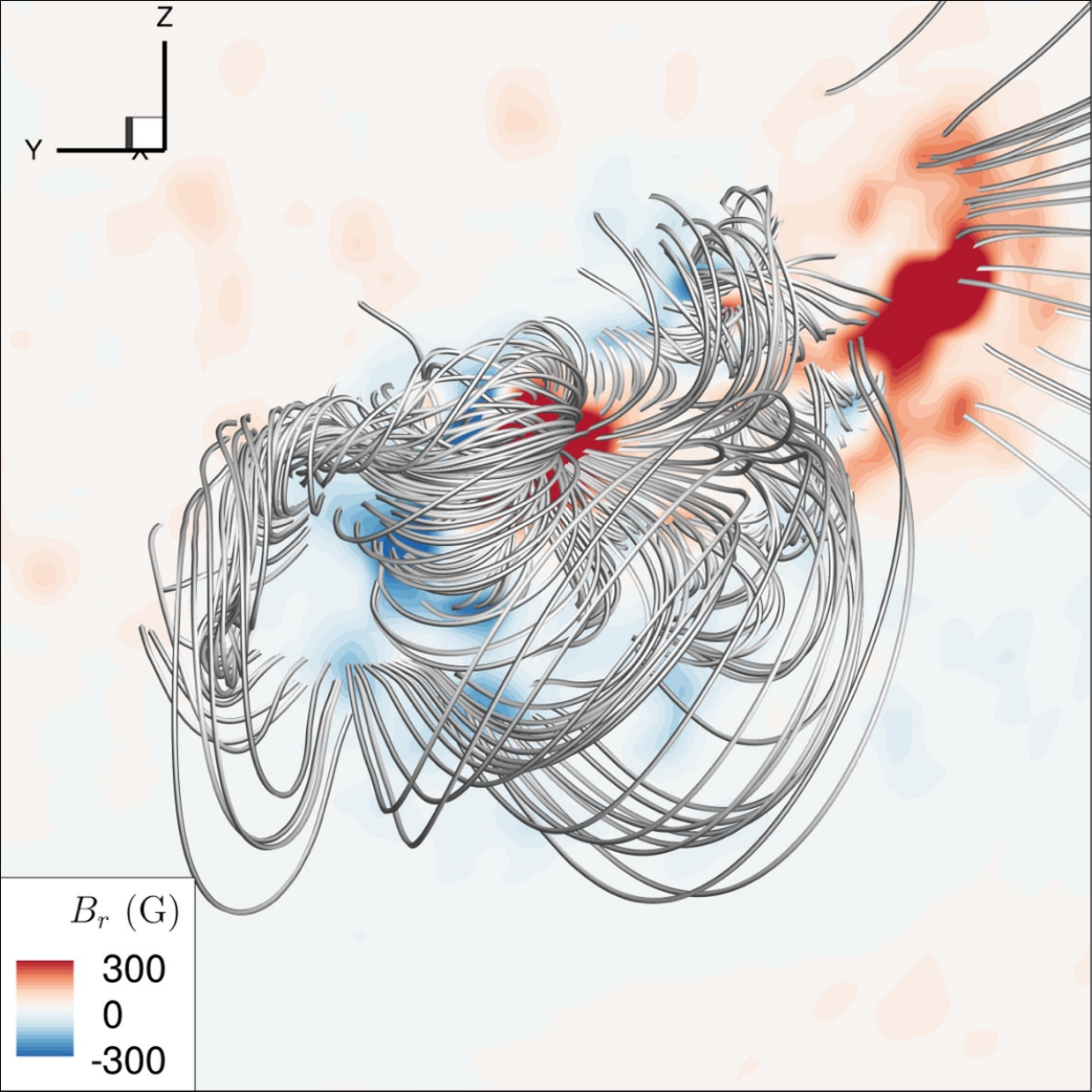}
         \caption{NLFFF field lines in AR 12297.}
         \label{fig:NLFFF_SIM_12297}
     \end{subfigure}
\vspace{-\lineskip}
        \begin{subfigure}[b]{0.35\textwidth}
         \centering
         \includegraphics[width=\textwidth]{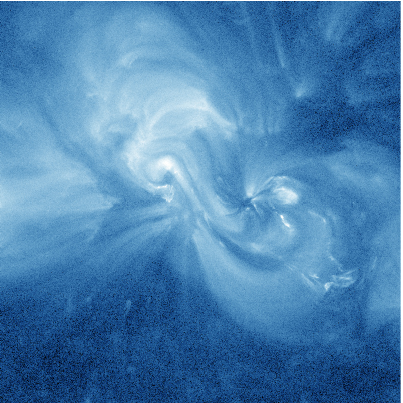}
         \caption{AIA 335\AA~observation of AR 11692.}
         \label{fig:NLFFF_AIA_11692}
     \end{subfigure}%
     \begin{subfigure}[b]{0.35\textwidth}
         \centering
         \includegraphics[width=\textwidth]{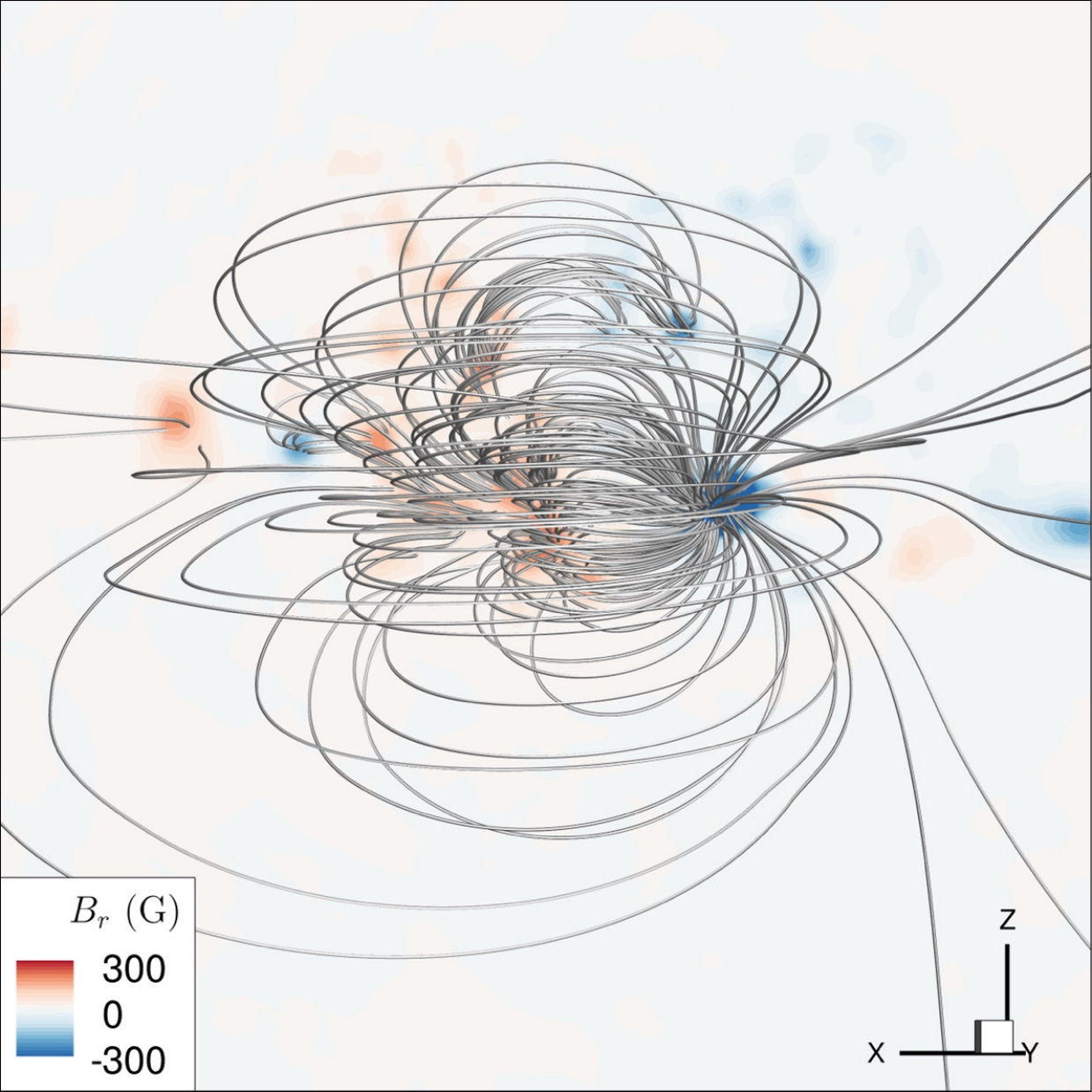}
         \caption{NLFFF field lines in AR 11692.}
         \label{fig:NLFFF_SIM_11692}
     \end{subfigure}
        \caption{Comparison between AIA 335\AA~observation and the NLFFF field lines for the three ARs. The observations are made at the same time as the vector magnetograms used as the boundary condition.}
        \label{fig:NLFFF_OBS_SIM}
\end{figure}

\begin{figure}[htb!]
     \centering
     \begin{subfigure}[b]{0.33\textwidth}
         \centering
         \includegraphics[width=\textwidth]{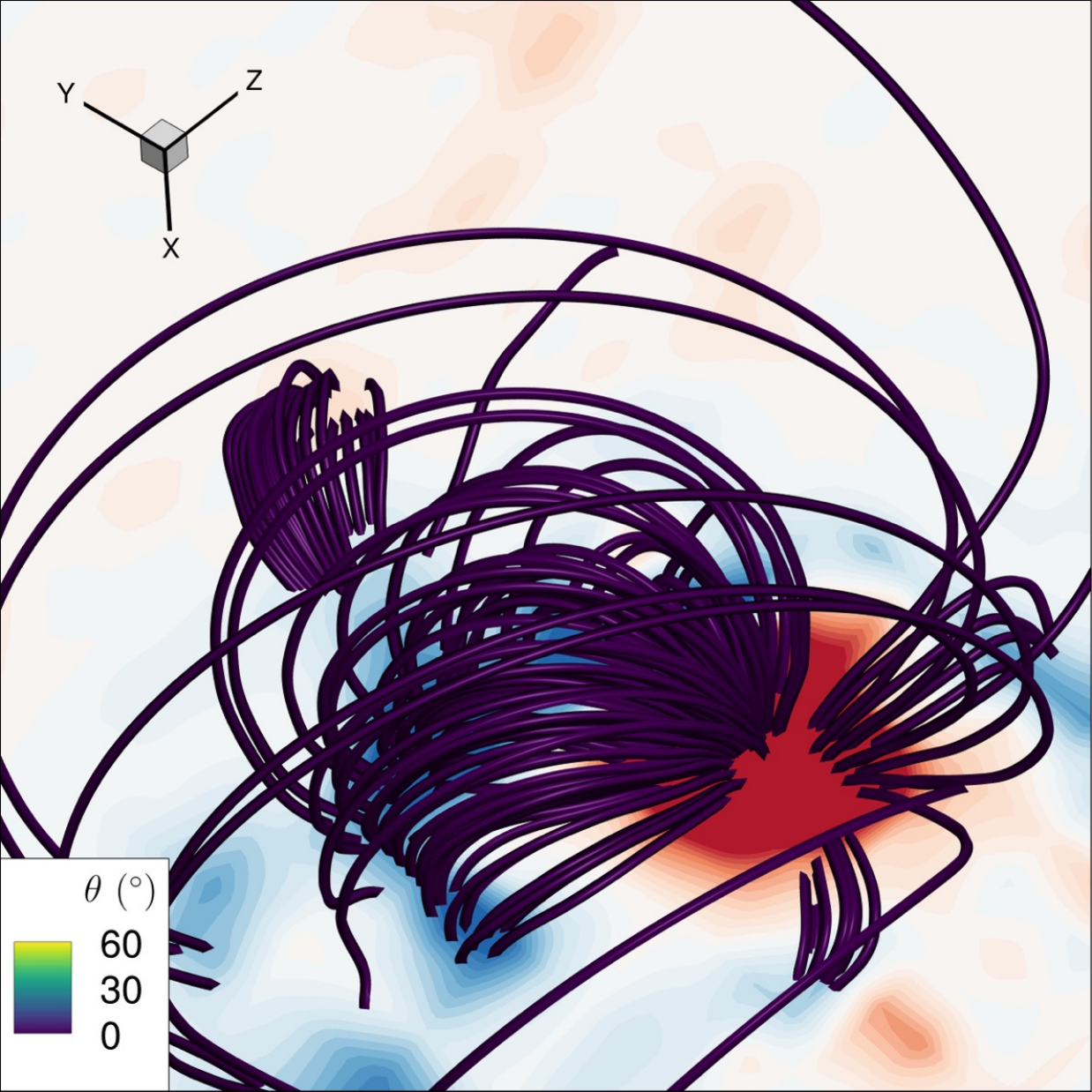}
         %\caption{$y=x$}
         %\label{fig:NLFFF_12297}
     \end{subfigure}%
     \begin{subfigure}[b]{0.33\textwidth}
         \centering
         \includegraphics[width=\textwidth]{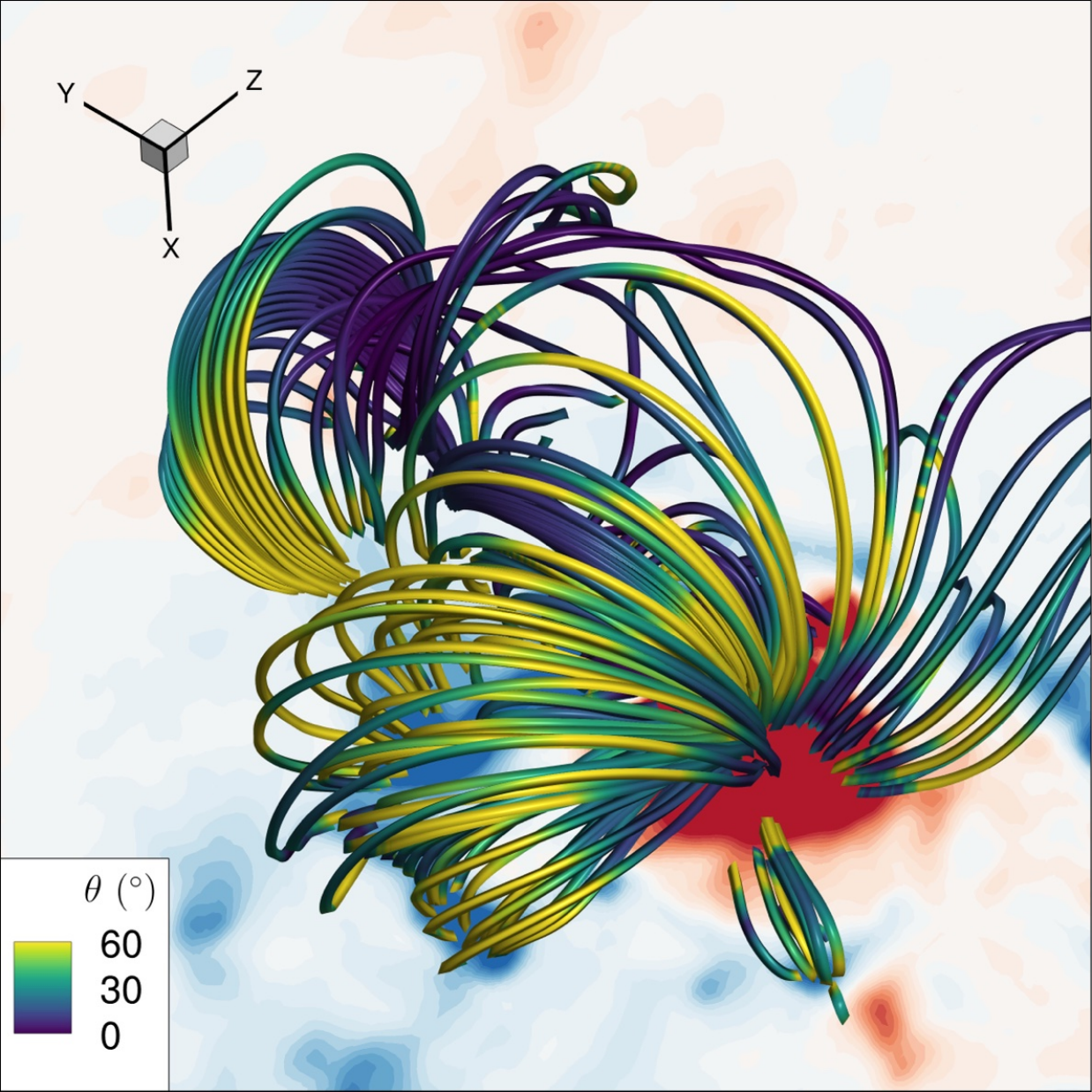}
         %\caption{$y=3\sin x$}
         %\label{fig:NLFFF_11692}
     \end{subfigure}%
     \begin{subfigure}[b]{0.33\textwidth}
         \centering
         \includegraphics[width=\textwidth]{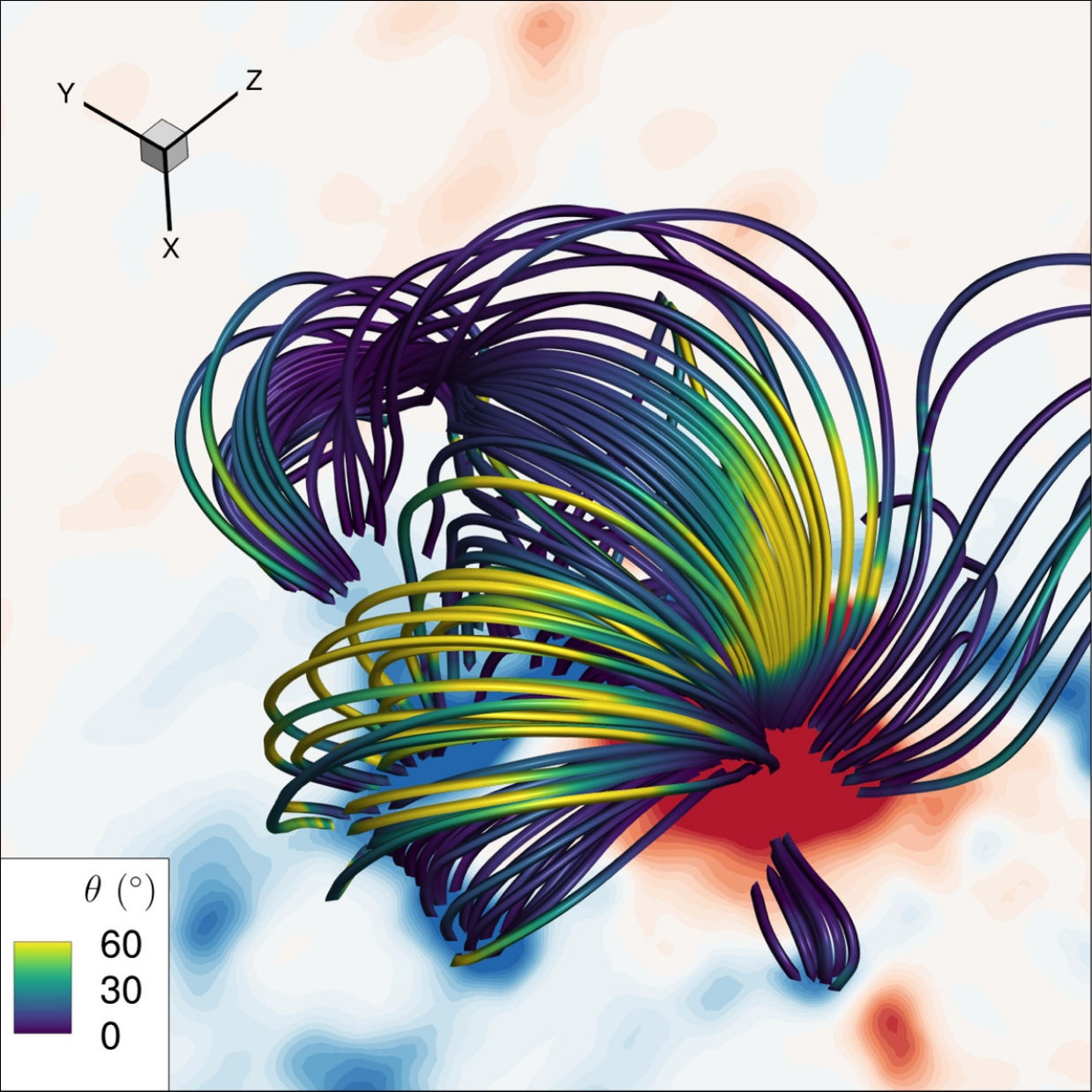}
         %\caption{$y=5/x$}
         %\label{fig:NLFFF_12158}
     \end{subfigure}
 \vspace{-\lineskip}
     \begin{subfigure}[b]{0.33\textwidth}
         \centering
         \includegraphics[width=\textwidth]{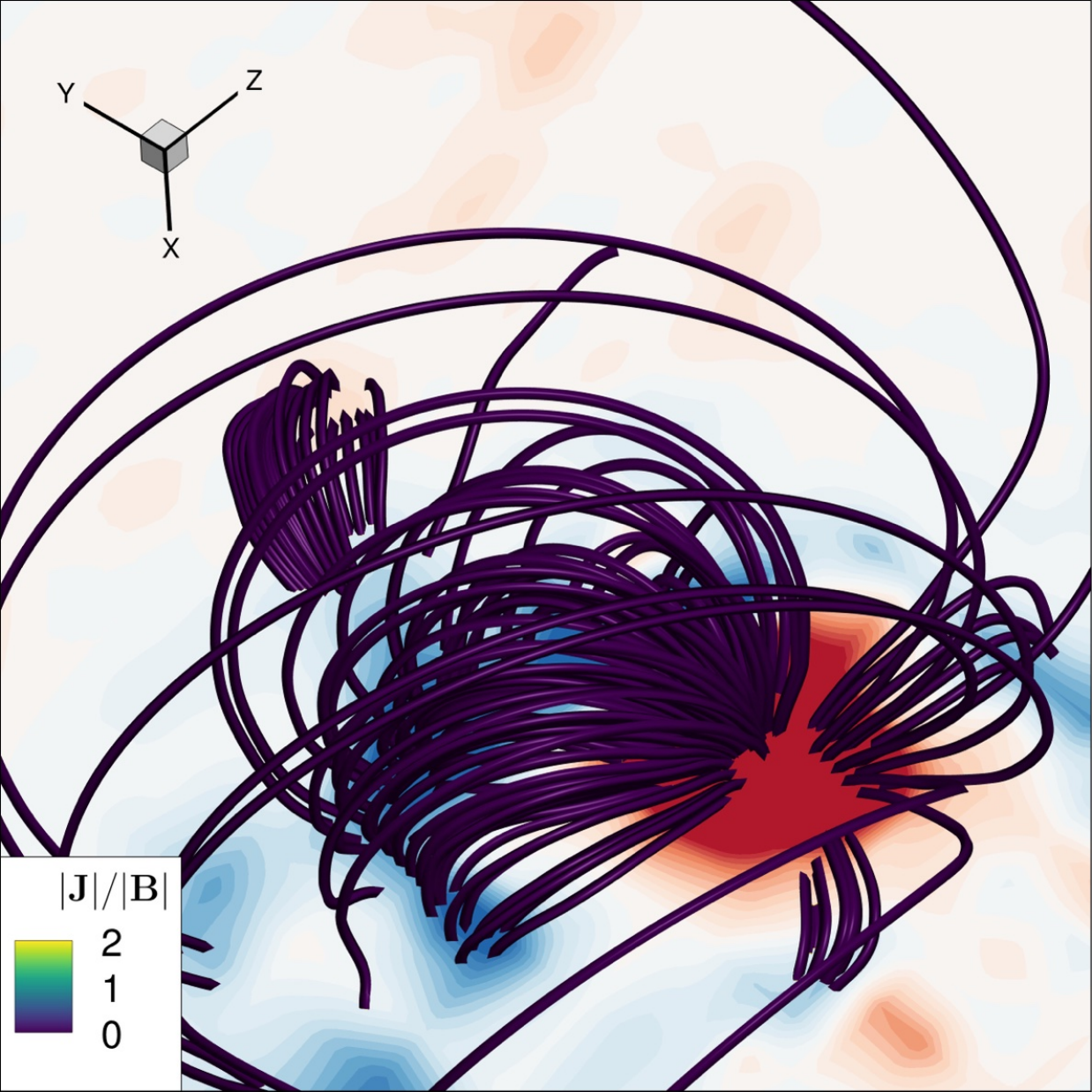}
         \caption{Initial condition, $\mathbf B_0$.}
         \label{fig:relaxation_1}
     \end{subfigure}%
     \begin{subfigure}[b]{0.33\textwidth}
         \centering
         \includegraphics[width=\textwidth]{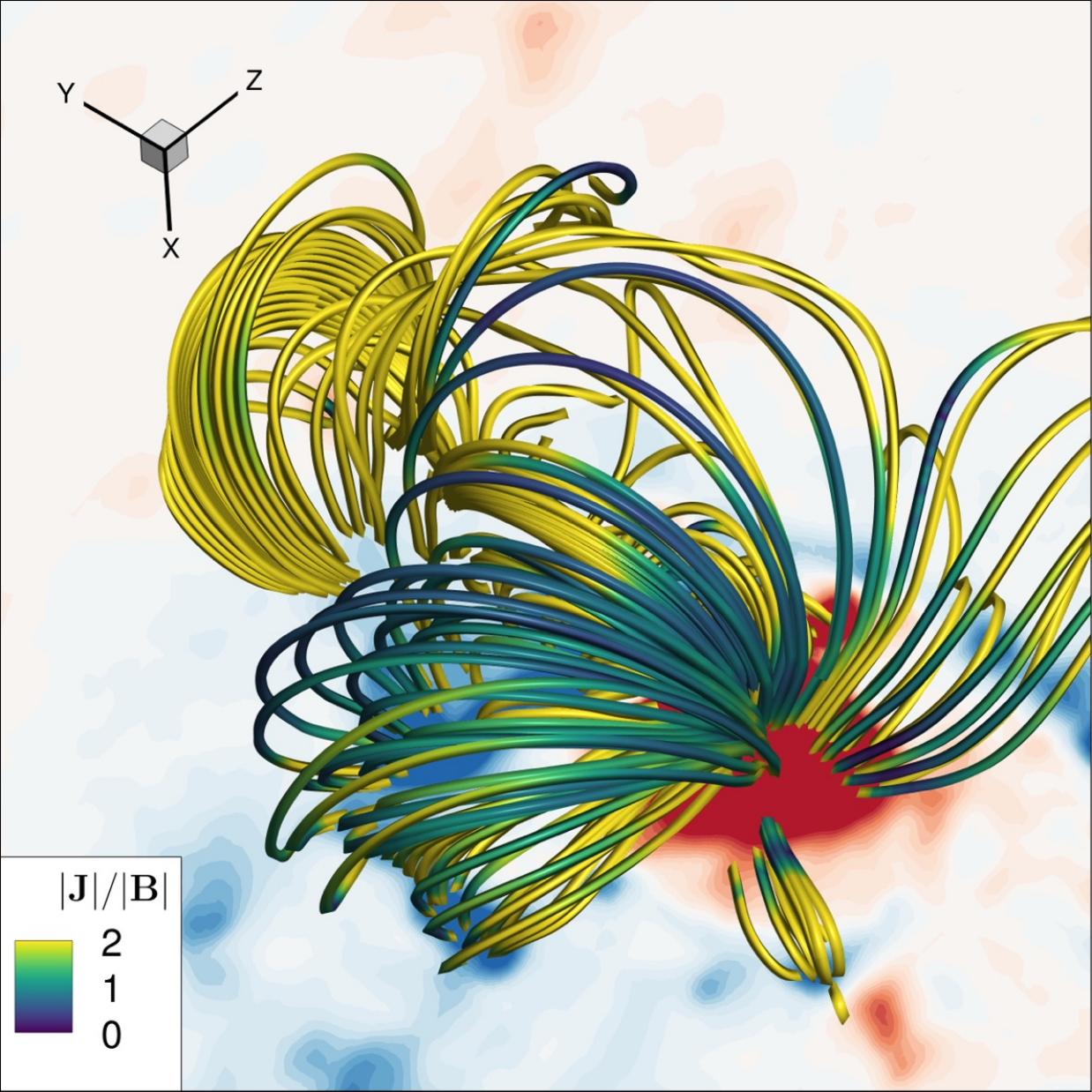}
         \caption{$6\times10^5$ steps. }
         \label{fig:relaxation_2}
     \end{subfigure}%
     \begin{subfigure}[b]{0.33\textwidth}
         \centering
         \includegraphics[width=\textwidth]{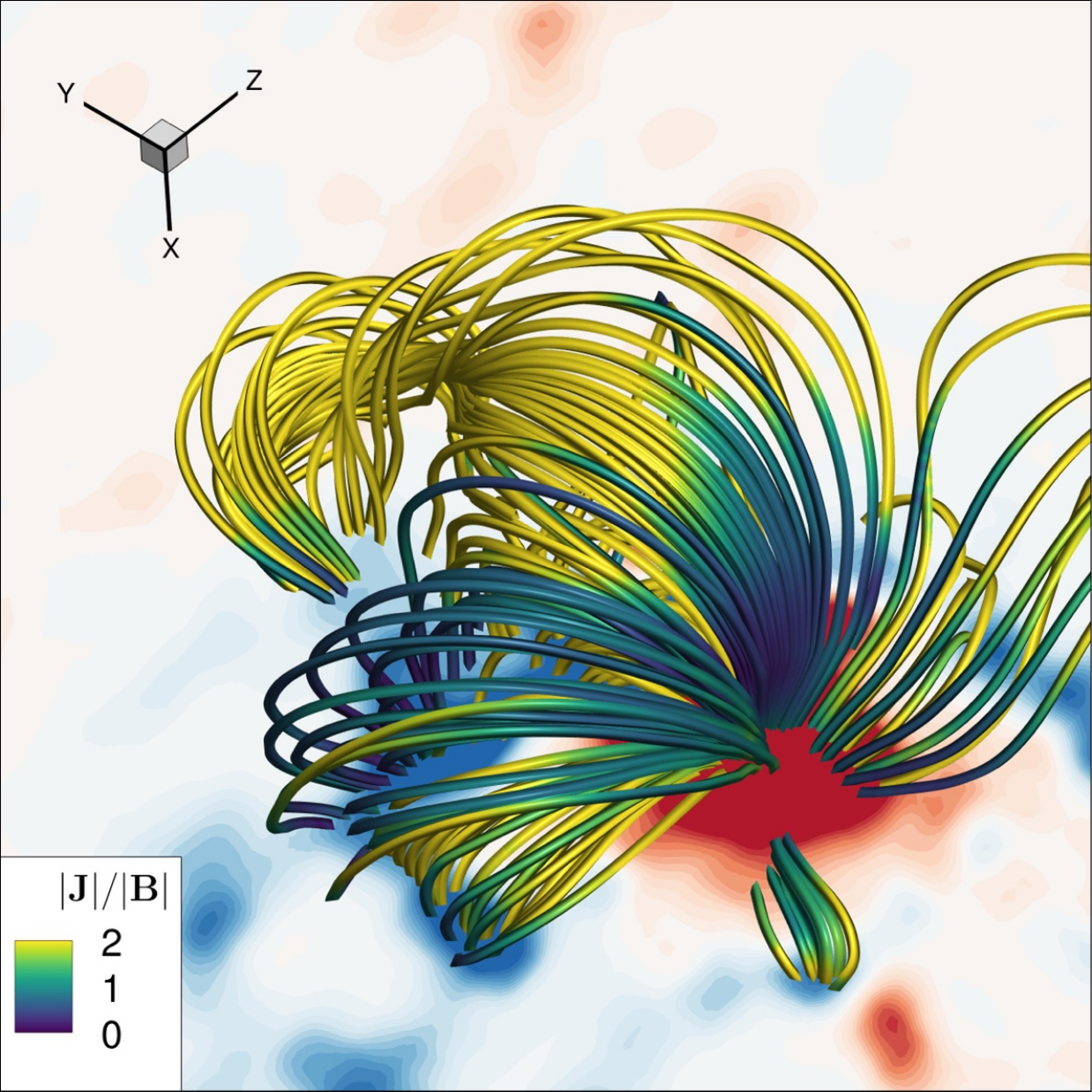}
         \caption{$9.2\times10^5$ steps.}
         \label{fig:relaxation_3}
     \end{subfigure}
        \caption{Evolution of the NLFFF solution in AR 12297 at 3 instants: (a) the initial condition, (b) the end of the stressing stage and (c) the end of the relaxing stage. The top row shows field lines colored by $\theta$ and the bottom row shows the same field lines colored by $|\mathbf J|/|\mathbf B|$.}
        \label{fig:NLFFF_relaxation}
\end{figure}

It is helpful to define a few quantities and discuss their mathematical properties.
The twist number along a magnetic field line, $T_w$, is given by \citep{Berger_2006,Liu_2016}
\begin{equation}
    \label{eqn_twist_number}
    T_w = %\frac{1}{4\pi}\int_L \frac{(\nabla \times \mathbf{B})\cdot \mathbf B}{|\mathbf B|^2}dl = 
    \frac{\mu_0}{4\pi} \int_L\frac{\mathbf J\cdot \mathbf B}{|\mathbf B|^2}dl,
\end{equation}
where $L$ is the path along a field line and $\mu_0$ is the vacuum permeability.
Let $t_w$ denote the integrand
\begin{equation}
t_w = \frac{\mathbf J\cdot \mathbf B}{|\mathbf B|^2},
\end{equation}
then $t_w$ is equivalent to the local twist number in unit $\mu \mathrm A/(\mathrm G \cdot\mathrm m^2)$.
Furthermore, in a perfect NLFFF solution $\mathbf J || \mathbf B$, so we have $t_w=\mu_0\alpha=\pm J/B$ from equation \ref{eqn_force_free_2}.
Taking the divergence of equation \ref{eqn_force_free_2}, we have $\nabla\alpha \cdot \mathbf B=0$, which suggests that $\alpha$ remains constant along field lines in a NLFFF.
Therefore, $t_w$ measures the strength of the field-aligned current and is an indicator of the force-freeness: constant, non-zero $t_w$ along field lines is a necessary condition for the local field to be force-free, whereas varying $t_w$ along field lines indicates the poor quality of the local NLFFF solution.

%The central region of the enlarged SHARP patch, which is a rectangular area of side length $20^\circ-30^\circ$, is resolved at $0.175^\circ$.
%The high-resolution region extends from 1\,$R_\odot$ to 1.25 \,$R_\odot$ in height.
%The rest of the patch is resolved at $0.35^\circ$ and extend to 1.75\,$R_\odot$.
To obtain the NLFFF solutions, we solve equations \ref{eqn_velocity_ours}-\ref{eqn_induction_ours} for the three vector magnetograms.
The stressing stage takes 600,000 steps, and the relaxing stage takes 320,000 steps.
The magnetic field at the end of the relaxing stage is used for subsequent simulations.
This process and the NLFFF solutions are demonstrated by 
Figures \ref{fig:NLFFF_3D}-\ref{fig:NLFFF_relaxation} and Table \ref{tab:NLFFF}.

Figure \ref{fig:NLFFF_3D} shows the 3-D NLFFF solutions for the three ARs.
For all three ARs, the solutions contain complex magnetic topologies.
In AR 12158, a low-lying MFR is enclosed by sheared magnetic loops.
Viewed from the top, they roughly resemble the sigmoid shape of the same active region found by \citet{Kilpua2021}.
In AR 12297, a MFR overlaps twisted field lines of opposite chirality over the PIL.
In AR 11692, magnetic loops dominate, but we find twisted field near the bottom (yellow) and in the center (blue).
%A comparison for AR 11692 is not included, as the agreement is not as good.

Figure \ref{fig:NLFFF_mfrun} shows the evolution of $\langle|f_i|\rangle$, $\sigma_J$, $R_f$, and $\epsilon_1$, defined in equations \ref{eqn_metric_div} through \ref{eqn_metric_emagnp}, during magneto-frictional runs.
At the end of the stressing stage, the four metrics converge or nearly converge.
In the relaxing stage, the solution loses most of the residual force $R_f$ while retaining most of the extra magnetic energy$\epsilon_1$.
The divergence $\mathbf B$ metric $\langle|f_i|\rangle$ increases by a factor of three but remains small, while the alignment metric $\sigma_J$ further decreases.
Note that in the relaxing stage, the free energy briefly increases before decreasing, which is probably due to transients caused by the sudden change at the bottom boundary.
%This is because the solution has not fully converged at the end of the stressing stage, as the rate of convergence diminishes with increasing step numbers.
%The solution in the computational domain not yet affected by the relaxed boundary condition continues to develop.
%Another reason is that the relaxed boundary condition in the relaxing stage, i.e. field lines are allowed to bend or twist in the relaxing stage due to the residual Lorentz force, may occasionally lead to larger tangential field.
%The free energy may increase as a result of the increased field.

Table \ref{tab:NLFFF} reports the values of these metrics at the end of the stressing stage and the relaxing stage.
The solutions have a divergence-free metric of $1.21$-$1.53\times10^{-4}$, an averaged sine of 0.48, 0.73, and 0.55, 16\%-22\% of the maximum residual force, and free energy 0.18-0.29 times the potential field energy.
In \citet{Guo:2016II}, $\langle|f_i|\rangle$ is $8.0$-$9.9\times10^{-4}$ and $\sigma_J$ is 0.51-0.52 for simulations on a spherical grid with AMR and a maximum resolution of $0.06^\circ$.
Compared to \citet{Guo:2016II}, we have better $\langle|f_i|\rangle$ metrics and similar $\sigma_J$ for AR 12158 and AR 12297, although we use a lower effective resolution of $0.088^\circ$.
The exception is $\sigma_J=0.73$ for AR 11692, which may be due to the topology of the active region that is difficult to model.
AR 11692 (see Figure \ref{fig:VM_11692}) features a single strong negative pole that concentrates most of the twist and loosely distributed positive poles with little twist.
The magneto-frictional method struggles to find a NLFFF solution given this topology, as the resolution we achieve is still not sufficient to resolve the strong and twisted pole.
Despite the large $\sigma_J$, the method builds up free energy in the solution for AR 11692, setting it up for a CME simulation.

Figure \ref{fig:NLFFF_OBS_SIM} compares AIA 335\,\AA~observations \citep{Lemen2012} and the field lines in the NLFFF solutions for the three active regions.
We find reasonably similar field line structures for AR 12158 and AR 12297 but are unable to reproduce the sigmoid shape found in the observations for AR 11692, which has the worst quality NLFFF in terms of $\sigma_J$.

Figure \ref{fig:NLFFF_relaxation} illustrates the evolution of the NLFFF field topology during magneto-frictional runs.
The field lines in the two rows are colored with two quantities that provide a comprehensive illustration of the evolution of the NLFFF.
The field lines in the upper row are colored with the angle between the local $\mathbf J$ and $\mathbf B$ ($\theta$).
Those in the bottom row are colored with the current per magnetic field fraction $|\mathbf J|/|\mathbf B|$.
The comparison of figures \ref{fig:relaxation_2} and \ref{fig:relaxation_3} to \ref{fig:relaxation_1} shows the drastic difference between the initial potential field and the NLFFF.
Comparing figures \ref{fig:relaxation_2} and \ref{fig:relaxation_3}, we notice that the relaxing stage greatly reduces $\theta$ close to the bottom boundary.
This is consistent with the rapid decrease of $R_f$ in the relaxing stage in figure \ref{fig:NLFFF_mfrun}.
In figure \ref{fig:relaxation_3}, $\theta$ is large for a few field lines, but those field lines do not carry large current, so they do not have a large impact on the quality of the NLFFF solution.

\subsection{The Pre-eruption Quasi-Steady State}
\begin{figure}[htb!]
     \centering
     \begin{subfigure}[b]{0.5\textwidth}
         \centering
         \includegraphics[width=\textwidth]{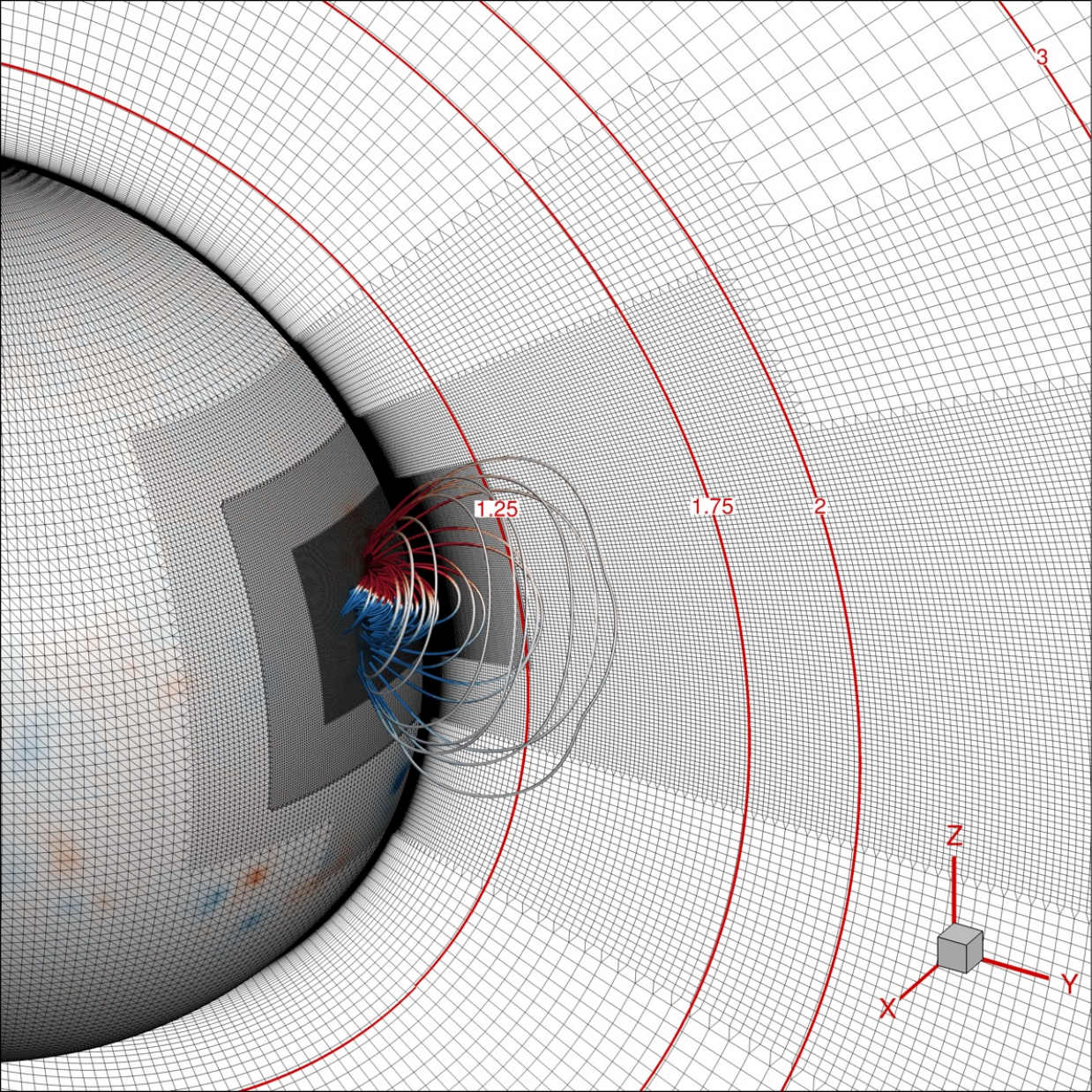}
         \caption{}
         \label{fig:preeruption_grid}
     \end{subfigure}%
     \begin{subfigure}[b]{0.5\textwidth}
         \centering
         \includegraphics[width=\textwidth]{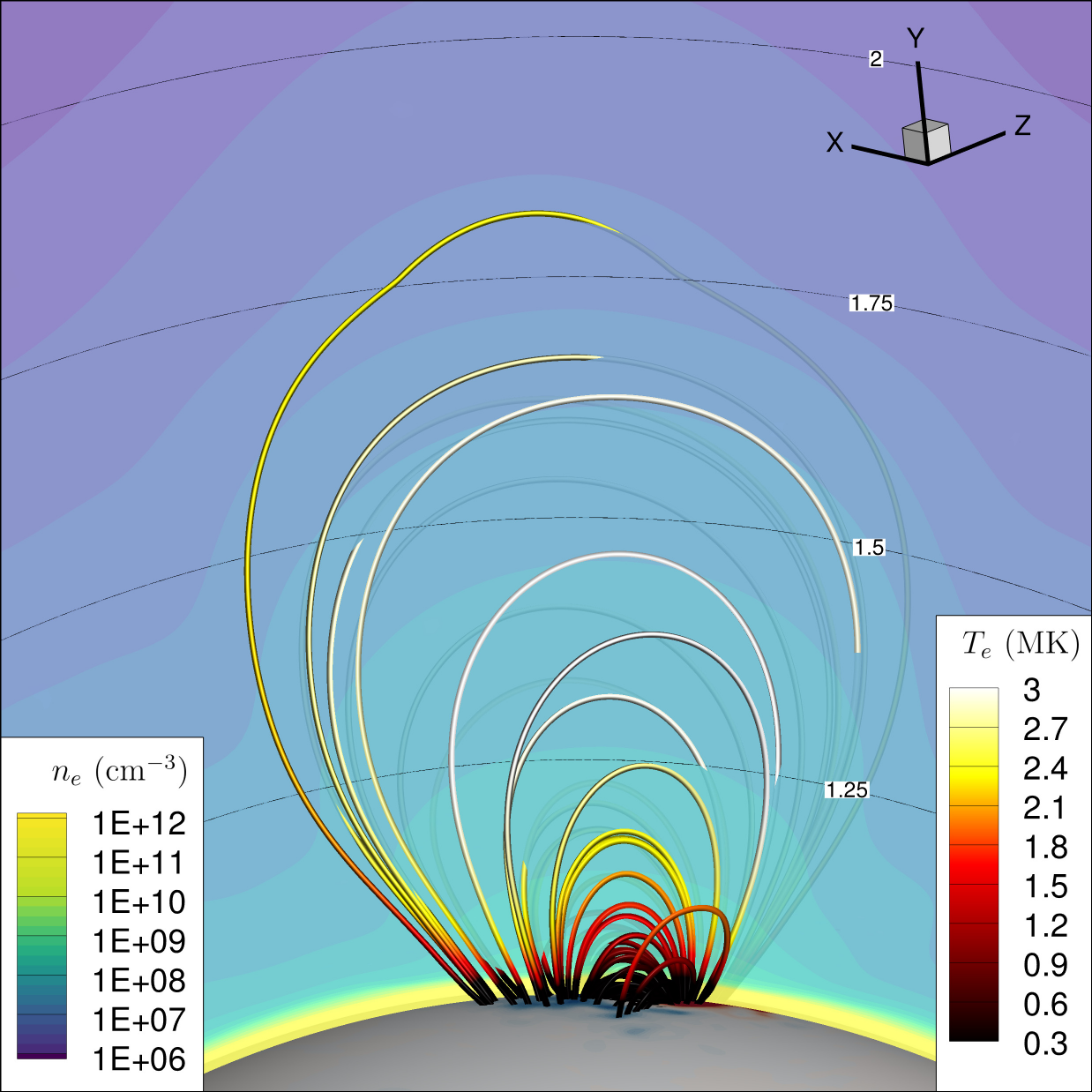}
         \caption{}
         \label{fig:preeruption_plasma}
     \end{subfigure}
        \caption{(a) The computational grid used to obtain the pre-eruption state. The field lines are from the NLFFF solution. They are colored with $B_r$ and roughly show the scale of NLFFF field lines. (b) Some magnetic field lines of the pre-eruption state on top of a slice through the active region. The field lines are colored with electron temperature, $T_e$; the slice shows contours of electron number density, $n_e$.}
        \label{fig:preeruption}
\end{figure}

\begin{figure}[htb!]
     \centering
     \begin{subfigure}[b]{\textwidth}
         \centering
         \includegraphics[width=0.25\textwidth]{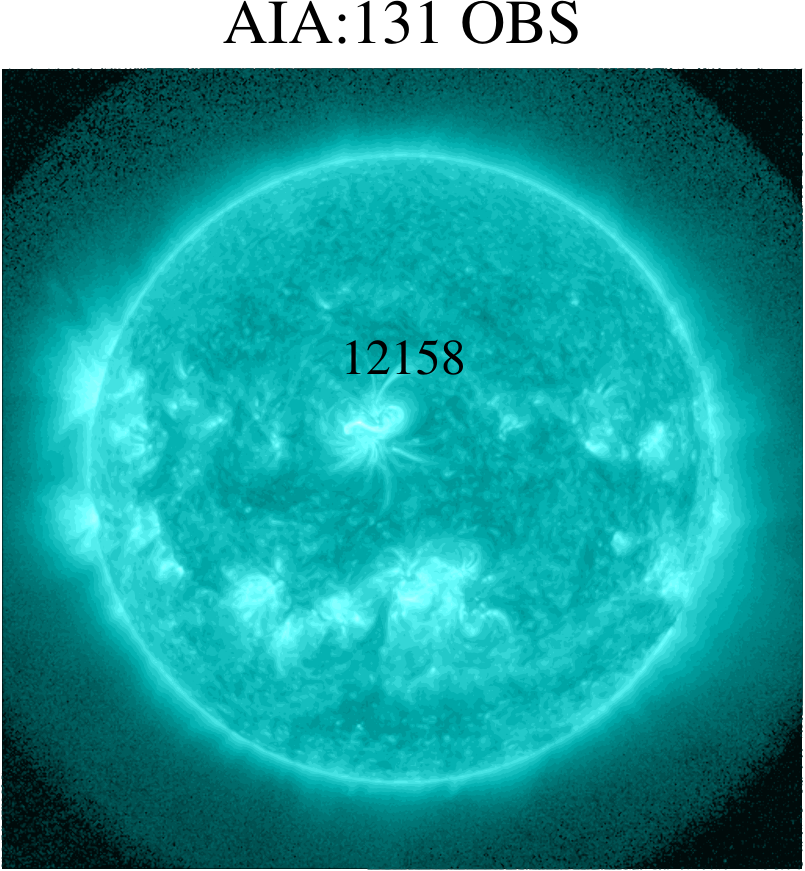}% 
          \includegraphics[width=0.25\textwidth]{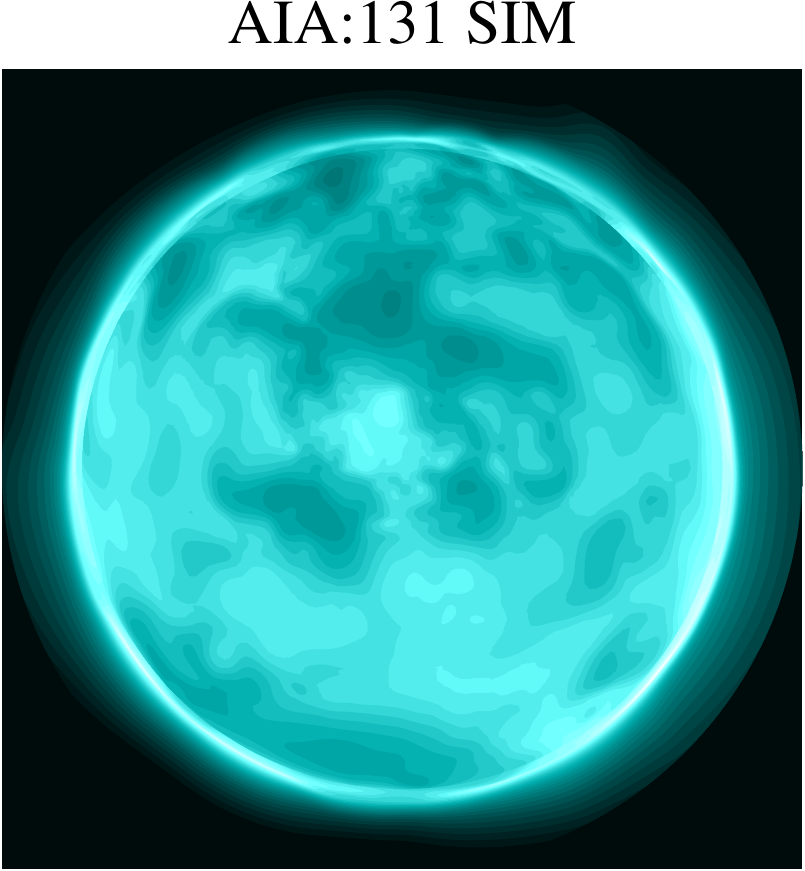}%
          \includegraphics[width=0.25\textwidth]{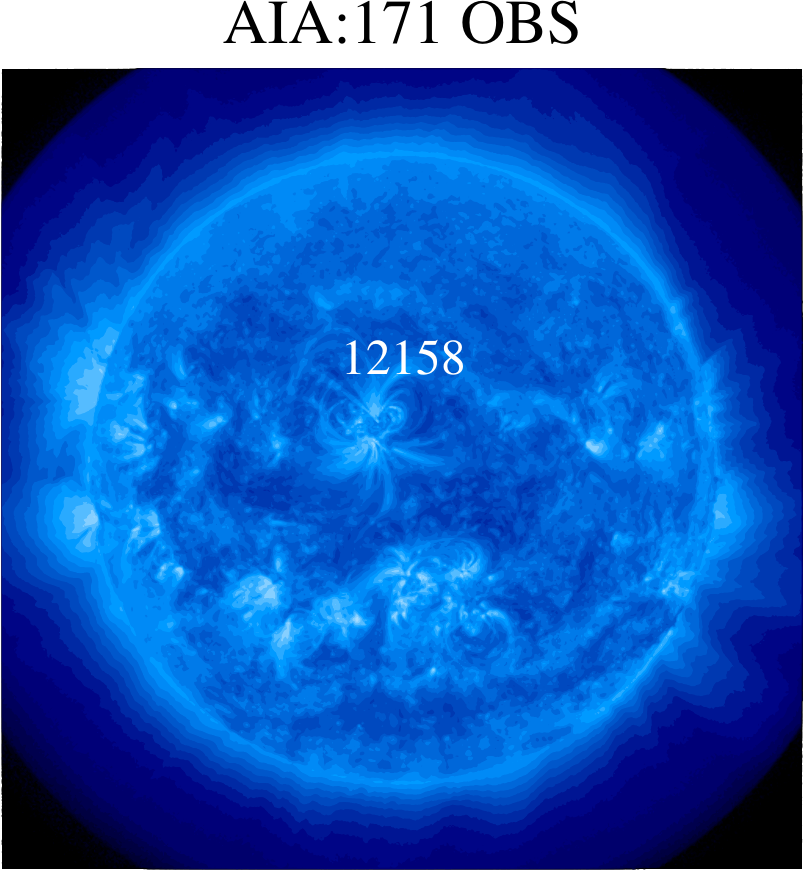}%  
          \includegraphics[width=0.25\textwidth]{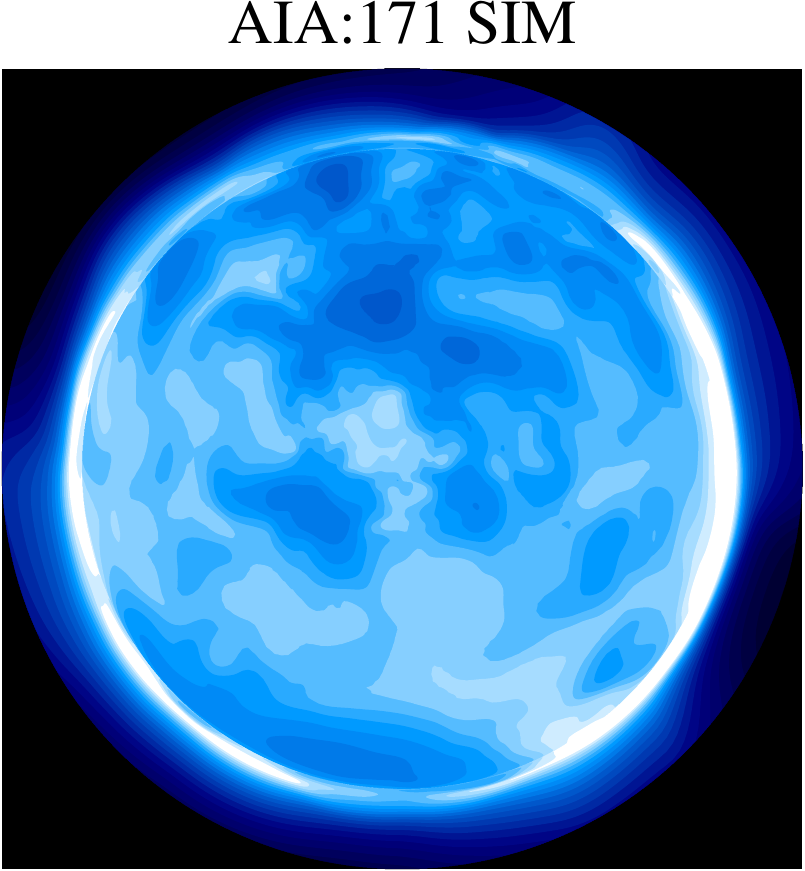}    
         \caption{Full disk observation and simulation before the 2014 CME from AR 12158.}
         \label{fig:preeruption_AR12158}
     \end{subfigure}
      \vspace{-\lineskip}
     \begin{subfigure}[b]{\textwidth}
         \centering
         \includegraphics[width=0.25\textwidth]{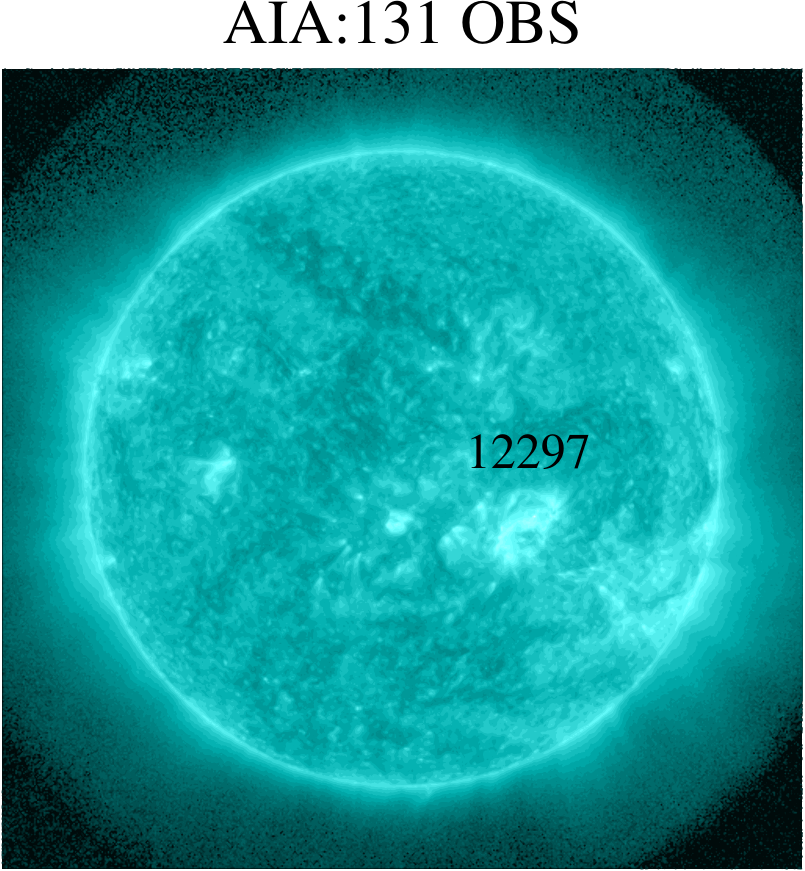}%
          \includegraphics[width=0.25\textwidth]{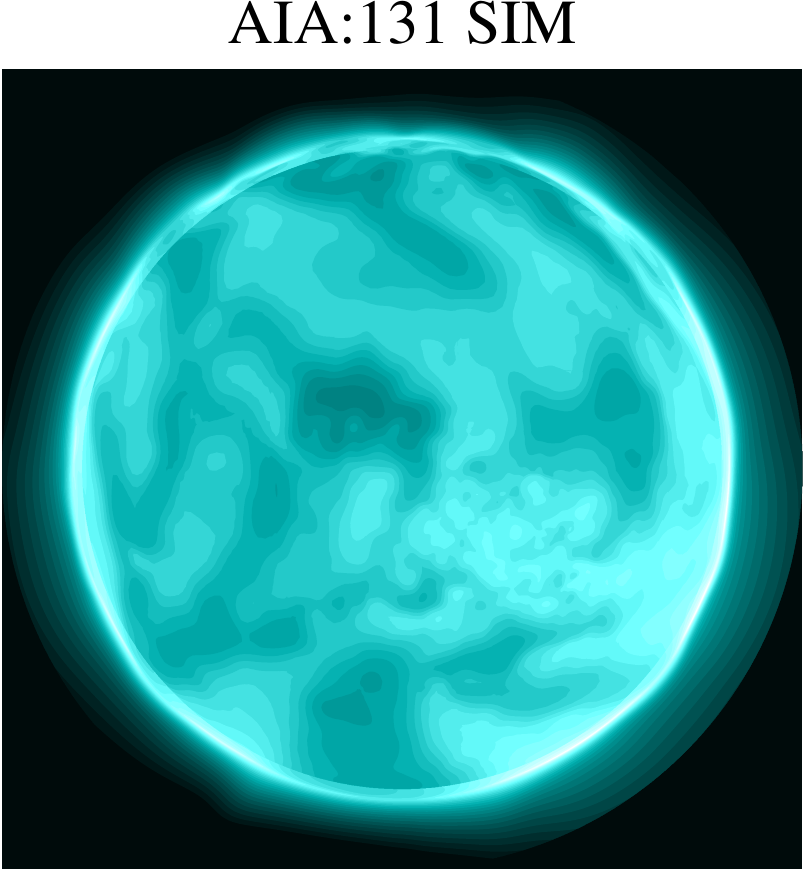}%
          \includegraphics[width=0.25\textwidth]{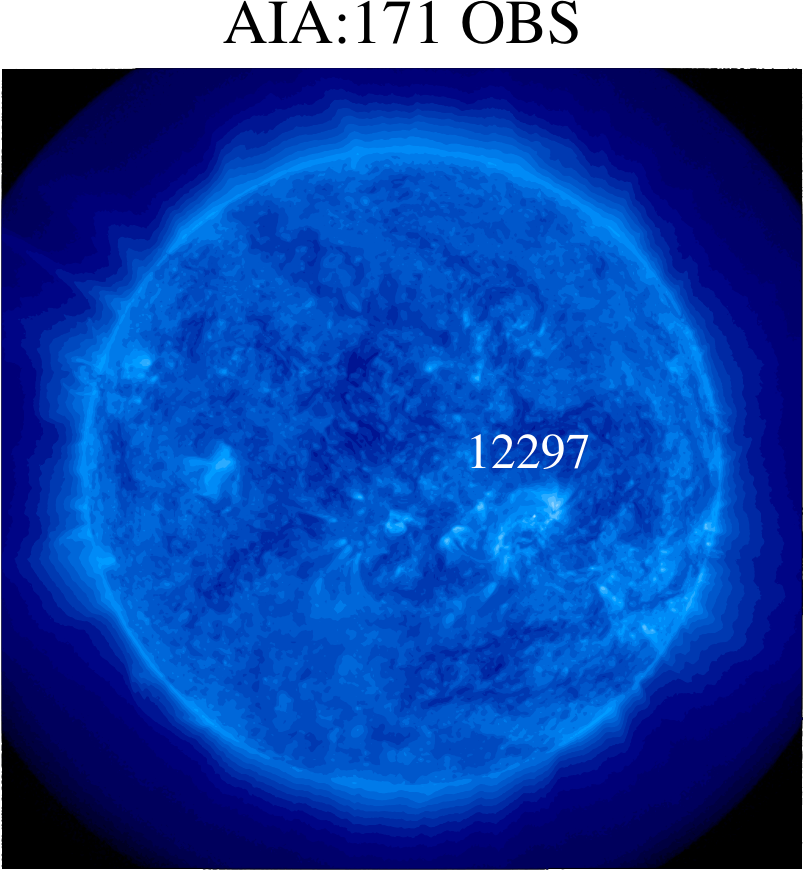}%
          \includegraphics[width=0.25\textwidth]{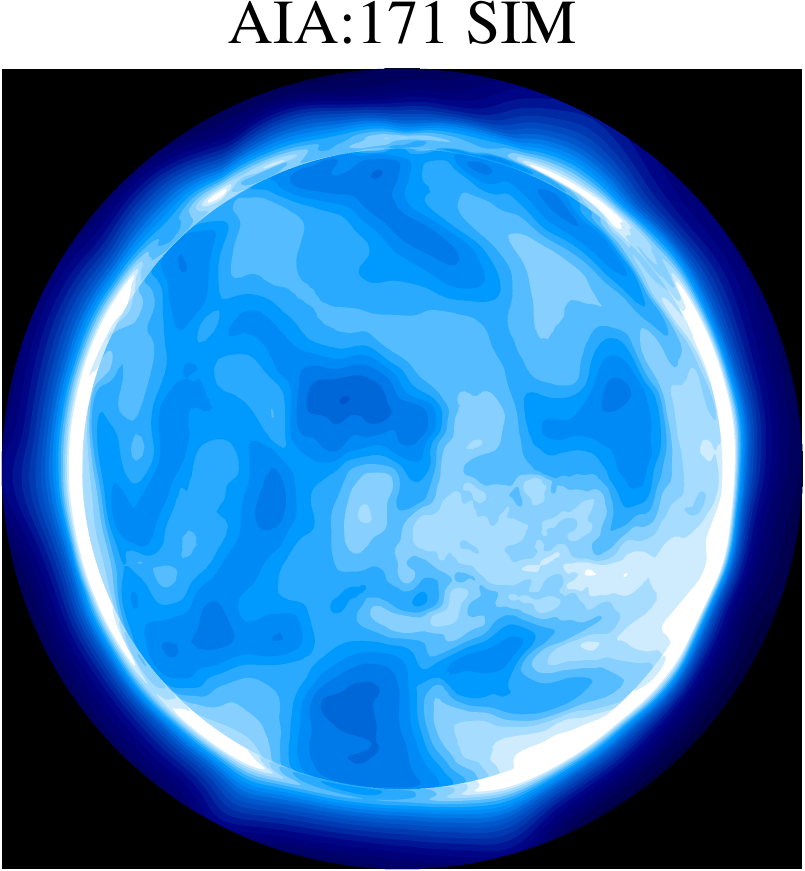}    
         \caption{Full disk observation and simulation before the 2015 CME from AR 12297.}
         \label{fig:preeruption_AR12297}
     \end{subfigure}
 \vspace{-\lineskip}
     \begin{subfigure}[b]{\textwidth}
         \centering
         \includegraphics[width=0.25\textwidth]{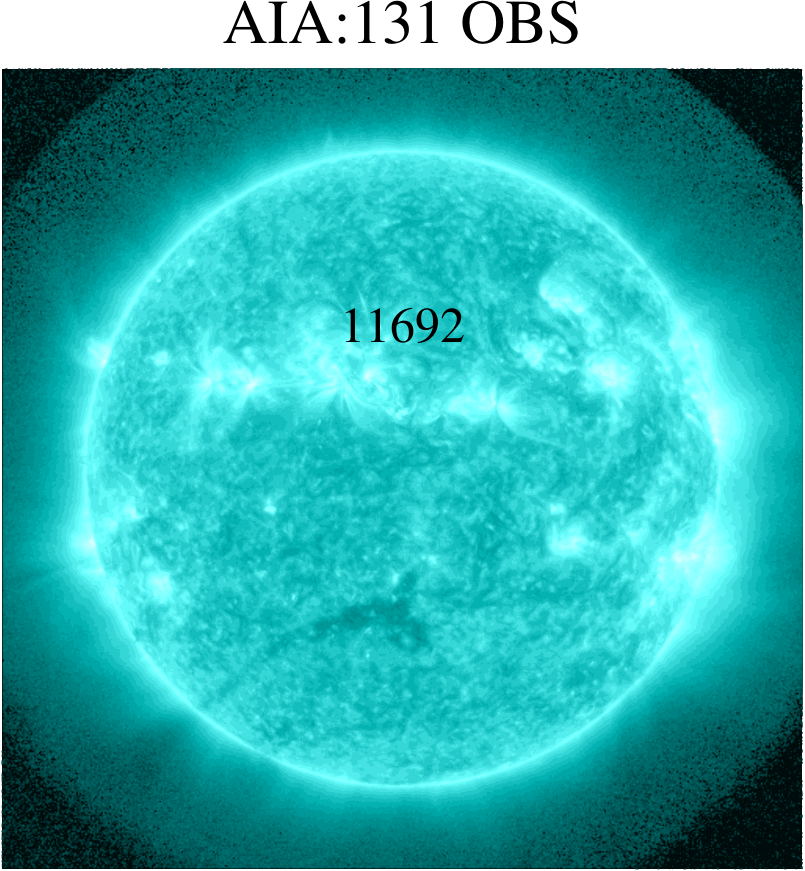}% 
          \includegraphics[width=0.25\textwidth]{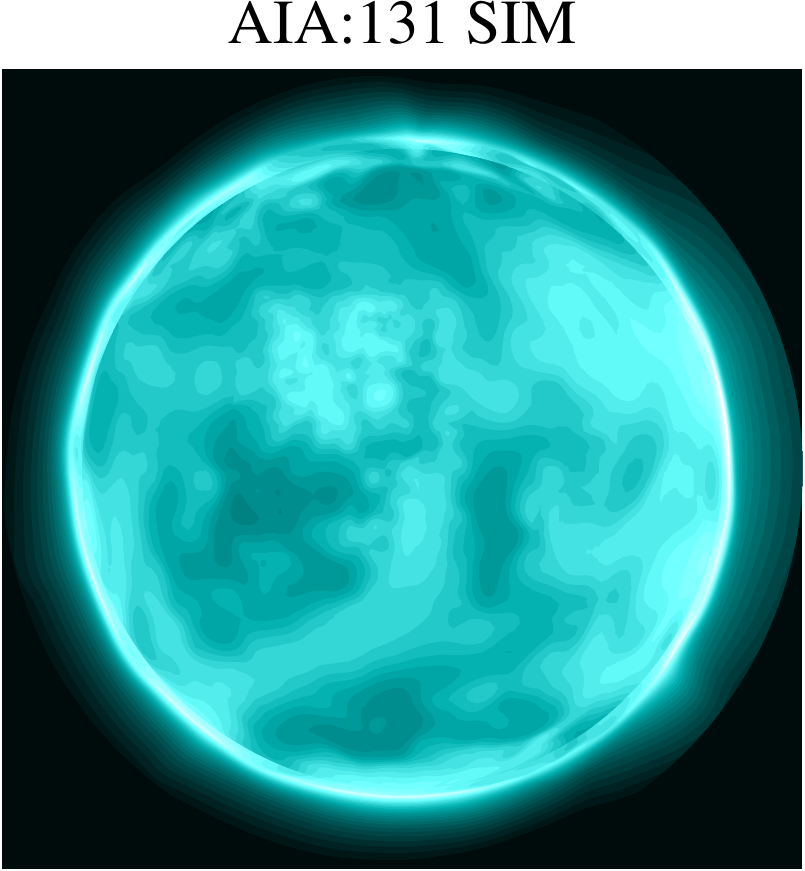}%   
          \includegraphics[width=0.25\textwidth]{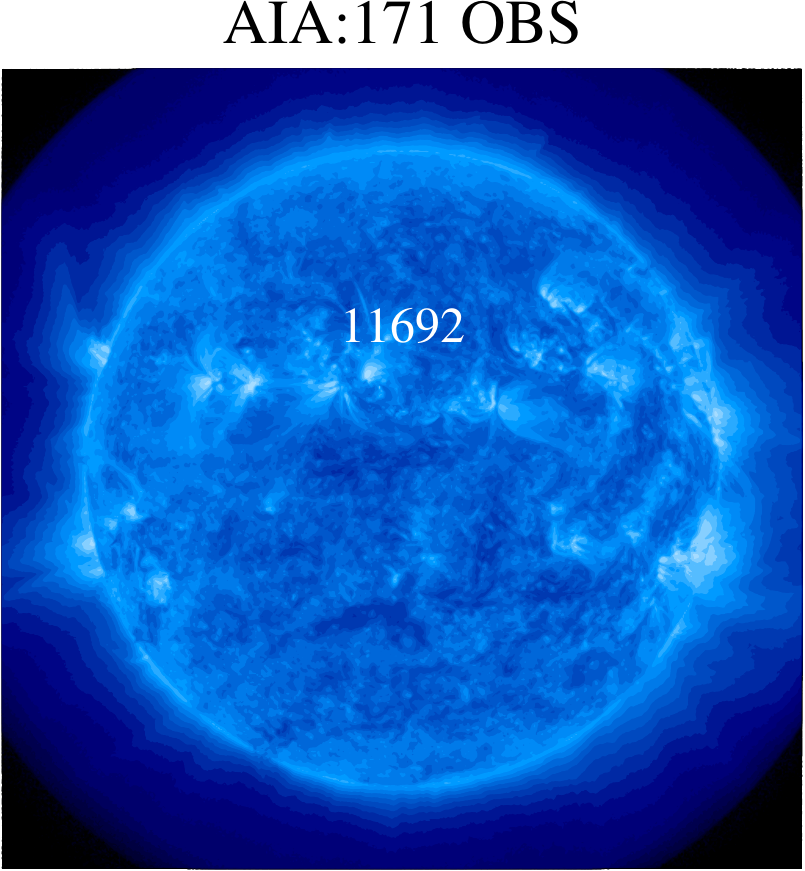}%  
          \includegraphics[width=0.25\textwidth]{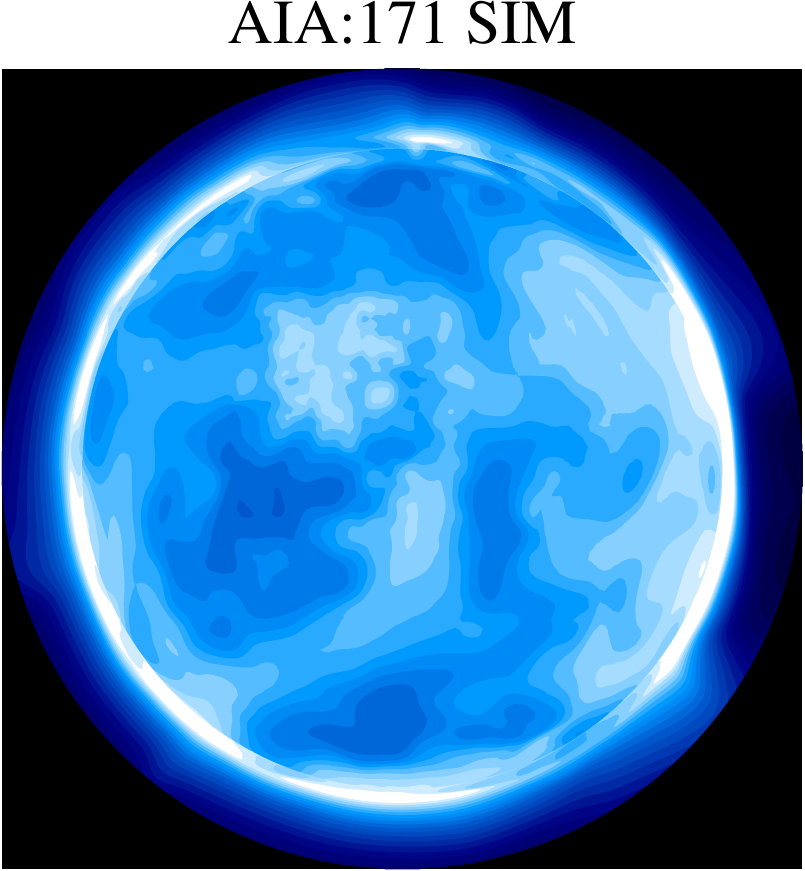}
         \caption{Full disk AIA observation and simulation before the 2013 CME from AR 11692.}
         \label{fig:preeruption_AR11692}
     \end{subfigure}
        \caption{LoS AIA 131 \AA~ and 171 \AA~ observations and synthetics for the pre-eruption states using a number density $n=2\times 10^{17}~\mathrm m^{-3}$ on the chromospheric boundary. The ARs of interest are labeled in the observations.}
        \label{fig:preeruption_EUV}
\end{figure}

In this subsection, we briefly examine the pre-eruption quasi-steady states obtained by using the NLFFF solutions in $\mathbf B_0$ in the AWSoM model.
For all 3 events, we use a number density $n=5\times 10^{18}~\mathrm m^{-3}$ and a temperature of $50,000$\,K on the chromospheric boundary.
This number density is deliberately chosen to be $25$ times larger than the default value for the AWSoM model, $n=2\times 10^{17}~\mathrm m^{-3}$, so that the initial phase of the subsequent time-accurate simulation can be $\sim5$ times faster due to the reduced Alfv\'en speed.
Increasing the number density at the inner boundary is our routine approach to improving the robustness of CME simulations.
It does not significantly impact the background solar wind at 1\,au, but it may increase the density of the ejecta.
The other important parameter that we tune is the Poynting flux.
We repeat the steady state run a few times with various Poynting flux values until we find a satisfactory agreement in the background solar wind at L1 between the simulation and in-situ measurements, following the approach of \citet{Jivani_2023}.
The optimal Poynting flux value is $3\times10^5~\mathrm W\mathrm m^{-2}\mathrm T^{-1}$ for all three events.
%We find matching speed and density in the background solar wind for AR 11692 and AR 12158, but are unable to do so for AR 12297.
%The best background solar wind we obtain for AR 12297 has a lower density and a slightly lower speed than observations.

Figure \ref{fig:preeruption_grid} shows the computational grid for the steady state of the 2014 event, which is also the initial grid for the time-accurate simulation.
The AMR criteria that define this grid have been discussed in section \ref{section_methods}.
Figure \ref{fig:preeruption_plasma} shows the filament channel formed by plasma surrounding the NLFFF field lines.
The distribution of electron temperature, $T_e$, along the field lines is similar to those previously found by \citet{Shi2024}.
The density is higher in regions with a strong magnetic field, which will further supply the eruption.
Figure \ref{fig:preeruption_EUV} compares the full-disk synthetic and observed AIA EUV images before the eruption.
As the active region is enlarged by a factor of two in our simulations, it also appears twice as large in the synthetics and resembles the shape of their observed counterparts.
Note that the simulations that produce the synthetics use the default inner boundary number density for AWSoM ($n=2\times 10^{17}~\mathrm m^{-3}$) and thus will not be reused in the following time-accurate simulations.
The reason for this choice is that the use of $n=5\times 10^{18}~\mathrm m^{-3}$ makes the rim of the disk optically too thick, which then causes the synthetic to lose details around those areas.
%These simulations are dedicated to illustrating the qualitative agreement of the pre-eruption steady state with the observed solar corona.

\subsection{Time-accurate CME Simulations}
For each event, the time-accurate CME simulation restarts from the saved states of its pre-eruption quasi-steady state run.
We return to using the potential field solution $\mathbf B_P$ without the NLFFF solution as the background field $\mathbf B_0$.
The approximate NLFFF solution $\mathbf B_N$ now captured by $\mathbf B_1$ erupts spontaneously, and we apply STITCH as needed to further energize the eruption.

The simulation in the SC domain lasts until 12 hours.
After 480\,s, 900\,s, and 2 hours, we switch on local time stepping for all grid cells below $r=1.01R_\odot$, $r=1.05R_\odot$, and $r=1.1R_\odot$ respectively.
The original SC grid, with a maximum resolution of $0.175^\circ$, is maintained until 1200\,s into the simulation; however, after this time, the resolution of the SC domain is coarsened to $0.35^\circ$ to speed up the simulation.
The IH domain starts after one hour of simulation (before the CME reaches the SC-IH boundary) and lasts for up to 4 days.
The coupling of the SC and IH domains takes place every 30\,s until the simulation in the SC domain stops at 12 hours, after the CME has fully transitioned into IH.

\subsubsection{The 2014 Event from AR 12158: STITCH-assisted Eruption}

In an AR whose magnetic field is a perfect NLFFF, one expects the solution to either remain stable or erupt gradually due to some instability. 
Our simulation for the 2014 event is a nearly ideal case, as the initial eruption due to force imbalance in an imperfect NLFFF is not strong enough to overcome the strapping forces from the background field.
We apply STITCH to all grid cells in the active region below $r=1.01R_\odot$ with $\zeta = 3.5\times 10^5~\mathrm{km}^2\mathrm{s}^{-1}$.
This is different from previous work \citep{vanderHolst2025}, which applies STITCH at $r=R_\odot$ only and uses $\zeta = 1.4\times 10^6~\mathrm{km}^2\mathrm{s}^{-1}$.
We mainly use STITCH to produce a shock rather than to form a filament channel over a longer period of time, because plasma has already condensed around the NLFFF in the pre-eruption state.
As a result of volumetric injection of helicity, it takes a much shorter time (within one minute) to generate a second, stronger shock.
The second shock coalesces with the first one, and the combined shock then propagates to L1 at a speed comparable to observations.

\begin{figure}[htb!]
     \centering
     \begin{subfigure}[b]{0.5\textwidth}
         \centering
         \includegraphics[width=\textwidth]{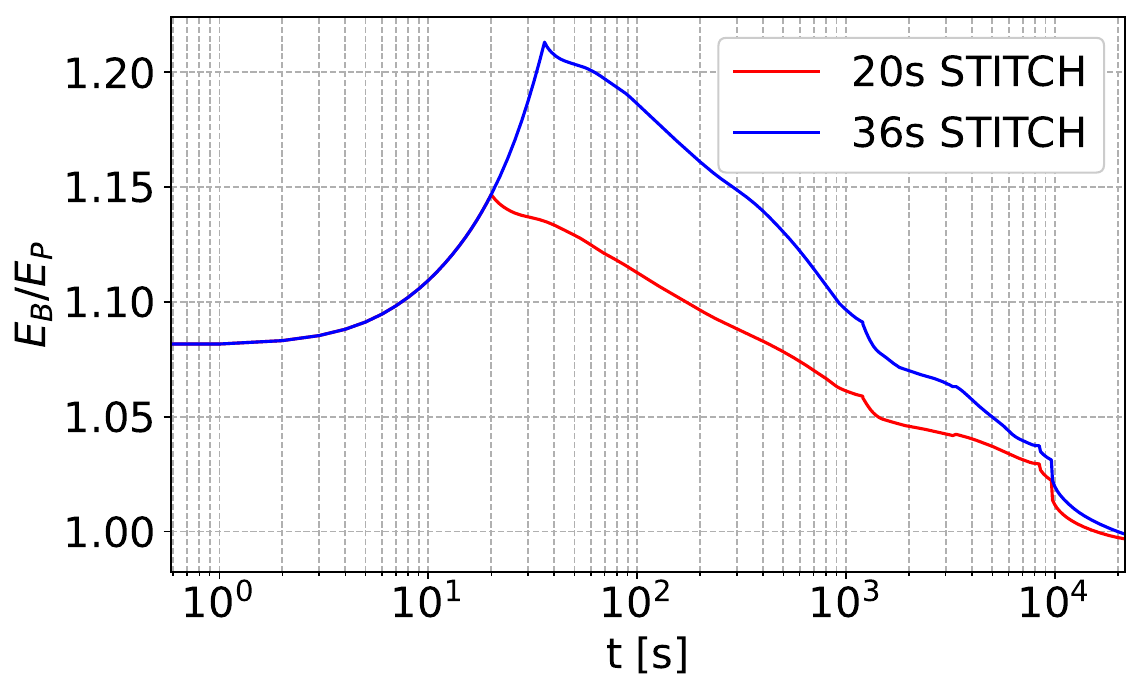}
         \caption{Total normalized magnetic energy vs. simulation time.}
         \label{fig:STITCH_eb_t}
     \end{subfigure}%
     \begin{subfigure}[b]{0.5\textwidth}
         \centering
         \includegraphics[width=\textwidth]{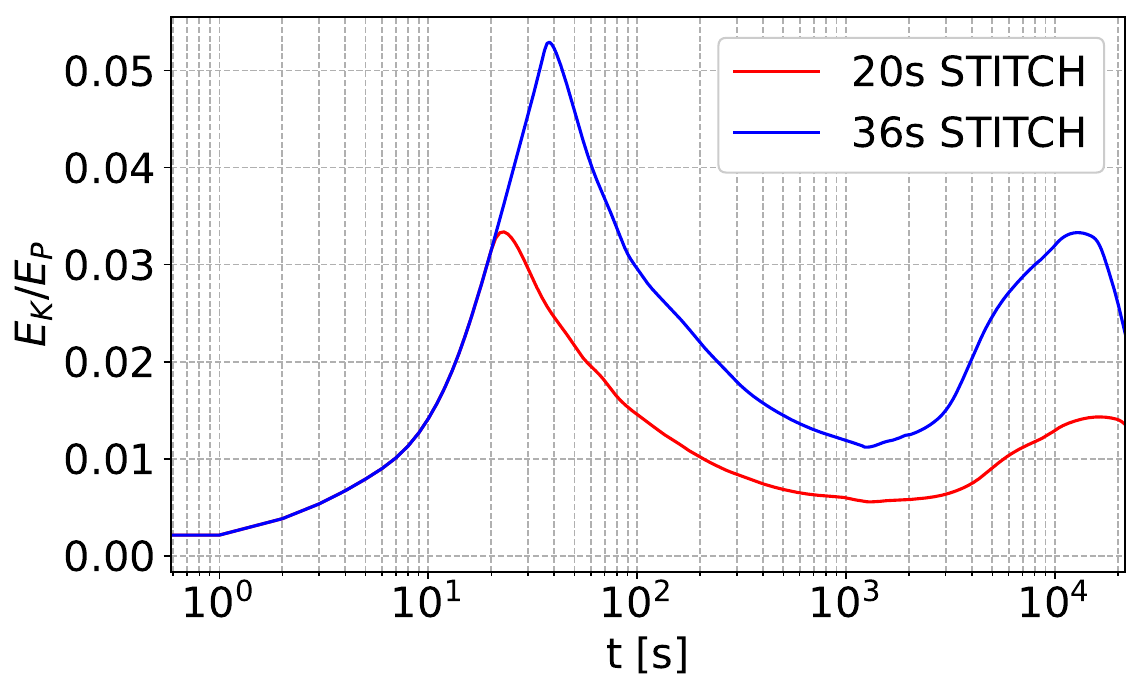}
         \caption{Normalized kinetic energy vs. simulation time.}
         \label{fig:STITCH_ek_t}
     \end{subfigure}
        \caption{Evolution of the magnetic and kinetic energy during time-accurate simulation for the 2014 event. The two curves envelope the ensemble of runs with STITCH duration between 20\,s and 36\,s.}
        \label{fig:STITCH_energetics}
\end{figure}

\begin{figure}[htb!]
     \centering
         \includegraphics[width=0.5\textwidth]{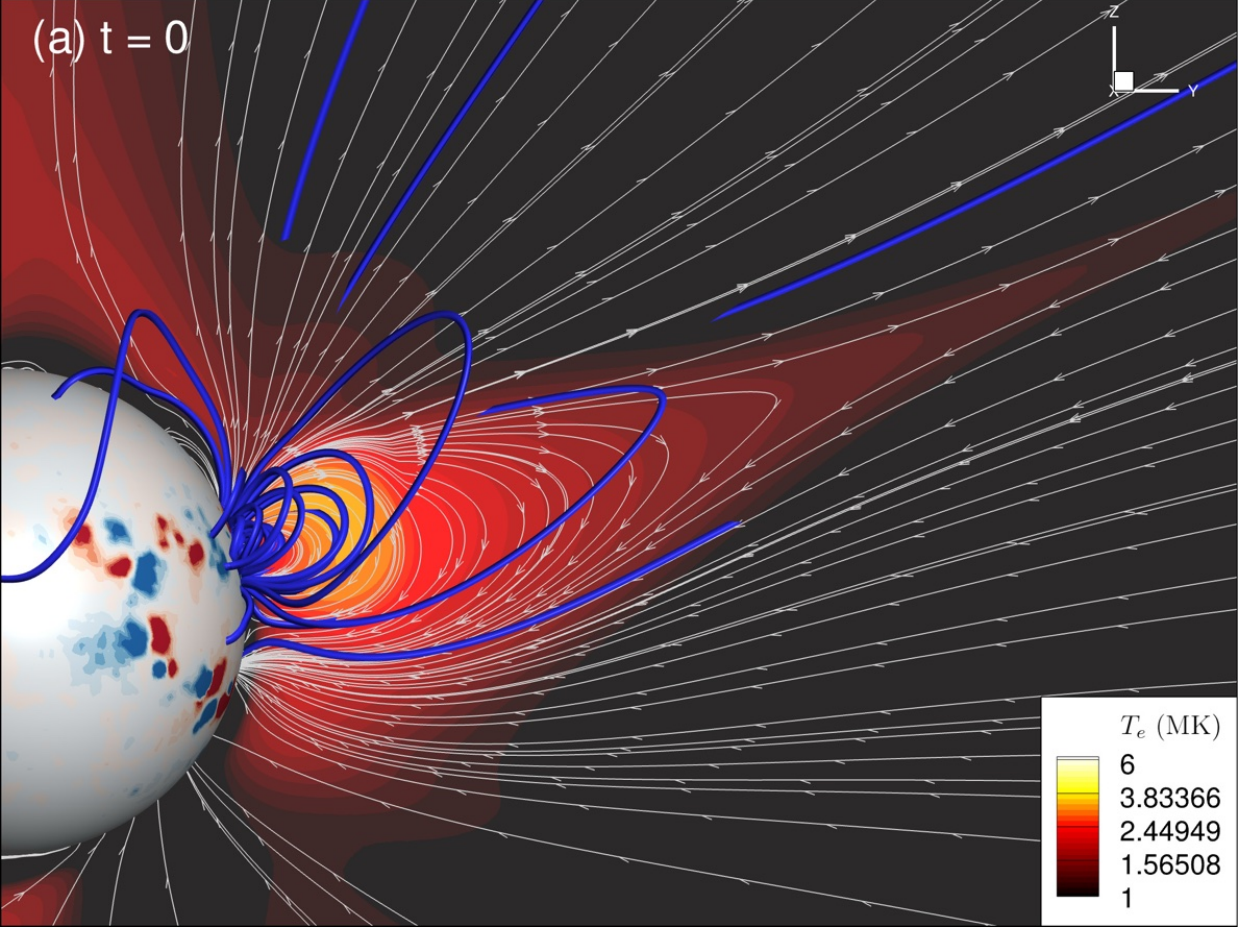}%
         \includegraphics[width=0.5\textwidth]{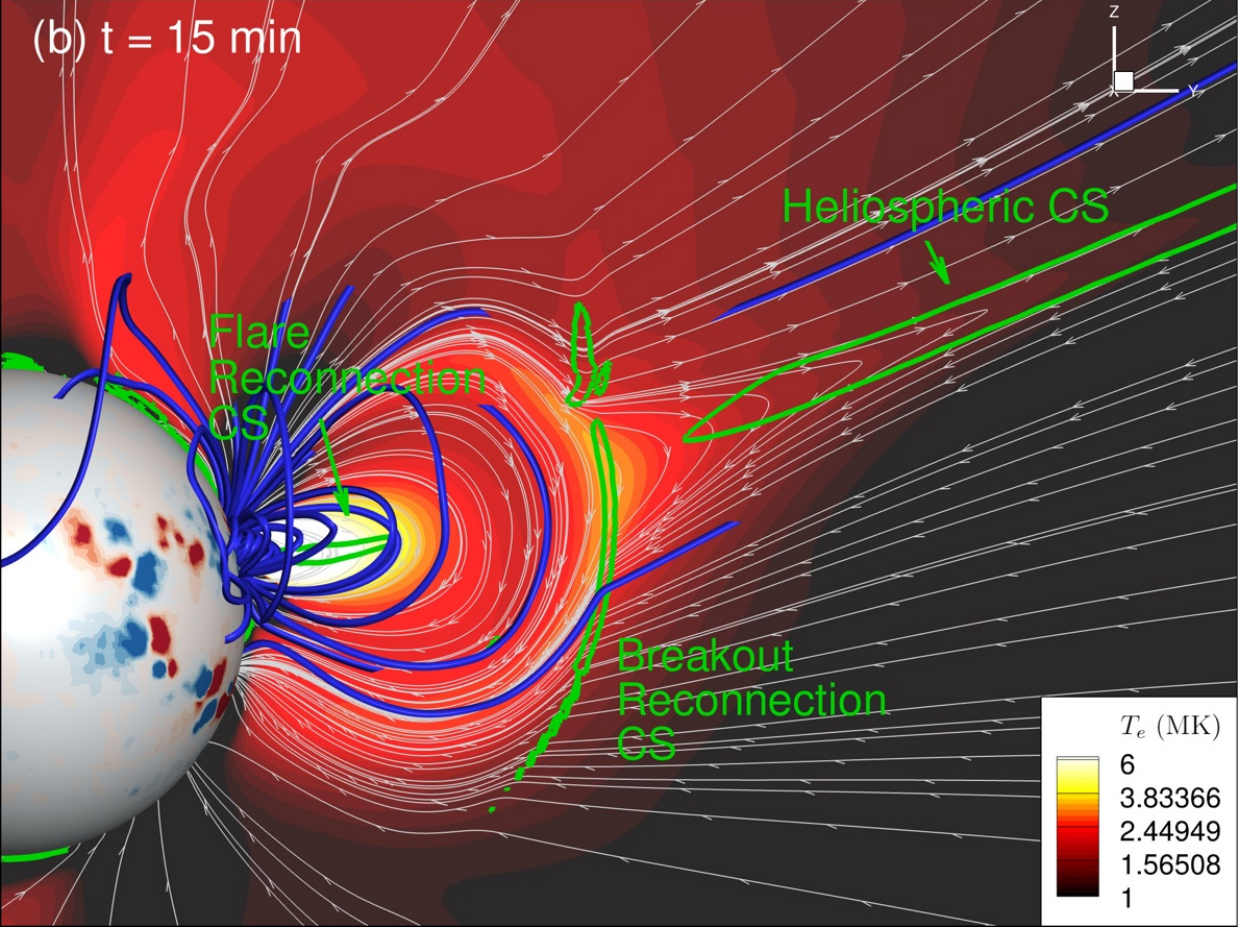}
         \vspace{-\lineskip}
         \includegraphics[width=0.5\textwidth]{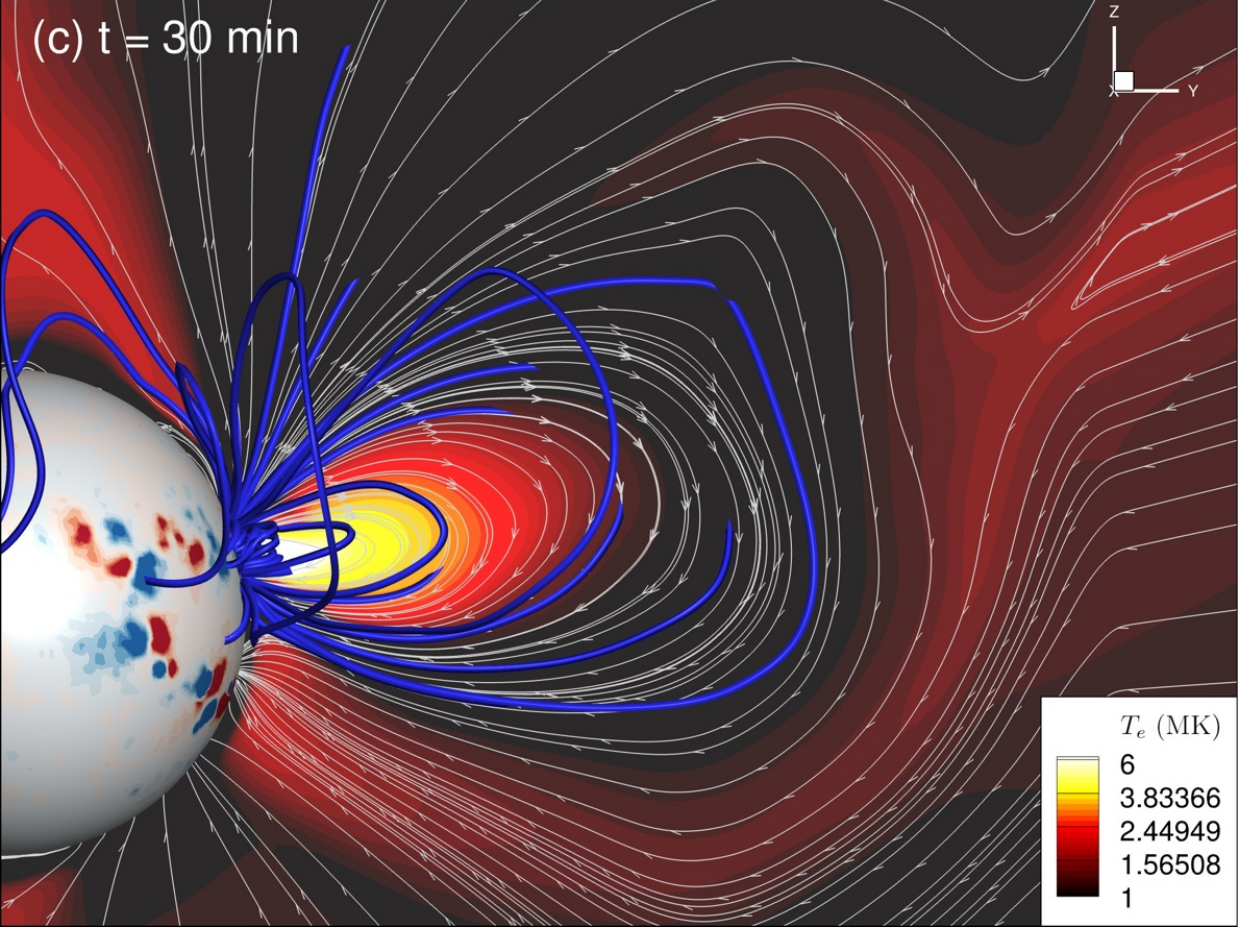}
        \caption{A slice through $90^\circ$ longitude near the center of the AR at three time instants. The slice shows contours of the electron temperature and magnetic field lines (white) in this plane. Selected 3-D field lines are colored in blue. Contours of $J/B=1.35$ $\mu \mathrm A/\mathrm G$ mark the locations of current sheets and are plotted in (b).}
        \label{fig:STITCH_CS}
\end{figure}

Figure \ref{fig:STITCH_energetics} shows the change in the total magnetic energy ($E_B$) and the kinetic energy ($E_K$) in the SC domain over time.
Both quantities are normalized with the total magnetic energy ($E_P$) of the background potential field \textit{without} the NLFFF solution ($\mathbf B_P$).
We run a few simulations with STITCH starting at $t=0$ and lasting 20 to 36 seconds.
The two simulations with 20 and 36 seconds of STITCH thus encompass an ensemble of those with a duration of STITCH in between.
We find that applying STITCH for 30 seconds produces the best matching shock strength.
Our simulation with 30 seconds of STITCH starts with $E_B=1.08E_P$ and peaks at $E_B = 1.19E_P$.
The maximum kinetic energy at $t=30$\,s is around $E_K=0.045E_P$.
Among the maximum free energy ($0.19E_P$) in the SC domain, 42\% comes from NLFFF while 58\% comes from STITCH.
In Figure \ref{fig:STITCH_eb_t}, an initial rapid increase in $E_B/E_P$ characterizes the accumulation of energy during volumetric STITCH.
A dip around $t=10^4$\,s indicates the CME's departure from the SC domain.
In Figure \ref{fig:STITCH_ek_t}, the temporary decrease in $E_K/E_P$ from the end of STITCH to around $1200$\,s before increasing again.
Figure \ref{fig:STITCH_CS} suggests that the trend in $E_K/E_P$ is a result of the topology of the background field and magnetic reconnection.
After the initial acceleration due to STITCH and force imbalance, the kinetic energy quickly decreases as the plasma moves through closed field lines in \ref{fig:STITCH_CS}(a).
The disappearance of the last closed field lines of the background field from \ref{fig:STITCH_CS}b to \ref{fig:STITCH_CS}c corresponds to the inflection point in $E_K/E_P$ around $t=1200$\,s.
After that, the flare reconnection site near the solar surface continues converting magnetic energy into kinetic energy and heat.
The ejecta gain more kinetic energy from flare reconnection than they lose as they move through open field lines, so $E_K/E_P$ increases again.
The flare reconnection (its current sheet out of plane) and the breakout reconnection form the paradigm configuration for reconnection in CME simulations.
They also account for most of the loss in $E_B/E_P$, which is converted to heat, a process also revealed by the contours of $T_e$.

\begin{figure}[htb!]
     \centering
%            \hfill
        \includegraphics[width=0.4\linewidth]{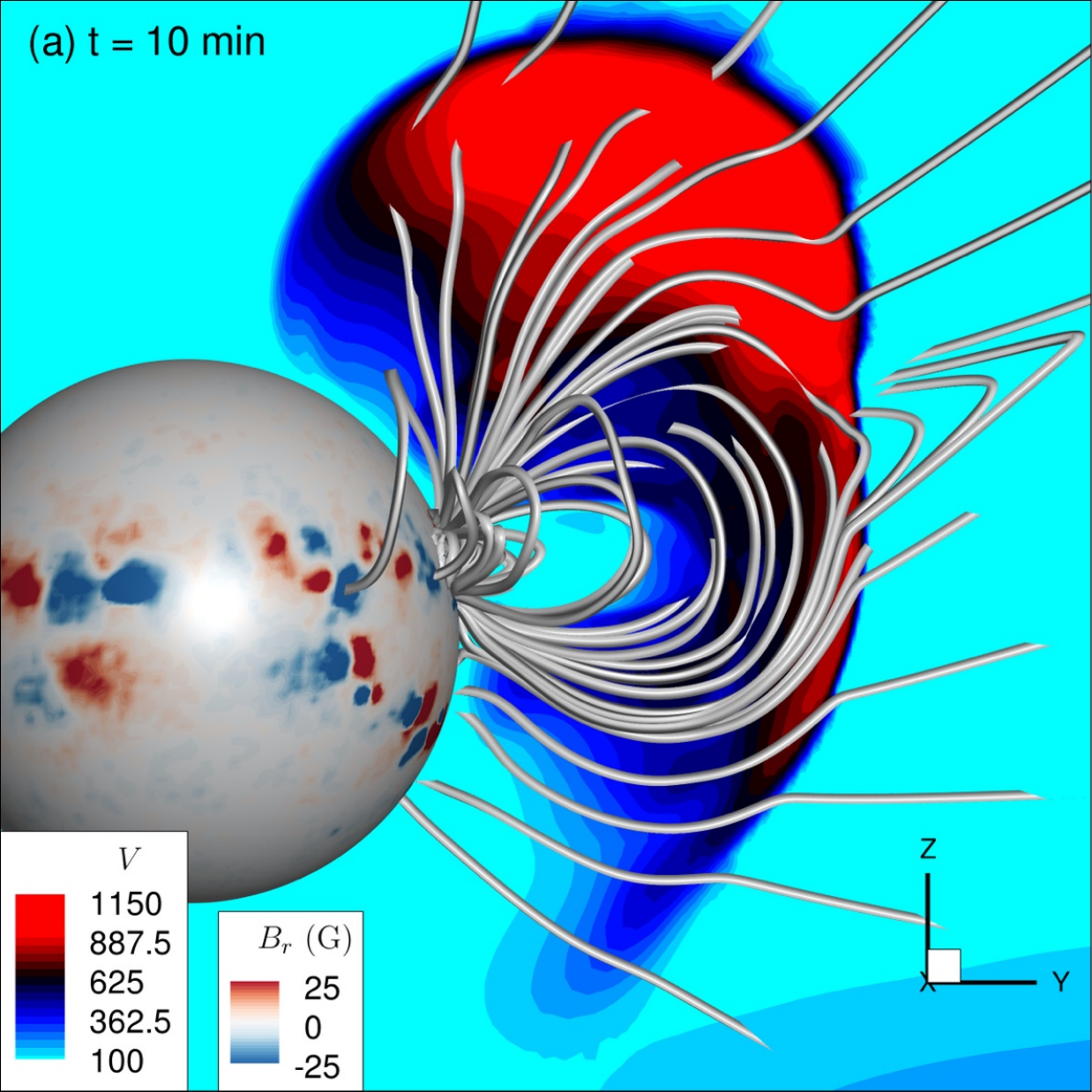}%
        \includegraphics[width=0.4\linewidth]{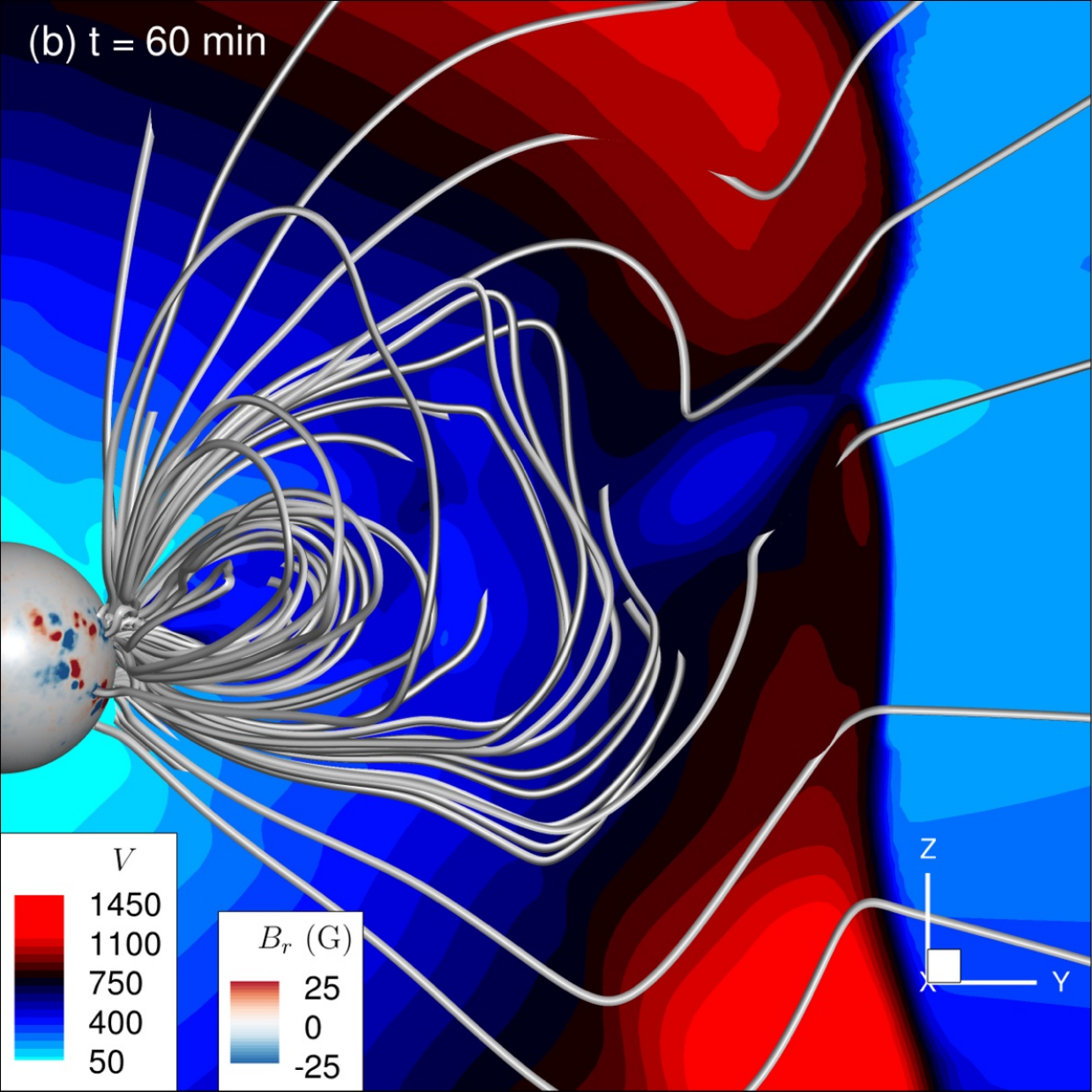}
        \vspace{-\lineskip}
        \includegraphics[width=0.4\linewidth]{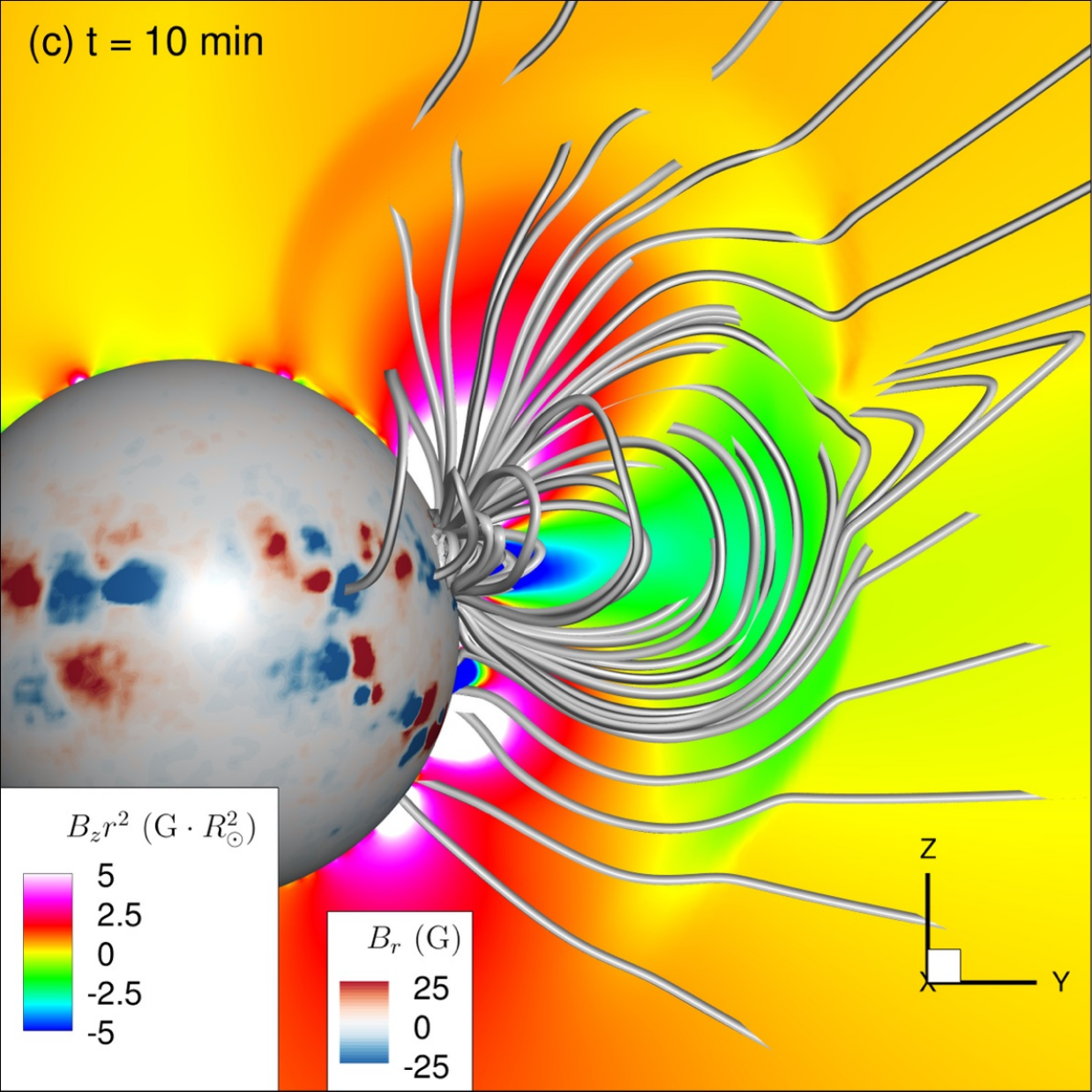}%
        \includegraphics[width=0.4\linewidth]{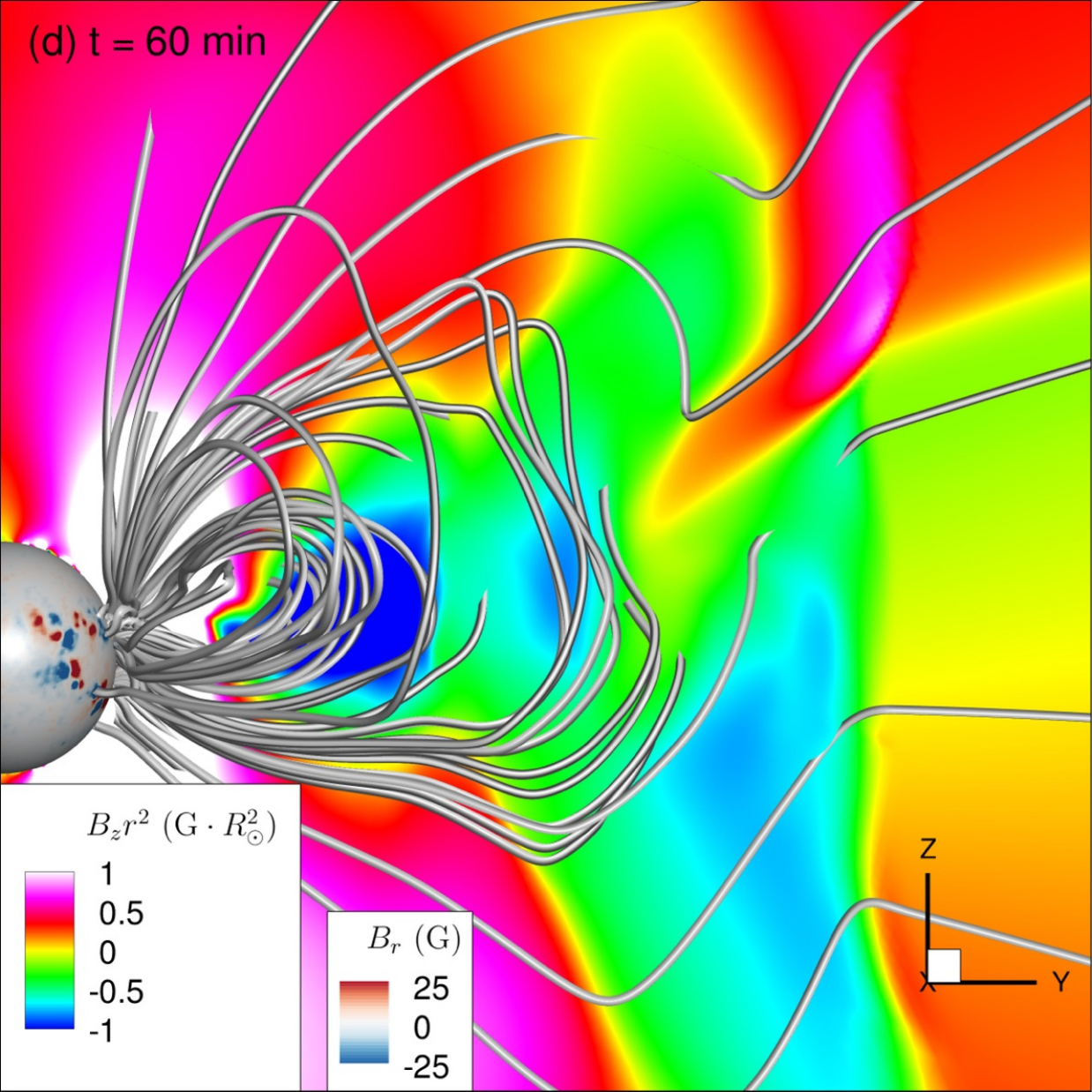}
        \vspace{-\lineskip}
        \includegraphics[width=0.4\linewidth]{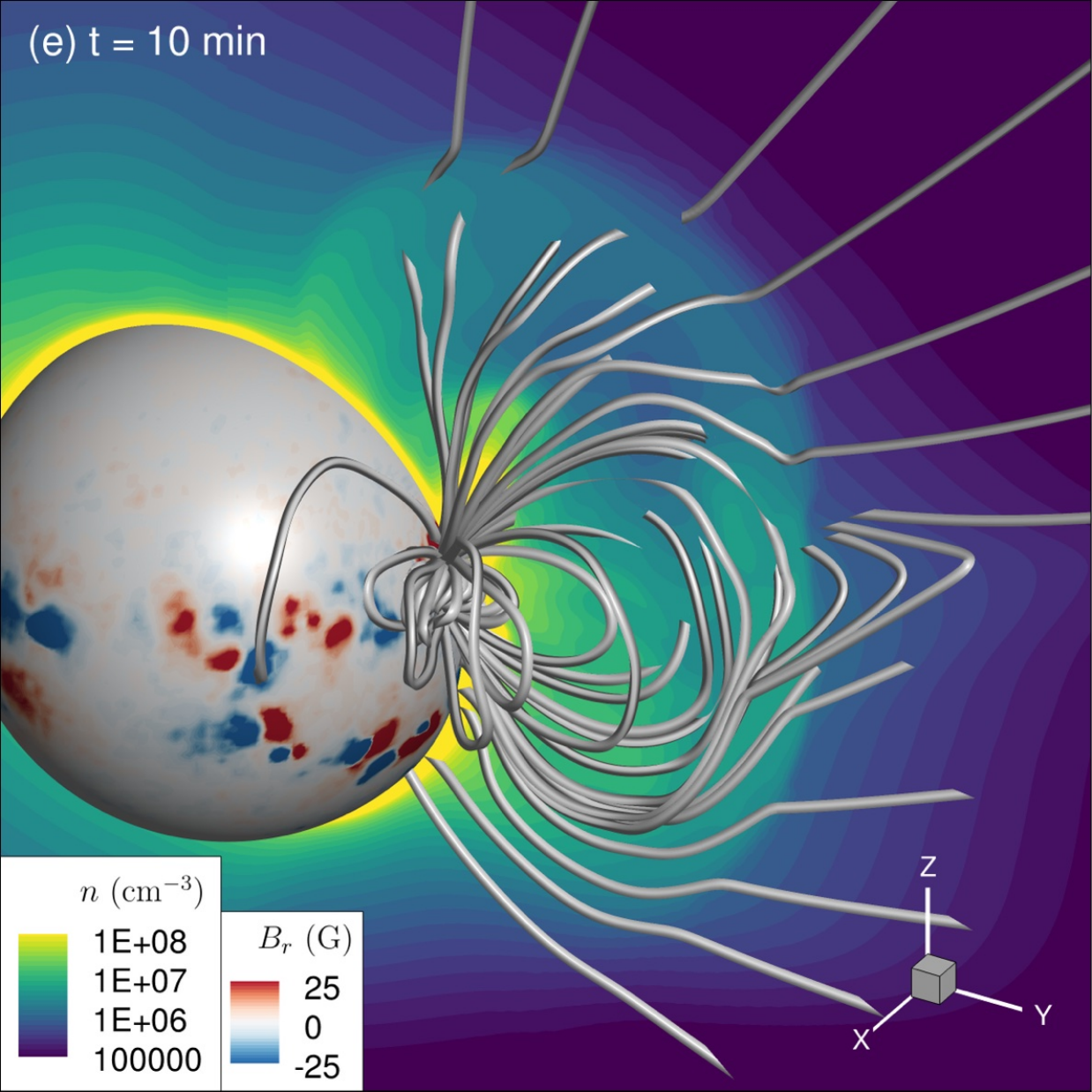}%
        \includegraphics[width=0.4\linewidth]{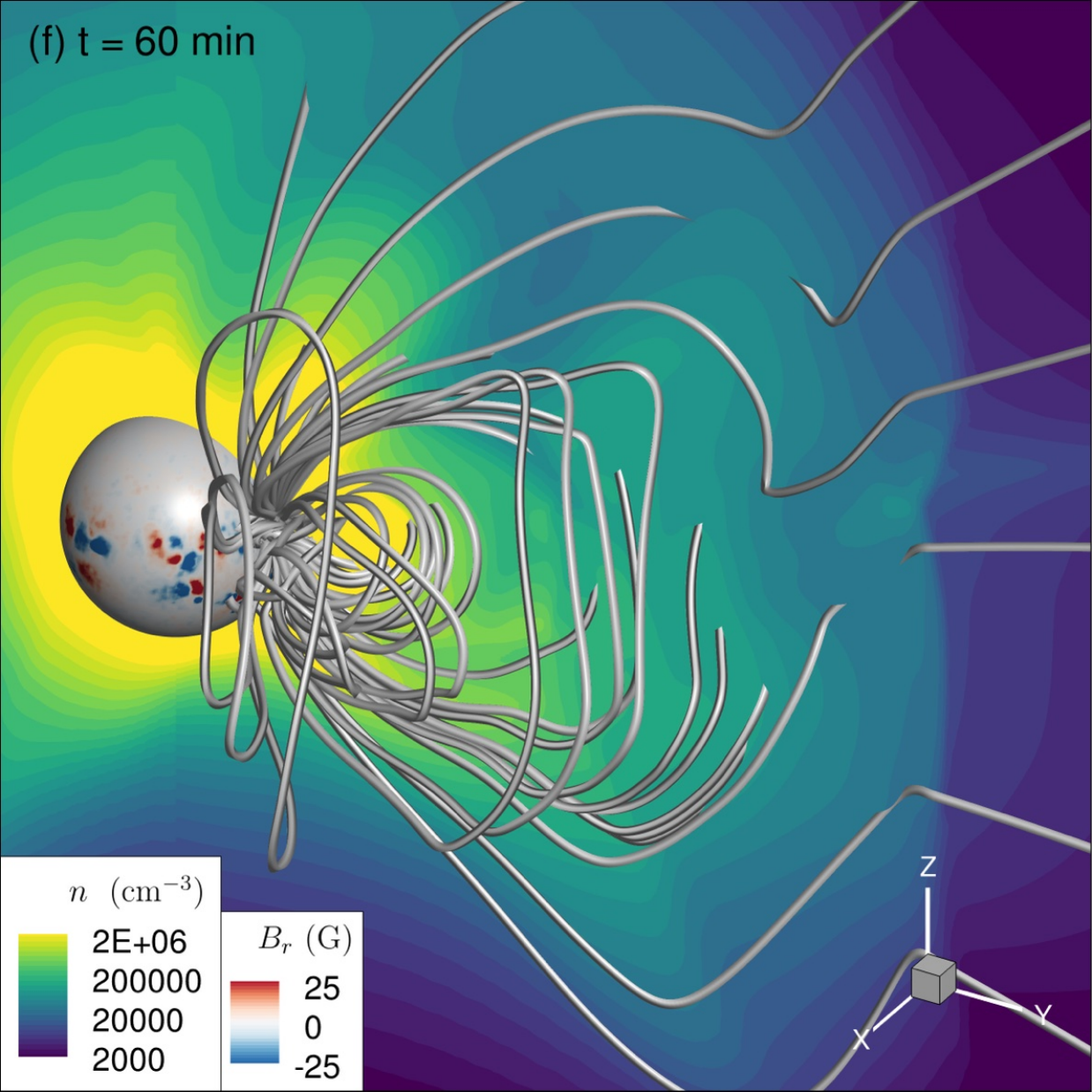}
%            \hfill
        \caption{Simulation for the 2014 event in SC at two time instants. A slice through $90^\circ$ longitude shows contours of speed in (a), (b), contours of $B_zr^2$ in (c), (d) and contours of number density, $n$ in (e), (f). The viewing angles in (e) and (f) are rotated. Note that the ranges for $V$, $n$ and $B_zr^2$ are chosen differently for the two instants. Magnetic field lines crossing the slice plane are not shown behind the plane.}
        \label{fig:STITCH_SC}
\end{figure}

\begin{figure}[htb!]
     \centering
%            \hfill
        \includegraphics[width=0.4\linewidth]{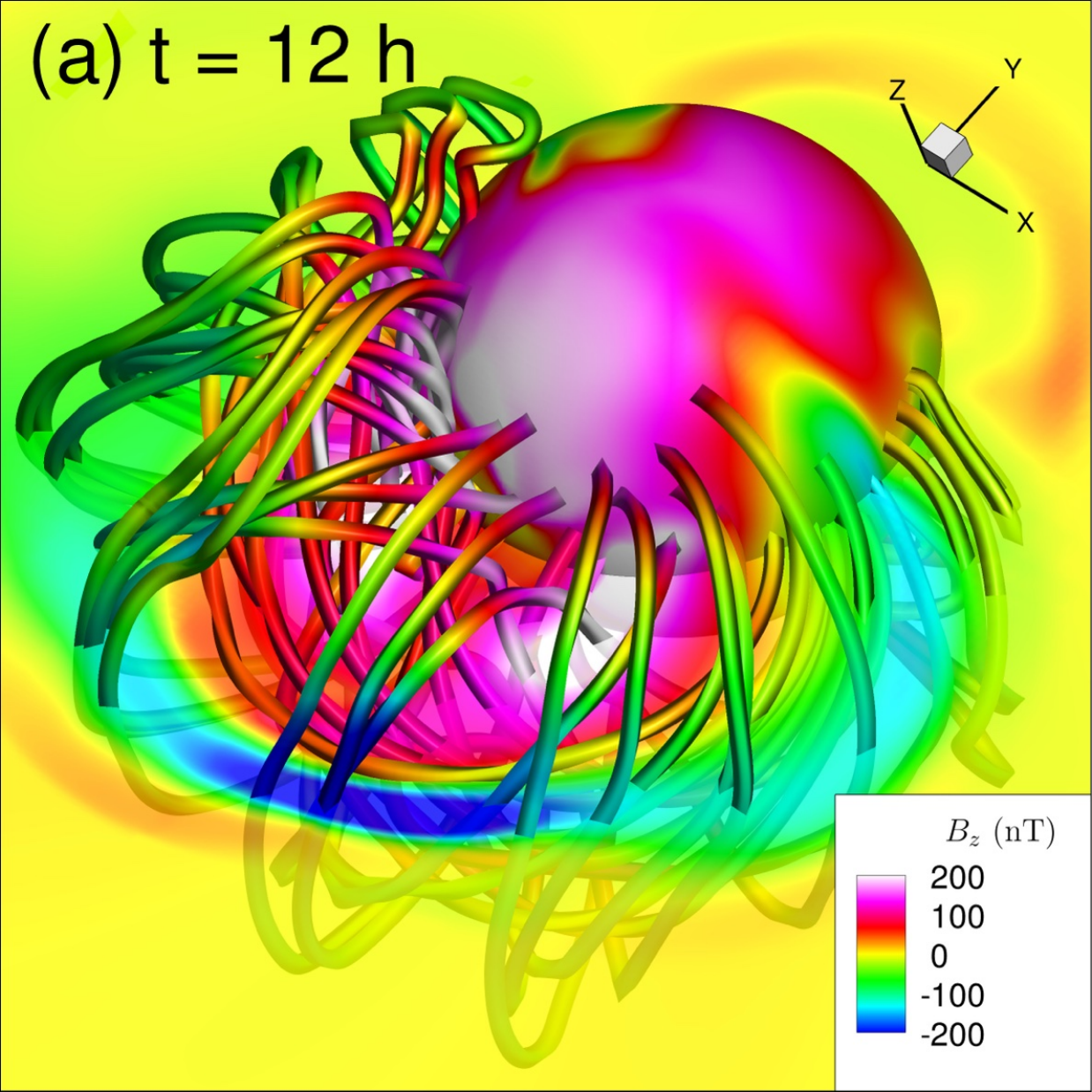}%
        \includegraphics[width=0.4\linewidth]{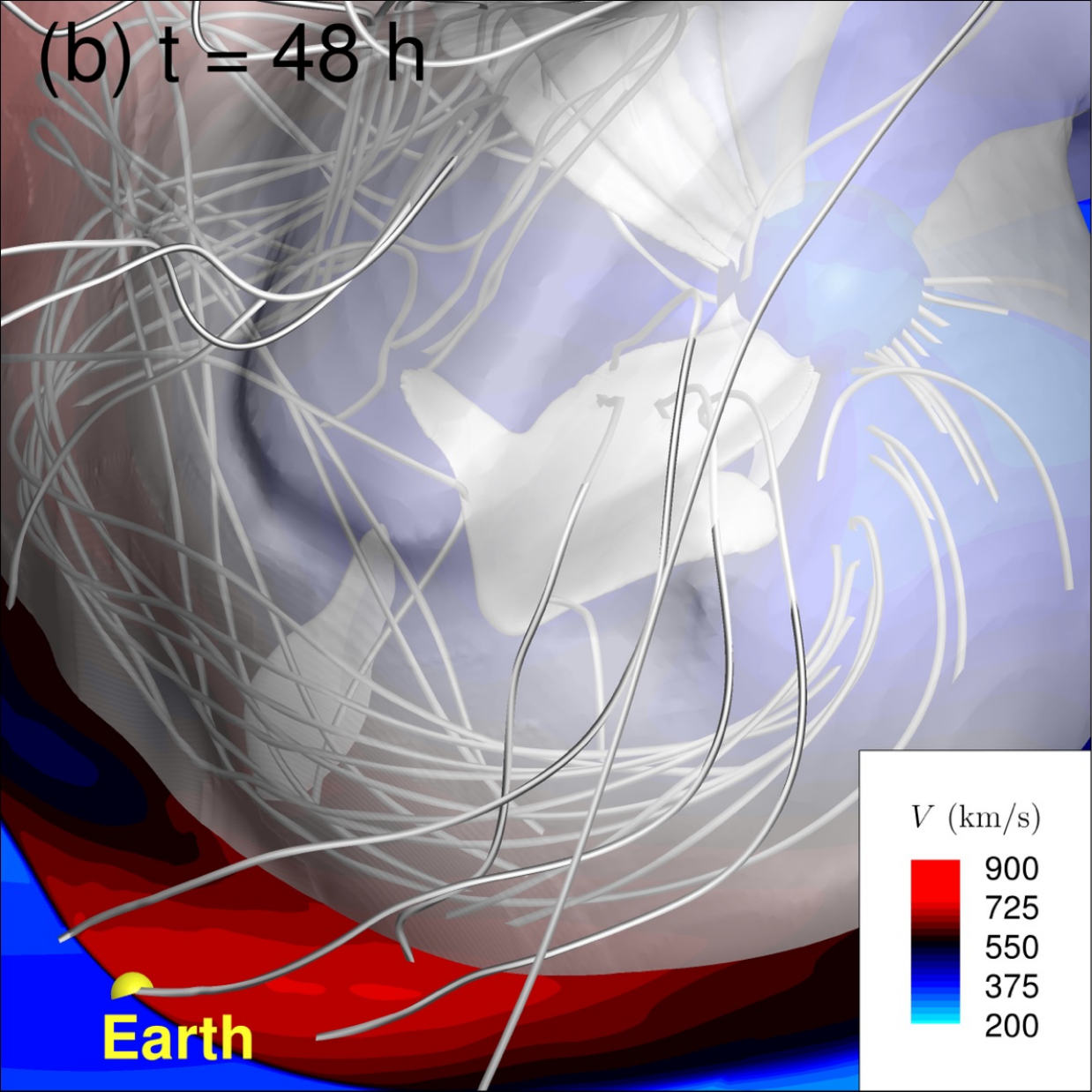}
\vspace{-\lineskip}
        \includegraphics[width=0.8\linewidth]{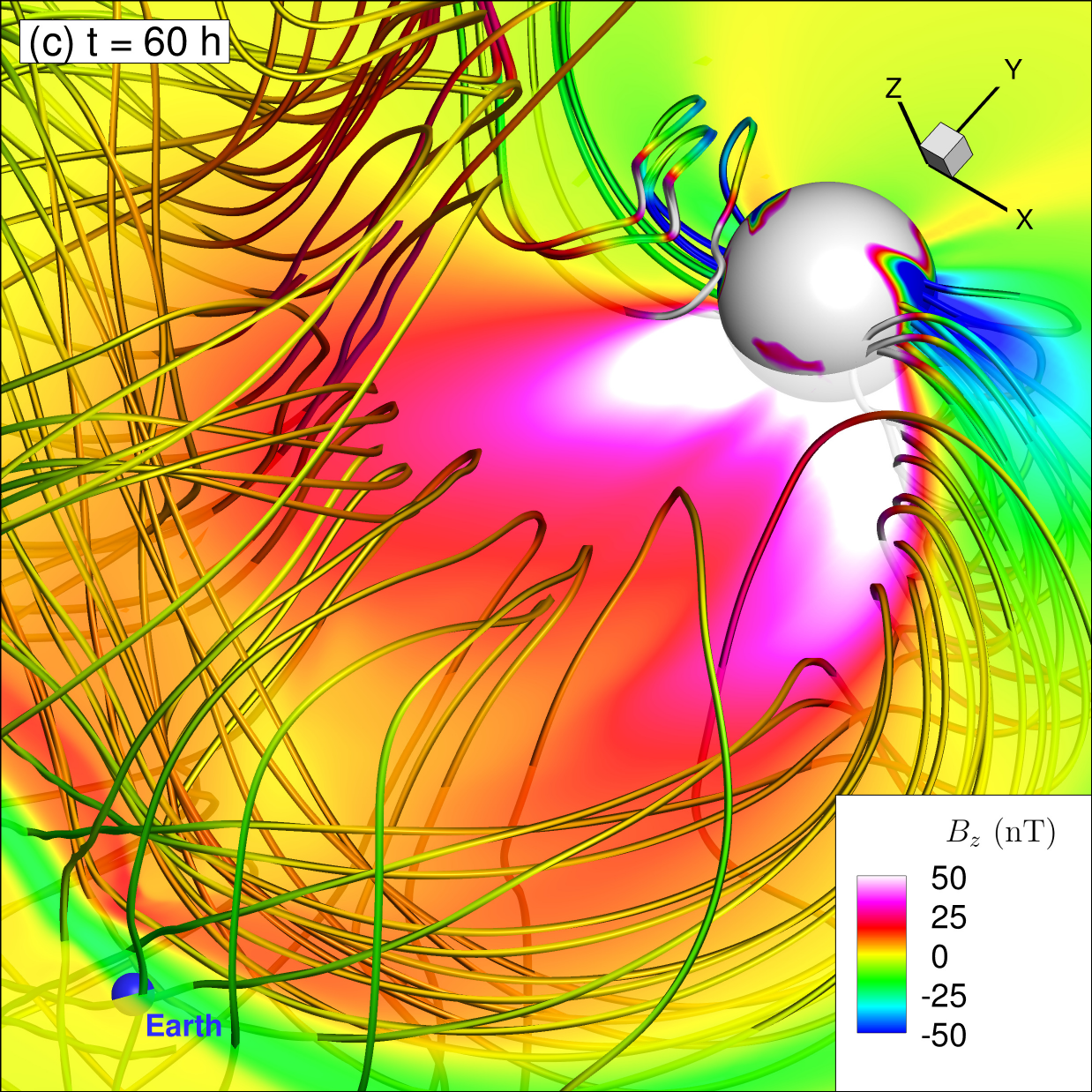}% 
%            \hfill
        \caption{Simulation for the 2014 event in IH at three time instants. (a) demonstrates the IMFR at 12 hours. (b) plots contours of plasma speed $V$ in the ecliptic plane and the plasma $\beta=1$ iso-surface at 48 hours. (c) shows Earth's crossing of the IMFR at 60 hours and contours of $B_z$. In (b) and (c), Earth is represented by a disproportionally large sphere in contrasting colors. A sphere of $24R_\odot$ is placed at the origin and is colored by contours of physical quantities near the inner boundary of IH.}
        \label{fig:STITCH_IH}
\end{figure}

Figures \ref{fig:STITCH_SC}, \ref{fig:STITCH_IH} demonstrate snapshots of the simulation for the 2014 event in the SC and IH domains, respectively.
Figure \ref{fig:STITCH_SC} shows the evolution of 3-D magnetic field lines that originate in the active region and contours of various physical quantities in a plane through $90^\circ$ longitude.
The magnetic field lines remain twisted during their transition.
The ejecta travel faster along open field lines, as shown in Figures \ref{fig:STITCH_SC}a and b, and develop into complicated structures as they interact with the background field, as shown in Figures \ref{fig:STITCH_SC}c and d.
Figure \ref{fig:STITCH_IH} displays the self-similar expansion of the interplanetary magnetic flux rope (IMFR) in the IH domain.
Figure \ref{fig:STITCH_IH}a shows the initial formation 12\,hours after the eruption.
Figure \ref{fig:STITCH_IH}b demonstrates the IMFR enclosed within the $\beta:=\frac{P}{B^2/(2\mu_0)}=1$ iso-surface at the time of arrival of the shock front.
Figure \ref{fig:STITCH_IH}c is taken at an instant during the transit of the CME through the location of Earth.
The simulation produces a well-developed flux rope that mostly propagates in the ecliptic plane.

\begin{figure}[htb!]
         \centering
%            \hfill
        \includegraphics[width=0.25\linewidth]{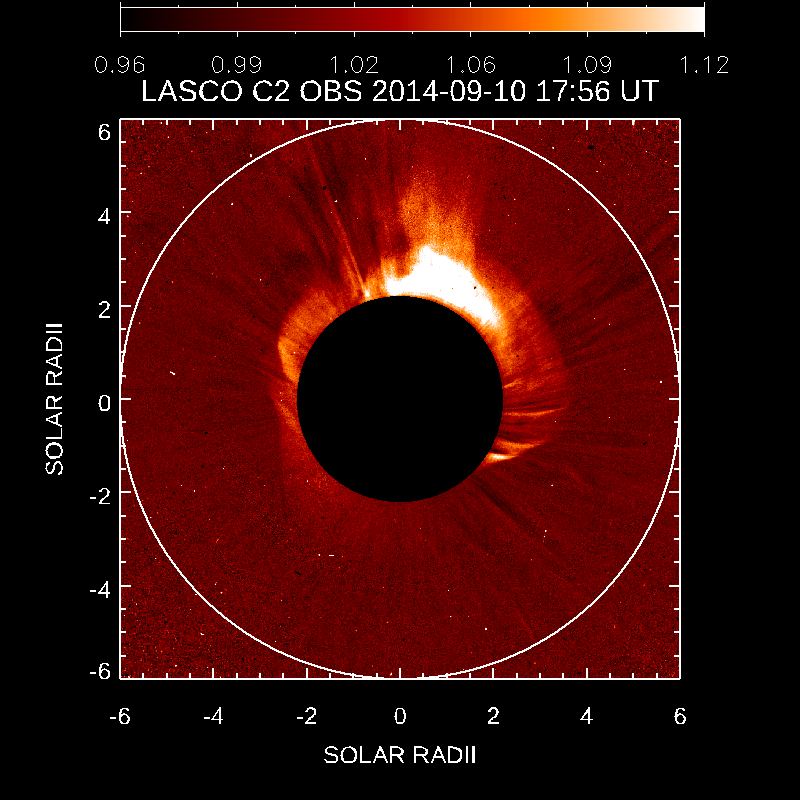}%
        \includegraphics[width=0.25\linewidth]{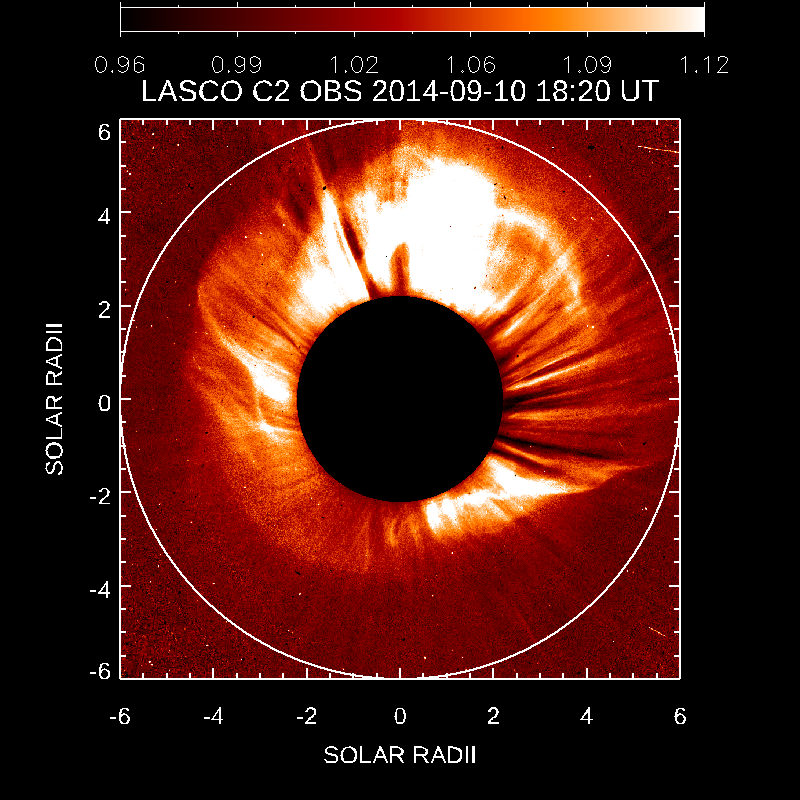}%
        \includegraphics[width=0.25\linewidth]{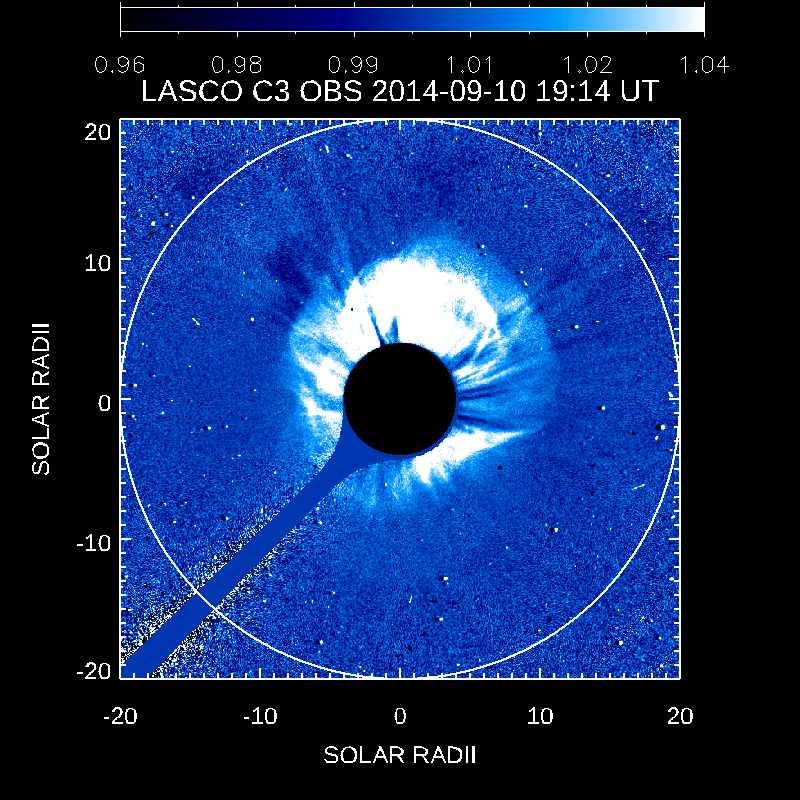}%
        \includegraphics[width=0.25\linewidth]{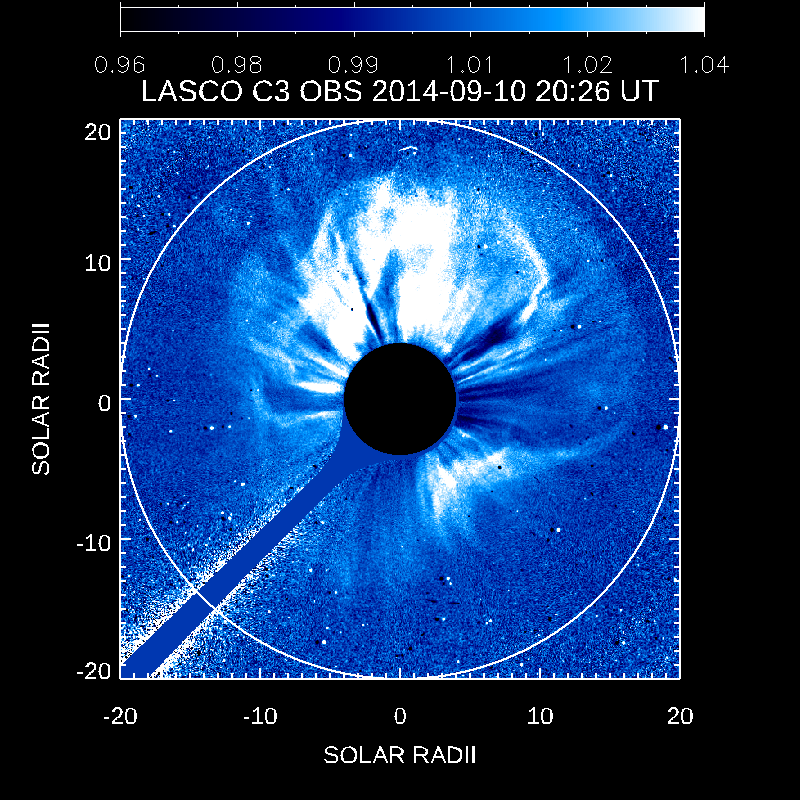}
\vspace{-\lineskip}
        \includegraphics[width=0.25\linewidth]{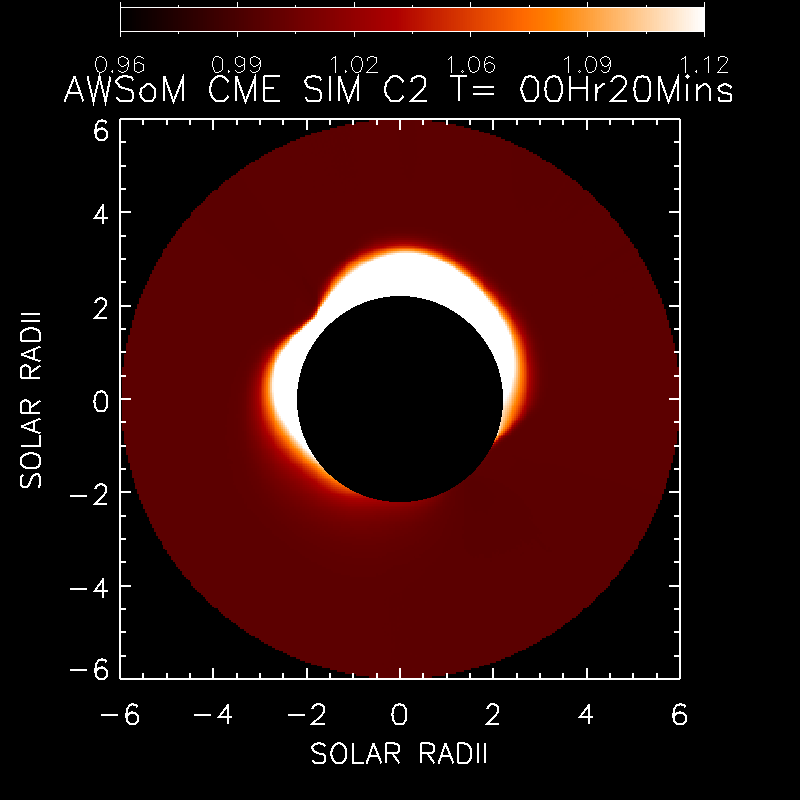}%
        \includegraphics[width=0.25\linewidth]{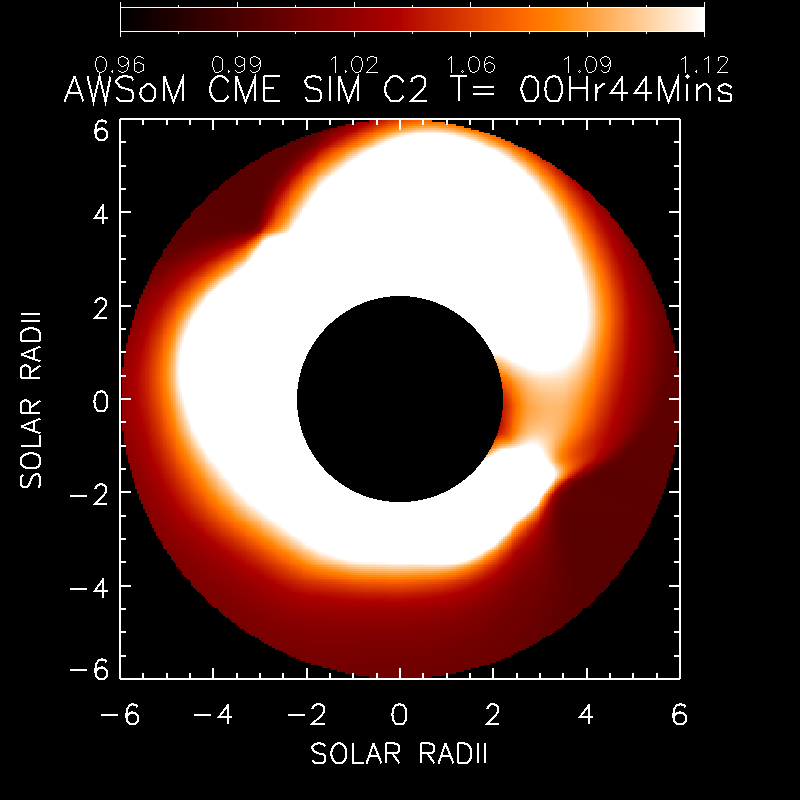}%
        \includegraphics[width=0.25\linewidth]{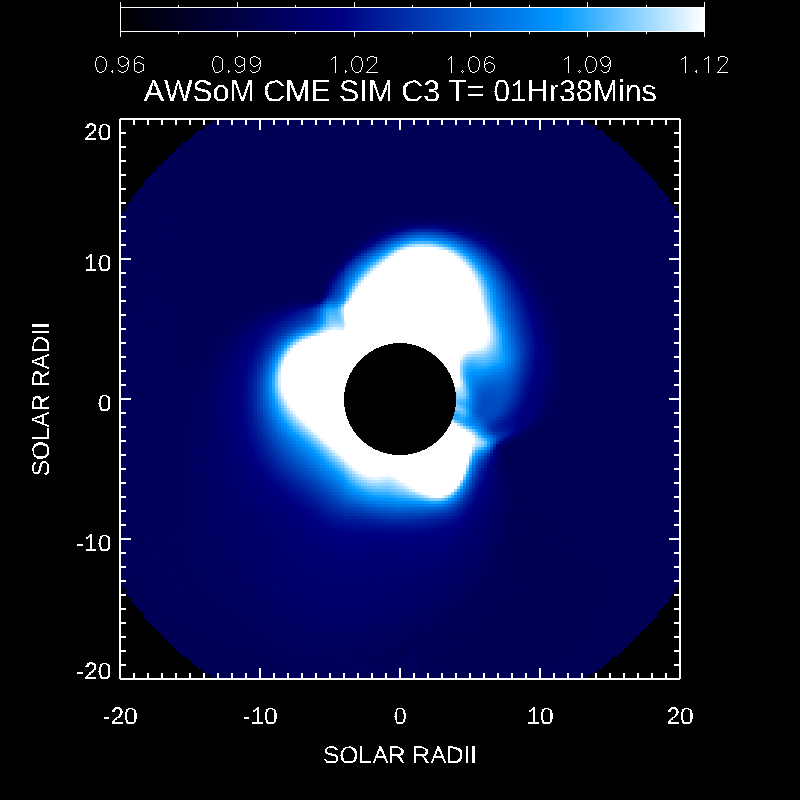}%
        \includegraphics[width=0.25\linewidth]{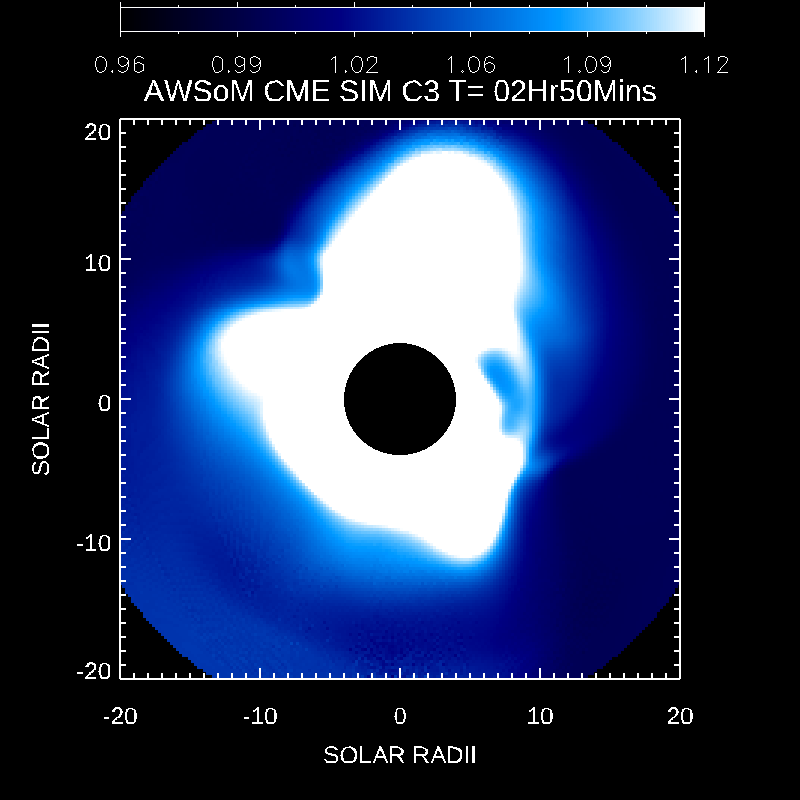}
%            \hfill
        \caption{Observed (top row) vs. synthetic (bottom row) LASCO C2/C3 white light images for the CME of the 2014 event. 
        The timestamps for observations and synthetic images in each column are synchronized.
        The color scales are the same for C2 images but different for observed and modeled C3 images.
        %
        %The time intervals between observations are 36, 78 and 144 minutes, while the intervals between model outputs are 34, 66 and 138 minutes.
        }
        \label{fig:STITCH_WL}
\end{figure}

\begin{figure}[htb!]
    \centering
    \includegraphics[width=0.8\textwidth]{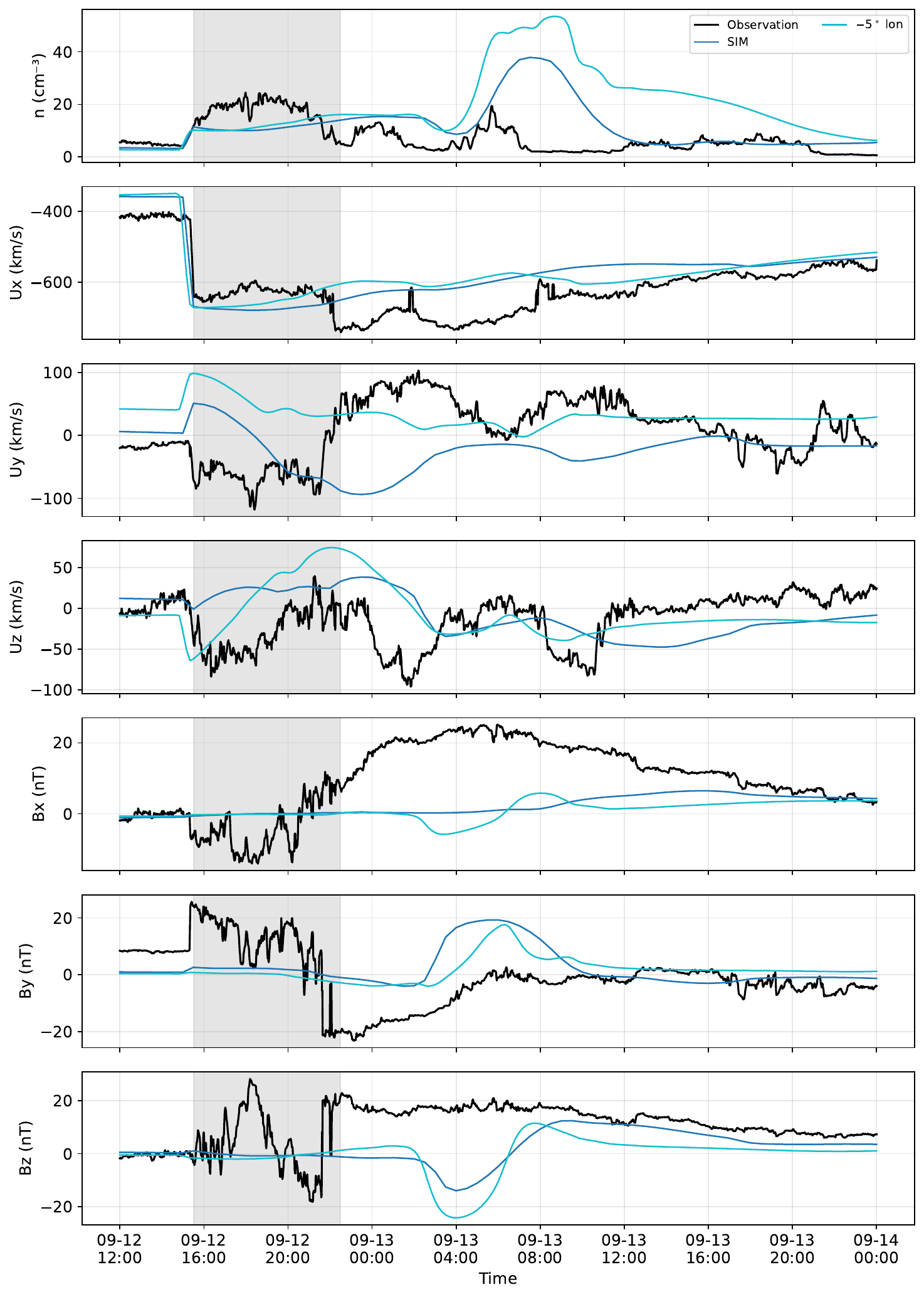}
    \caption{Observed vs. simulated plasma quantities in GSM coordinates at 1\,au for the 2014 event: $n$ is the number density, $U_x$, $U_y$ and $U_z$ the solar wind velocity components, and $B_x$, $B_y$ and $B_z$ the magnetic field components. The shade represents an educated guess of the sheath crossing. We plot the curve for the real location of Earth and the one for a $5^\circ$ shift retrograde along Earth's orbit. Both curves from simulation are shifted by 1.5\,h to match the shock arrival.}
    \label{fig:STITCH_1AU}
\end{figure}

We compare synthetic white light images of the CME with LASCO C2/C3 observations in Figure \ref{fig:STITCH_WL} and simulated plasma quantities vs. in-situ measurements at L1 in Figure \ref{fig:STITCH_1AU}.
The first row of Figure \ref{fig:STITCH_WL} shows observations at 4 instants, while the second row shows simulations at 4 synchronized instants: the time gaps between any two columns are the same for observations and simulations.
For C2 images, we use the same color scale, but for C3 we use different scales for observations and simulations to avoid saturating interesting details.
The outlines of the CME in the synthetic images match those of the observations very well, which suggests good agreement in the linear speed.
On the same scale, the synthetics are more saturated than the observations, especially in directions where the enhancement is weaker in the observations.
One probable reason is the much higher number density at the chromospheric boundary that we use, which makes the ejecta denser and optically thicker everywhere.
%\textcolor{red}{Are you sure? Do you have comparison for simulation with normal inner density?}
%This is also accounted for by a low-resolution background solar wind away from the AR: as the synthetic white light image strongly depends on the pre-eruption background, we do not expect to reproduce observations 
In figure \ref{fig:STITCH_1AU}, in addition to the simulated plasma quantities recorded along Earth's trajectory, we also plot those along a hypothetical trajectory where we find the maximum $|B_z|$.
This trajectory is $5^\circ$ retrograde to Earth's orbit.
%Its data are interpolated from simulated state variables on a shell with 1\,au radius, which we save during simulation.
The shock in the simulation arrives 1.5 hours later than in the observation, so we shift both curves from the simulation accordingly to match the arrival of the shock.
%The number density from simulation, and thus the temperature, is much higher as a result of the increased density on the chromospheric boundary.
We plot the extent of the sheath region in a light shade, which is estimated by identifying the high-frequency oscillations that follow the shock arrival or the beginning of a rotating magnetic field.
We do not expect the MHD states from the simulation to match observations in the sheath region, as the AWSoM model does not resolve the turbulence in the interplanetary magnetic field that gives rise to the oscillatory magnetic field within the sheath -- the AWSoM model only represents the time-averaged turbulence energy content.
The approximate sheath region corresponds to a crossing time of about seven hours in the observations.
In our simulation, around seven hours after the shock front, the magnetic field vector starts to rotate, indicating the arrival of the IMFR.
The IMFR in the simulation is capable of producing a southward $B_z$ with a magnitude of 20\,nT at 1\,au and a recovery phase $B_z$ that closely matches the observation.
The simulated $B_z$ matches the observation in magnitude but has the opposite sign from the observed IMFR.
The agreement in the components of the solar wind velocity is determined by how well the simulation reproduces the shock and how precisely we sample the shock front.
In the Geocentric Solar Magnetospheric (GSM) coordinate system, $U_x$ is the dominant component of the solar wind velocity.
The simulation is able to closely match $U_x$ after the shock and during the recovery phase.
Furthermore, the IMFR crossing in the simulation witnesses a higher number density than the observation, indicating an over-abundance of protons as a result of the increased density at the chromospheric boundary.
%Our simulation is also capable of matching the maximum southward $B_z$ but at a time later than the observations and the $B_z$ during the recovery phase.

\subsubsection{The 2015 Event From AR 12297}

\begin{figure}[htb!]
     \centering
%            \hfill
        \includegraphics[width=0.4\textwidth]{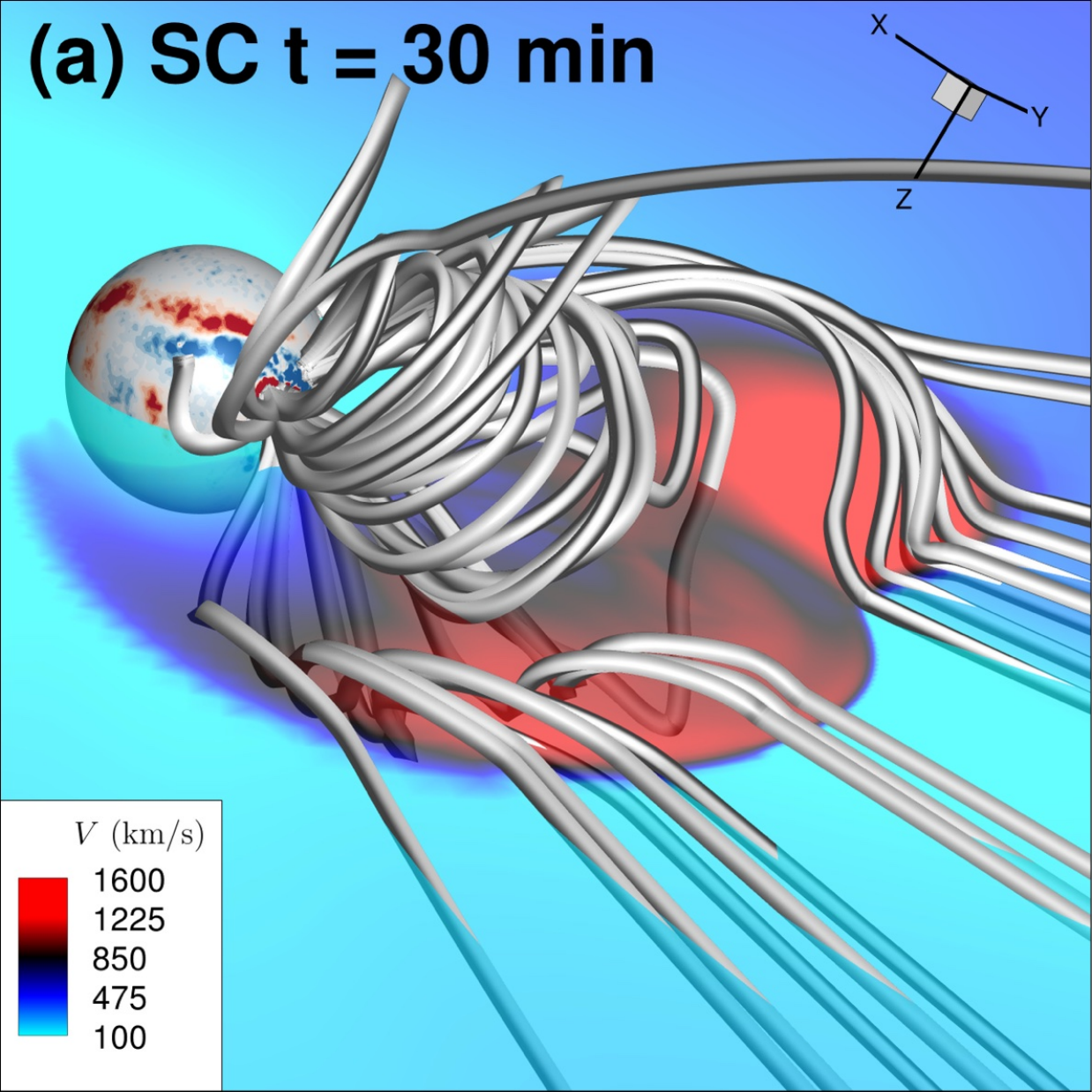}%
        \includegraphics[width=0.4\textwidth]{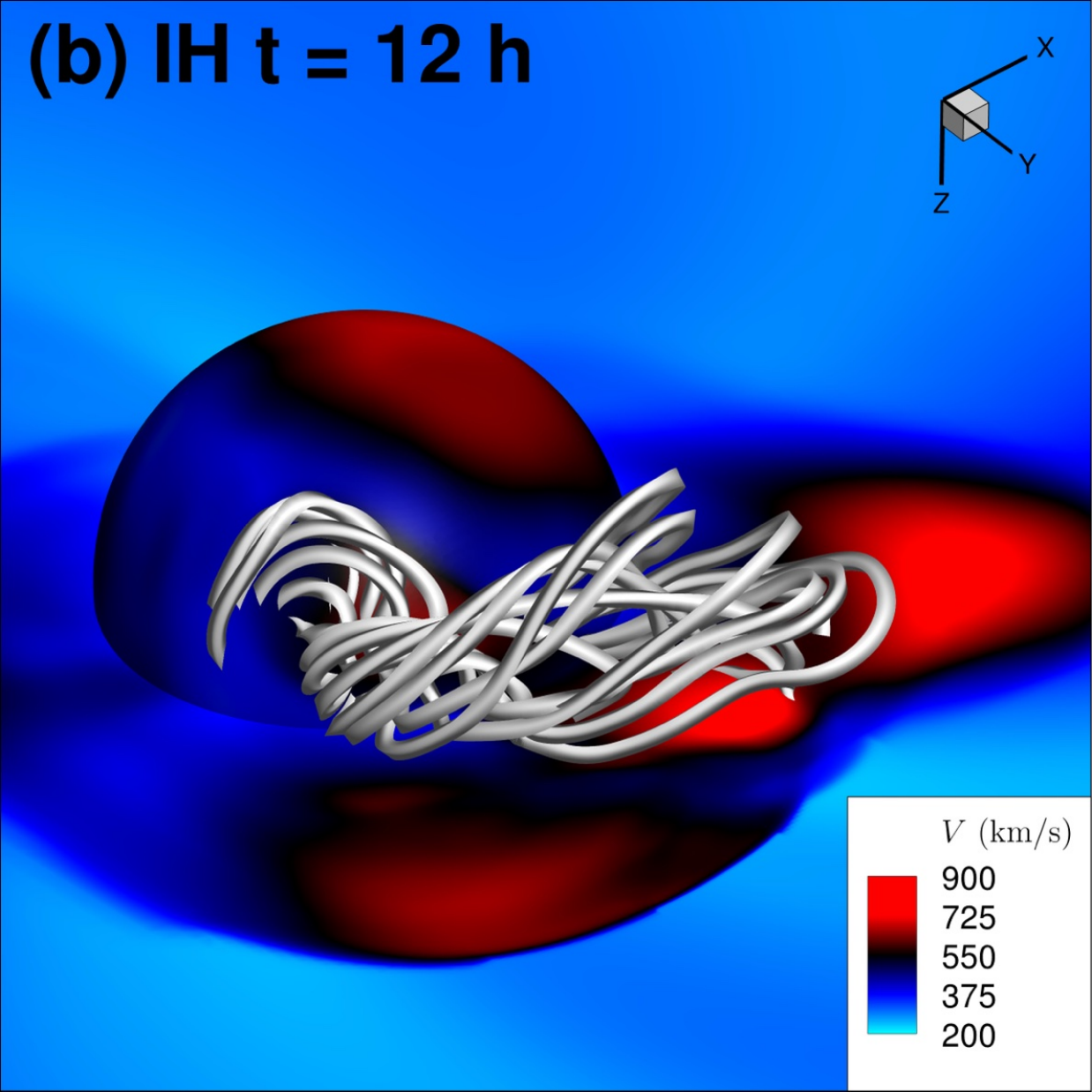}
\vspace{-\lineskip}
        \includegraphics[width=0.8\textwidth]{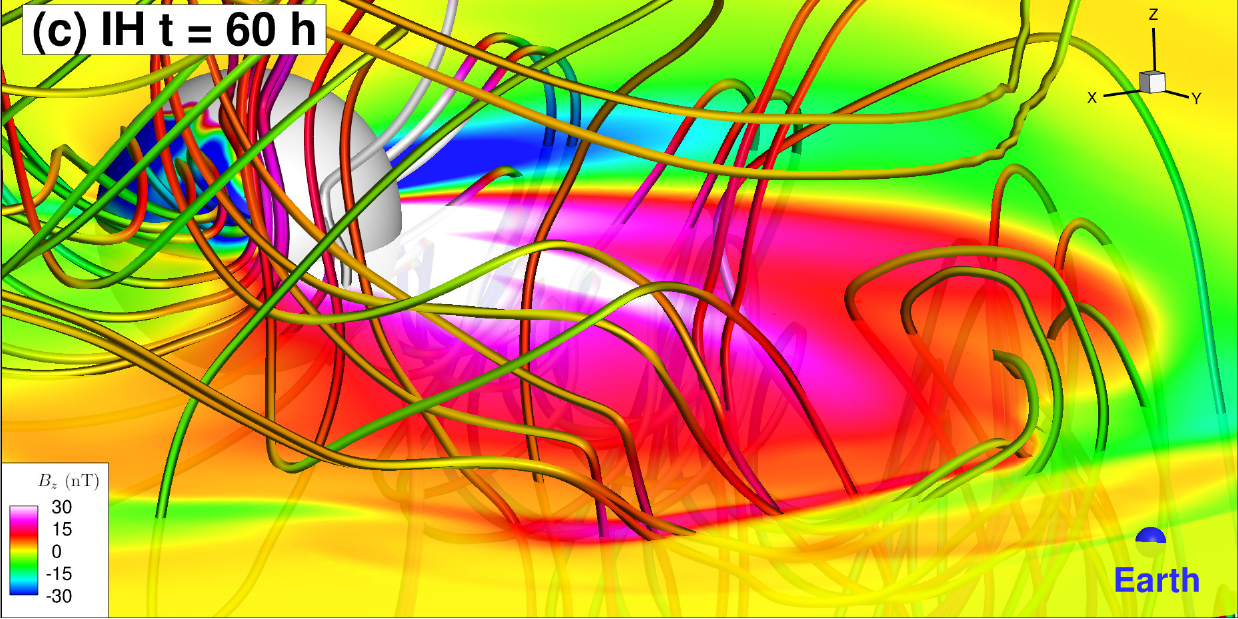}% 
%            \hfill
        \caption{3-D field lines under/over the ecliptic plane in SC and IH for the 2015 event simulation. The ecliptic plane is colored with contours of solar wind speed, $V$, or $Bz$. The three snapshots are taken at 30 minutes, 12 hours and 60 hours simulation time. Note that in (a), (b) the viewing angle is from below (south of) the ecliptic plane, while in (c) it is from above (north of) the ecliptic plane. (a) is in SC while (b) and (c) are in IH. The large sphere where the field lines originate is the Sun in (a) and a sphere of 24$R_\odot$ in (b) and (c). Earth is represented by a disproportionally large sphere in (c).}
        \label{fig:AR12297_3D}
\end{figure}

Although the NLFFF solution for the 2015 event has similar metrics for the divergence-free condition $\langle|f_i|\rangle$, averaged sine of angle $\theta_J$, and residual force $R_f$ as for the 2014 event, it produces a spontaneous eruption strong enough to propagate to Earth.
 In Figure \ref{fig:AR12297_3D} are three snapshots of the propagation of the CME in the simulation.
The ecliptic plane serves as a reference for the direction in which the CME travels.
Figure \ref{fig:AR12297_3D}a, taken below the ecliptic plane in the SC domain, shows early signs that the CME is traveling southward relative to Earth.
In the IH domain, Figure \ref{fig:AR12297_3D}b is a sphere with a radius of $24R_\odot$, just outside the inner boundary of IH, from below the ecliptic plane.
The southward motion of the CME continues and develops into an IMFR.
In Figure \ref{fig:AR12297_3D}, an instant is shown when the ICME has arrived at Earth from above the ecliptic plane.
The southward propagation of the CME causes the location of Earth to miss the main body of the IMFR, which is visible below the ecliptic plane.
As the location with maximum $|B_z|$ in this plane is far from Earth, we expect to find a hypothetical trajectory out of the ecliptic plane to examine the strongest magnetic field we can capture at 1\,au.

\begin{figure}[htb!]
         \centering
         \includegraphics[width=0.25\textwidth]{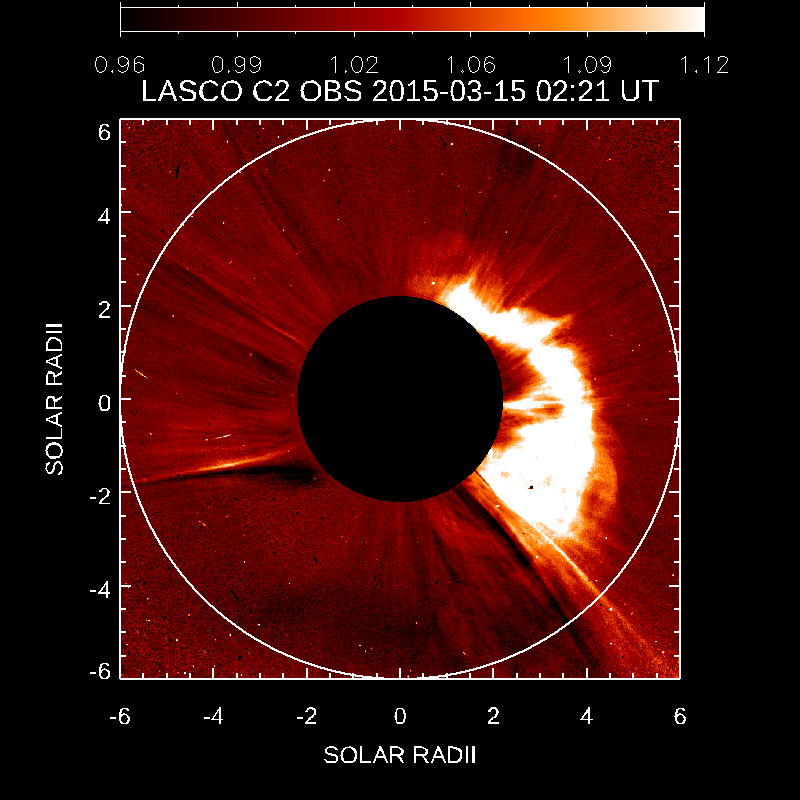}%
        \includegraphics[width=0.25\textwidth]{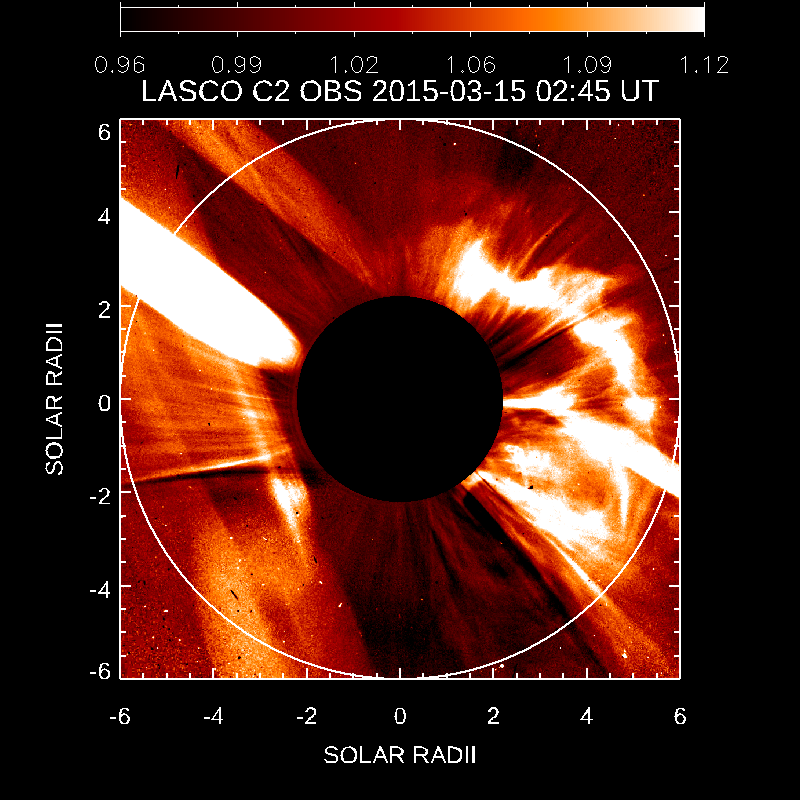}%
        \includegraphics[width=0.25\textwidth]{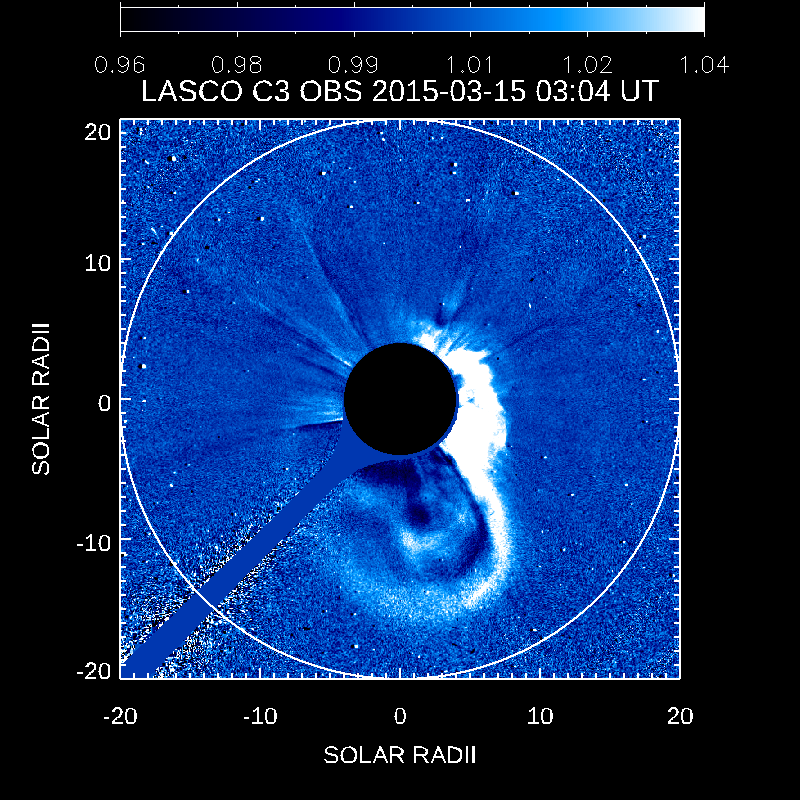}%
        \includegraphics[width=0.25\textwidth]{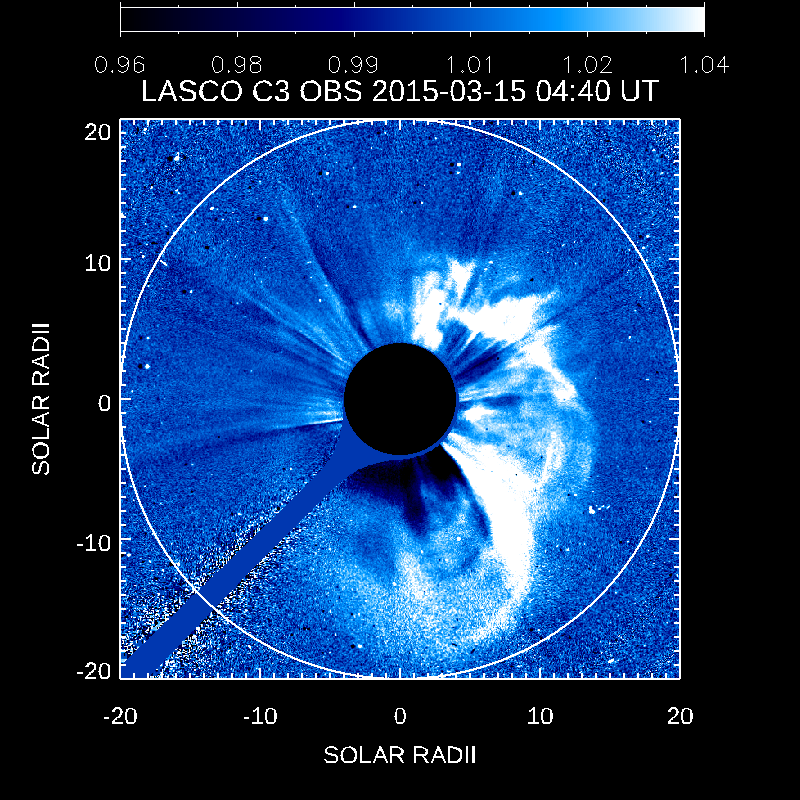}
\vspace{-\lineskip}
        \includegraphics[width=0.25\textwidth]{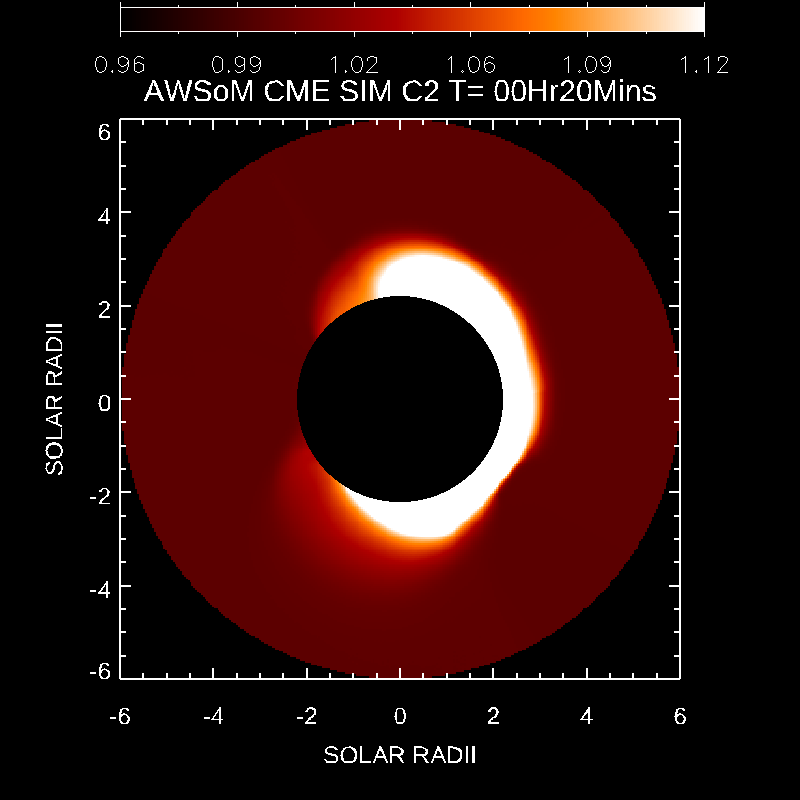}%
        \includegraphics[width=0.25\textwidth]{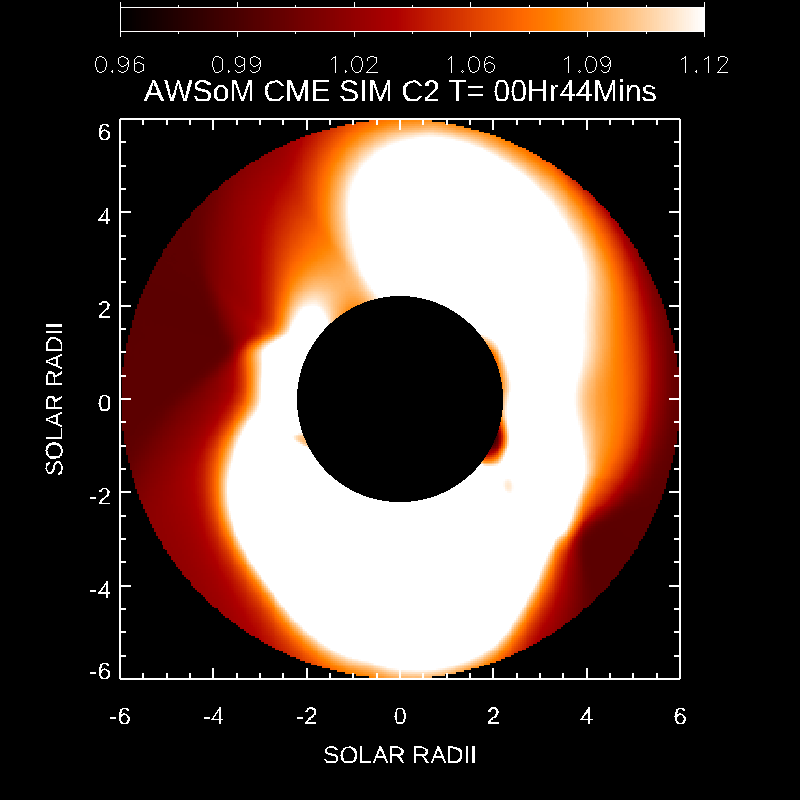}%
        \includegraphics[width=0.25\textwidth]{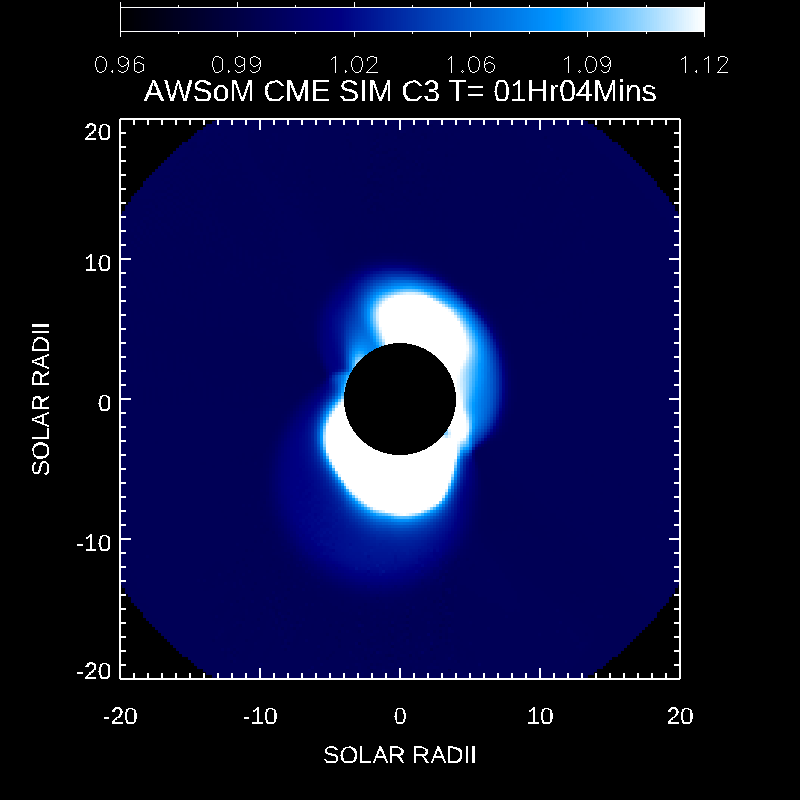}%
        \includegraphics[width=0.25\textwidth]{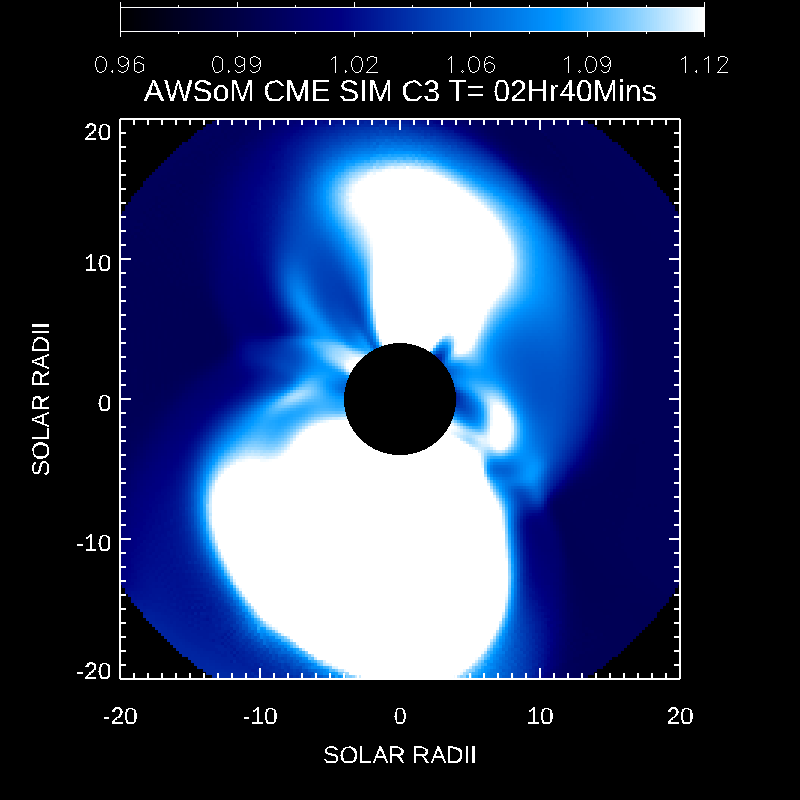}
         \caption{Observed (top row) vs. synthetic (bottom row) LASCO C2/C3 white light images for the CME of the 2015 event. The timestamps for observations and synthetics are synchronized. Note that C2 images share the same scale while C3 images use different color scales in observations and synthetic images.}
         \label{fig:AR12297_WL}
\end{figure}

\begin{figure}[htb!]
     \centering
         \includegraphics[width=0.8\textwidth]{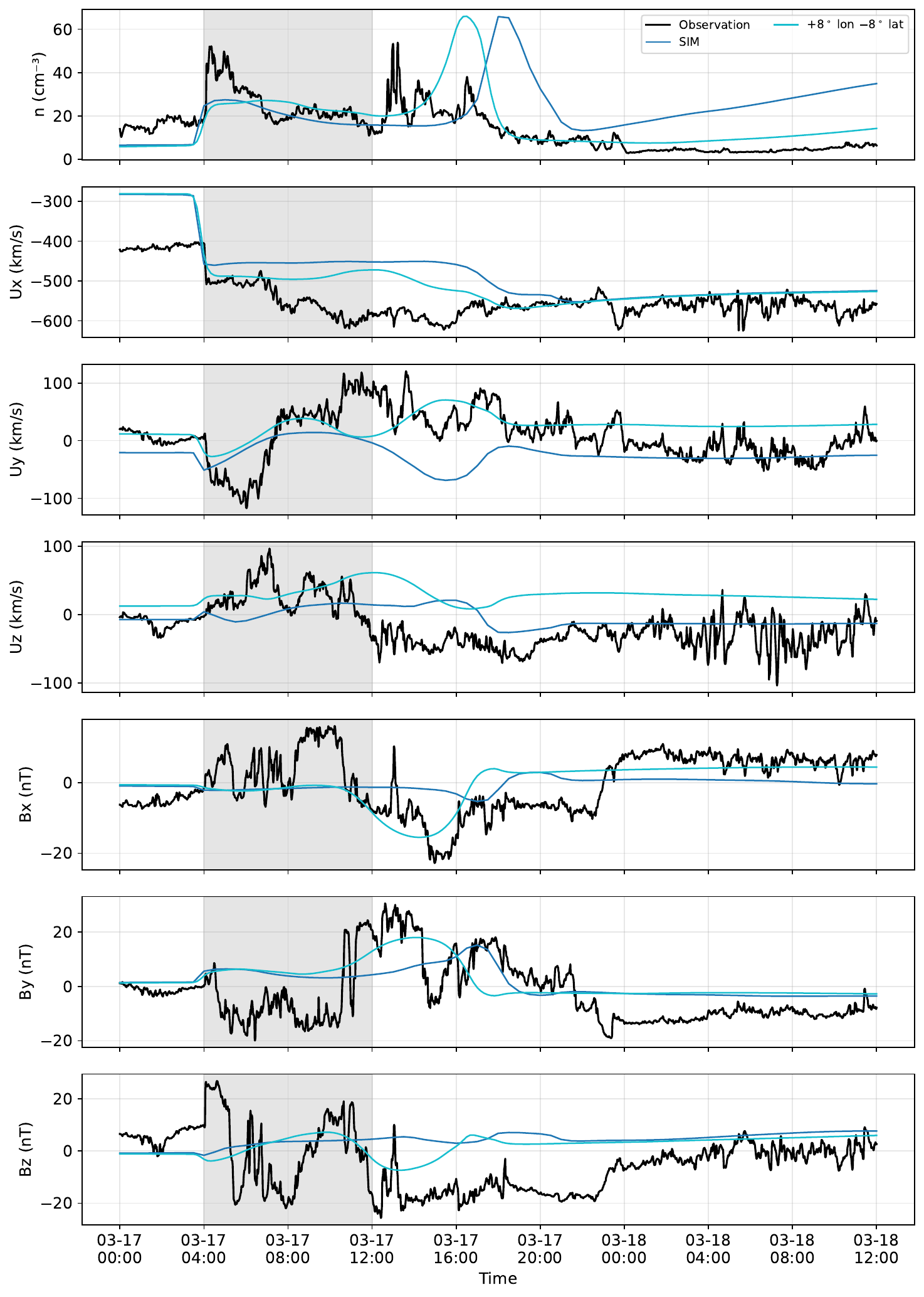}
         \caption{Observed vs. simulated plasma quantities in GSM coordinates at L1 for the 2015 event : $n$ is the number density, $U_x$, $U_y$ and $U_z$ the solar wind velocity components, and $B_x$, $B_y$ and $B_z$ the magnetic field components. The shade represents an educated guess of the sheath crossing. We plot the curve for the real location of Earth and the one for a $8^\circ$ prograde, $8^\circ$ southwards shift. They are shifted by 11 hours and 9 hours respectively.}
         \label{fig:AR12297_1AU}
\end{figure}

Figure \ref{fig:AR12297_WL} shows the observed and synthetic C2/C3 white light images at four synchronized instants of time. The first column shows reasonable agreement between the observed and simulated C2 images. 
In the second column, the observations seem to have major errors on the left half of the image. 
The extent on the right side is comparable with the simulation. 
Except for an overestimated northward extent, synthetic C3 images show a high resemblance in shape to the observed ones, such as fan-shaped ejecta to the west limb and the southward-traveling oval shape.
Similarly to the 2014 event, the enhancement is larger in the simulation, most likely due to an over-abundance of material provided by the high density we set at the chromospheric boundary.
Figure \ref{fig:AR12297_1AU} compares the solar wind states from the simulation with in-situ measurements.
In addition to the real trajectory of Earth, we plot the solar wind along a hypothetical trajectory that is $8^\circ$ prograde and $8^\circ$ southward relative to the location of Earth.
This corresponds to a $11^\circ$ shift away from the real trajectory.
The curve that corresponds to the real location of Earth only shows the arrival of a weak IMFR, with only a significant change in $B_y$.
As demonstrated by Figure \ref{fig:AR12297_3D}, the reason is that the body of the IMFR in our simulation travels southward relative to the ecliptic plane and misses Earth.
On the other hand, we find better agreement along the hypothetical trajectory.
As before, we indicate the approximate sheath region in light shade, which lasts eight hours in the observation.
In the simulations, the magnetic field starts to rotate after roughly the same amount of time after the shock arrival, marking the beginning of the IMFR.
The signs in IMFR $B_x$, $B_y$, and $B_z$ all match observations.
The magnitude of the southward $B_z$ from the simulation is about half of the observed value, while the magnitudes of $B_x$ and $B_y$ are closely matched.
Despite difficulties in exactly reproducing a pre-eruption background solar wind, the jump in the three velocity components and their variations over time indicates that the simulation produces a shock that is very plausible.
The recovery phase $U_x$, $B_z$, and $B_x$ agree well, while the number density is higher due to the chromospheric boundary condition.

\subsubsection{The 2013 Event From AR 11692}

\begin{figure}[htb!]
     \centering
%            \hfill
        \includegraphics[width=0.4\linewidth]{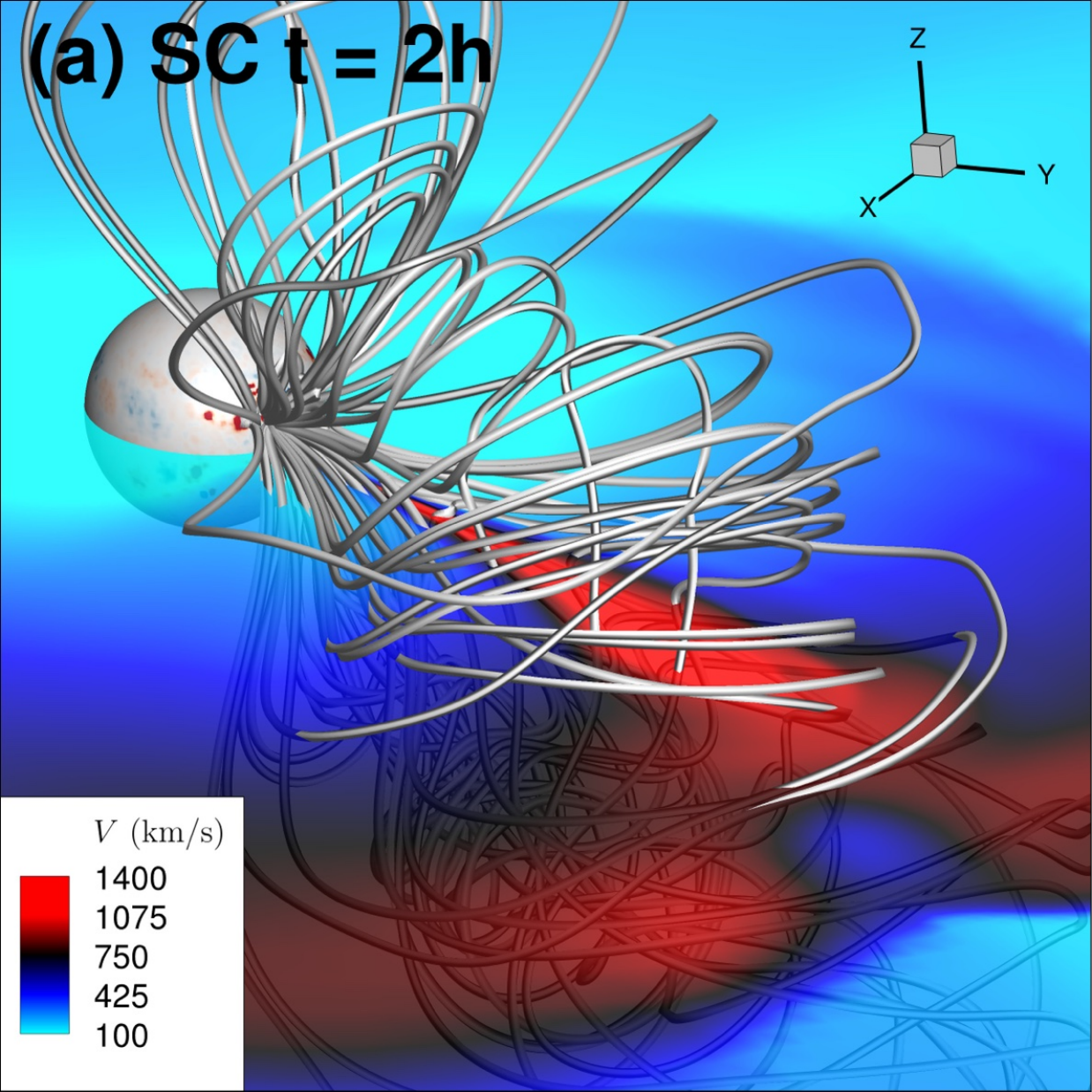}%
        \includegraphics[width=0.4\linewidth]{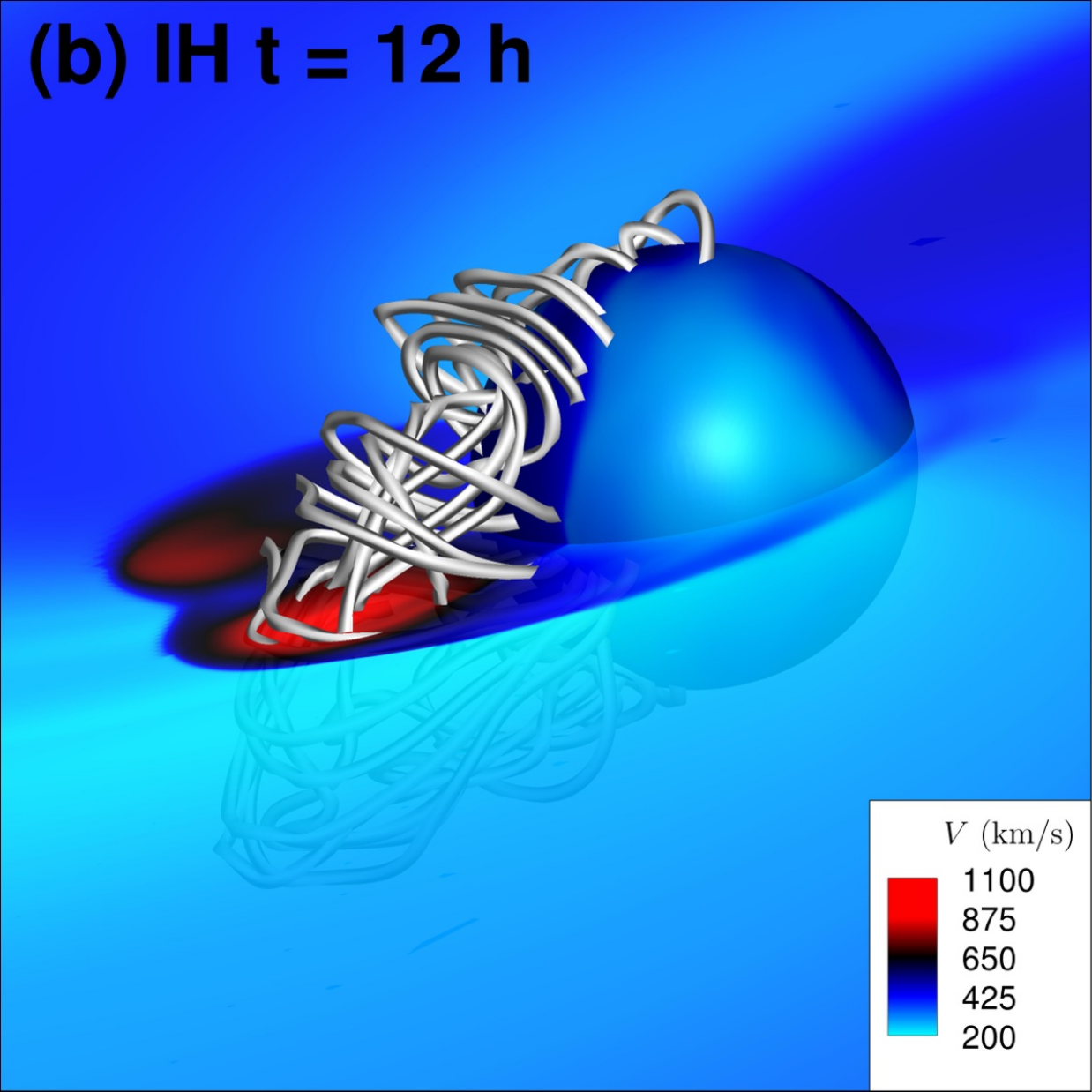}
\vspace{-\lineskip}
        \includegraphics[width=0.8\linewidth]{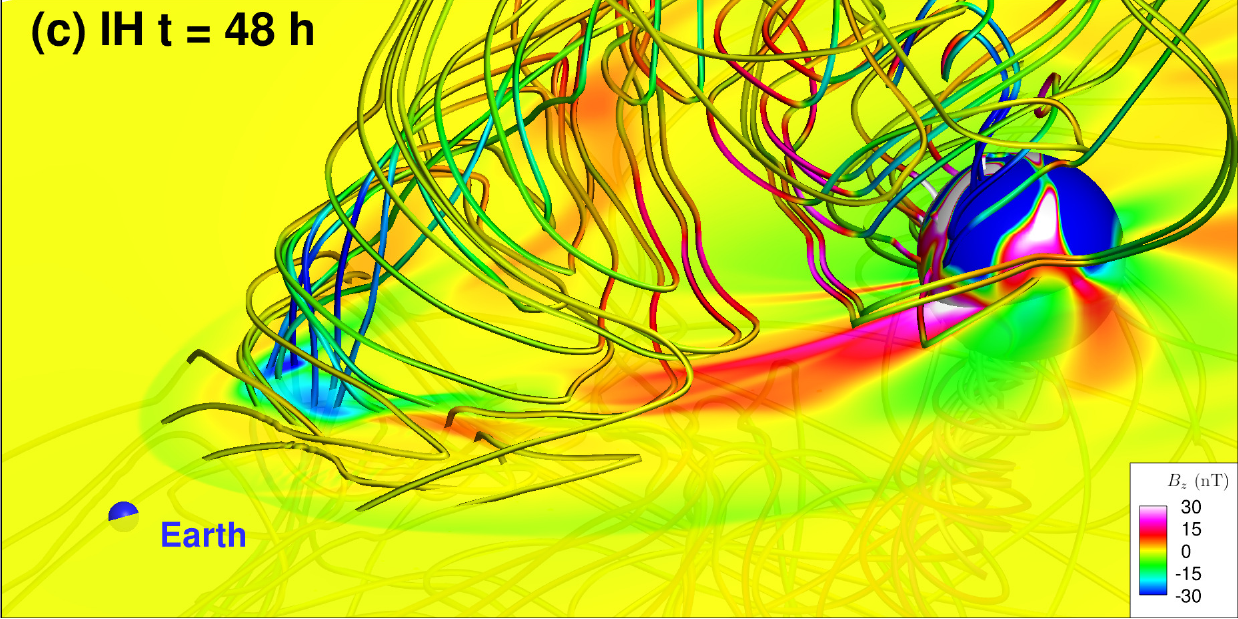}% 
%            \hfill
        \caption{Snapshots of the 2013 event simulation in the SC and IH domains. The slice is the ecliptic plane in all three panels. (a) shows the MFR in SC at 2 hours and the SE-NW axis of the MFR (i.e. the MFR body is nearly perpendicular to the ecliptic plane). (b) shows the perpendicular IMFR as it appears in the IH domain initially, (c) shows the IMFR in the IH domain at a later time and its position relative to Earth. Earth is represented by a disproportionally large sphere. The larger sphere at the origin represents the Sun in (a) and a $24R_\odot$ sphere in (b) and (c).}
        \label{fig:AR11692_3D}
\end{figure}

The NLFFF solution for the 2013 event also produces a strong and spontaneous shock due to the initial force imbalance in the approximate NLFFF.
To experiment with different shock strengths, we run simulations with $c_B=1.25$ and $c_B=1.0$ (see Section \ref{sec:preprocessing}).
The results presented in this section are mostly from the $c_B=1.25$ simulation because it results in a closer CME arrival time.
Figure \ref{fig:AR11692_3D} shows three snapshots during the simulation with $c_B=1.25$.
Figure \ref{fig:AR11692_3D}a shows the MFR in the SC domain at two hours and its attitude relative to the ecliptic plane.
Interestingly, this MFR has a SE-NW axis, which has also been reported in previous studies \citep{Wu_2016,pal_2017}.
This MFR orientation continues as the IMFR initially forms at 12 hours in the IH domain, as shown in Figure \ref{fig:AR11692_3D}b.
Figure \ref{fig:AR11692_3D}c demonstrates the IMFR as it approaches 1\,au at 48 hours and the position of its front relative to Earth.
The shock front is marked by the weak and arc-shaped green contour that leads the IMFR in the ecliptic plane.
Sheath material carrying little magnetic field is found between the shock front and the IMFR.
The location with maximum southward $B_z$ is a few degrees retrograde relative to Earth.

\begin{figure}[htb!]
         \centering
         \includegraphics[width=0.25\textwidth]{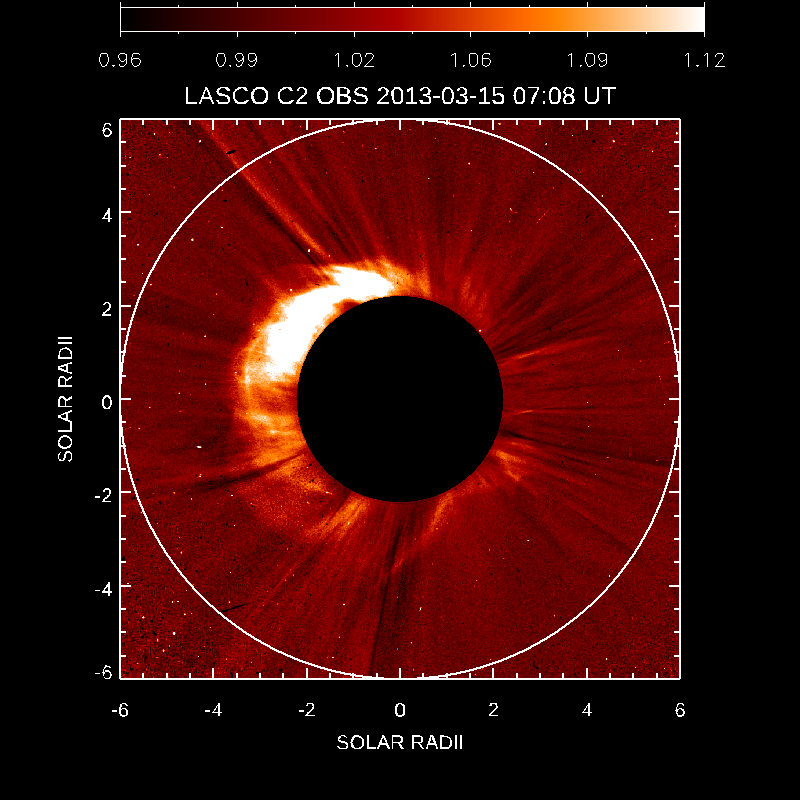}%
        \includegraphics[width=0.25\textwidth]{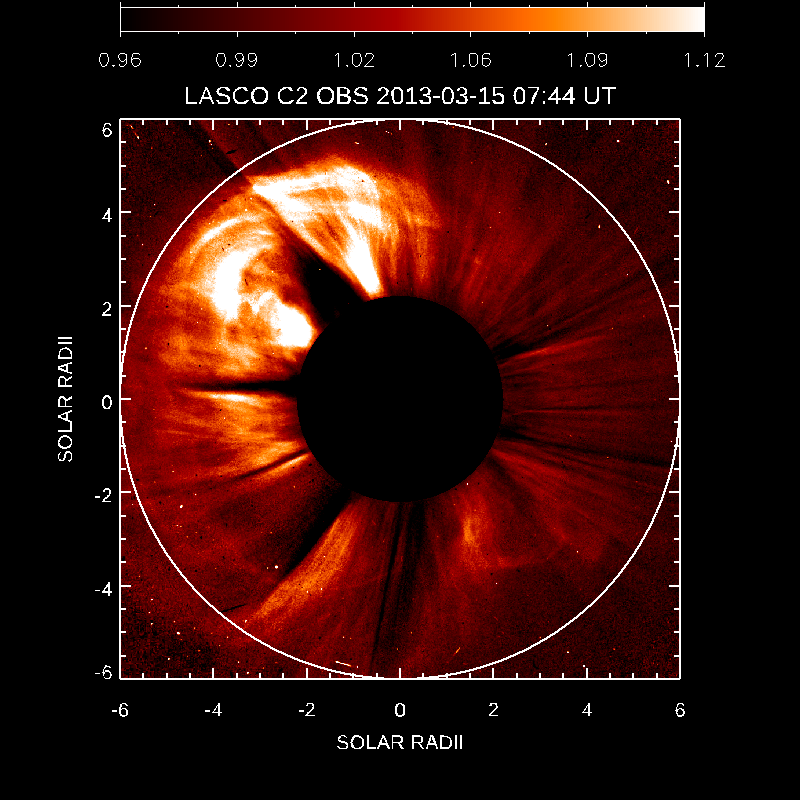}%
        \includegraphics[width=0.25\textwidth]{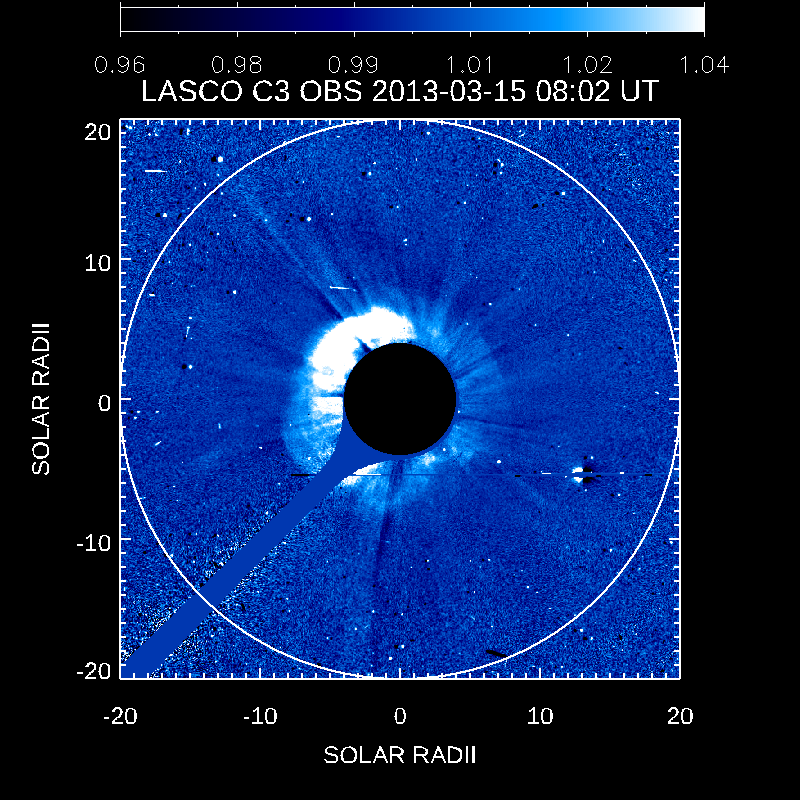}%
        \includegraphics[width=0.25\textwidth]{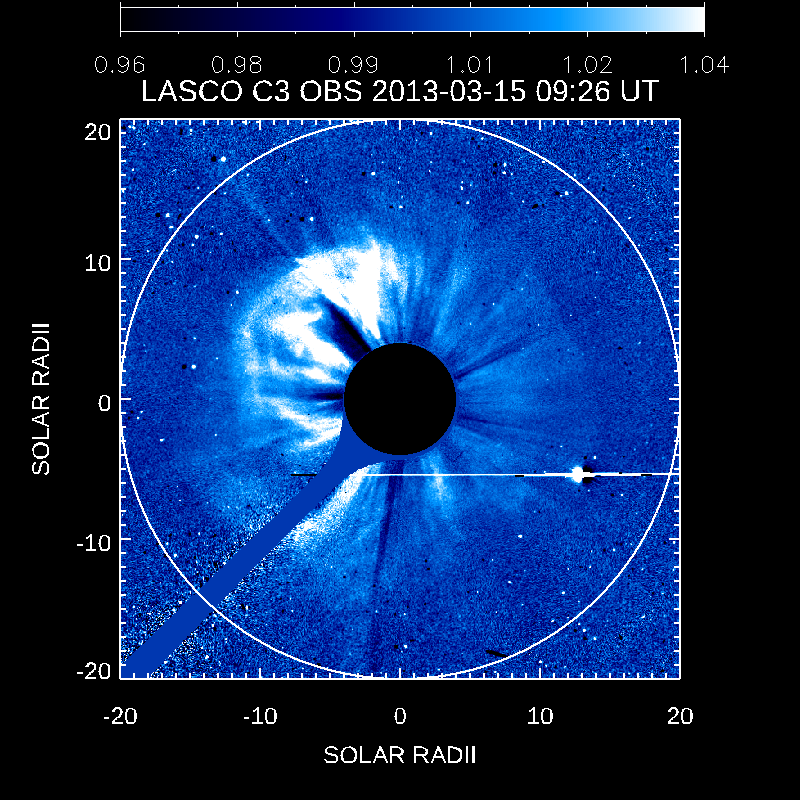}
\vspace{-\lineskip}
        \includegraphics[width=0.25\textwidth]{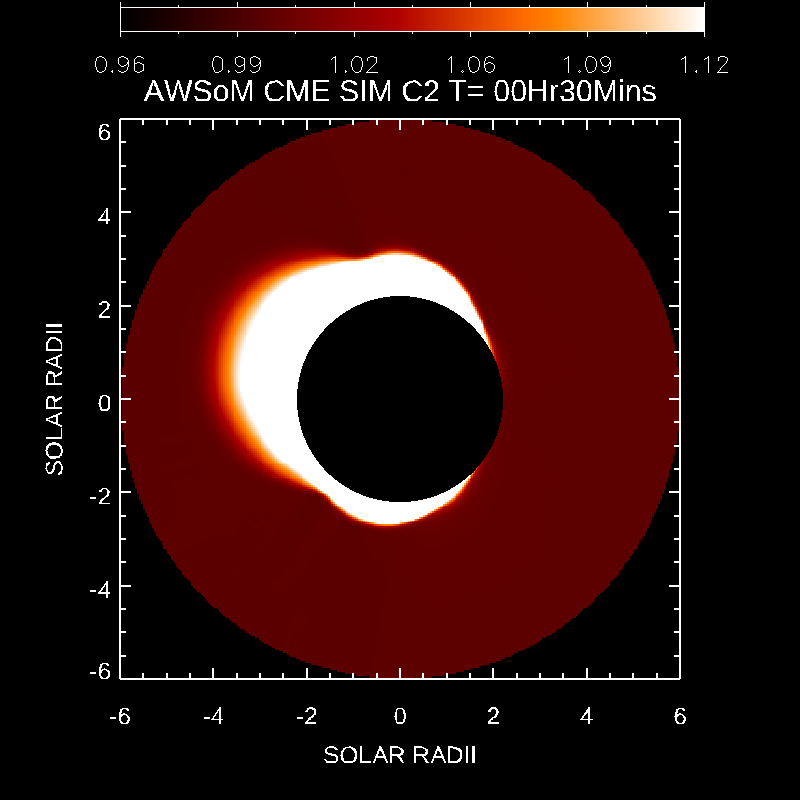}%
        \includegraphics[width=0.25\textwidth]{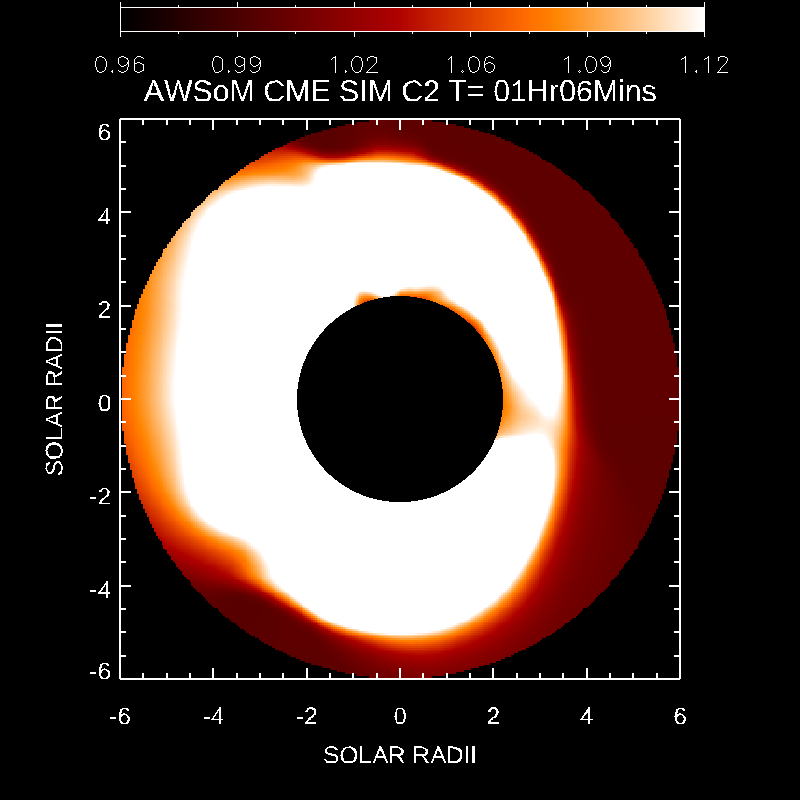}%
        \includegraphics[width=0.25\textwidth]{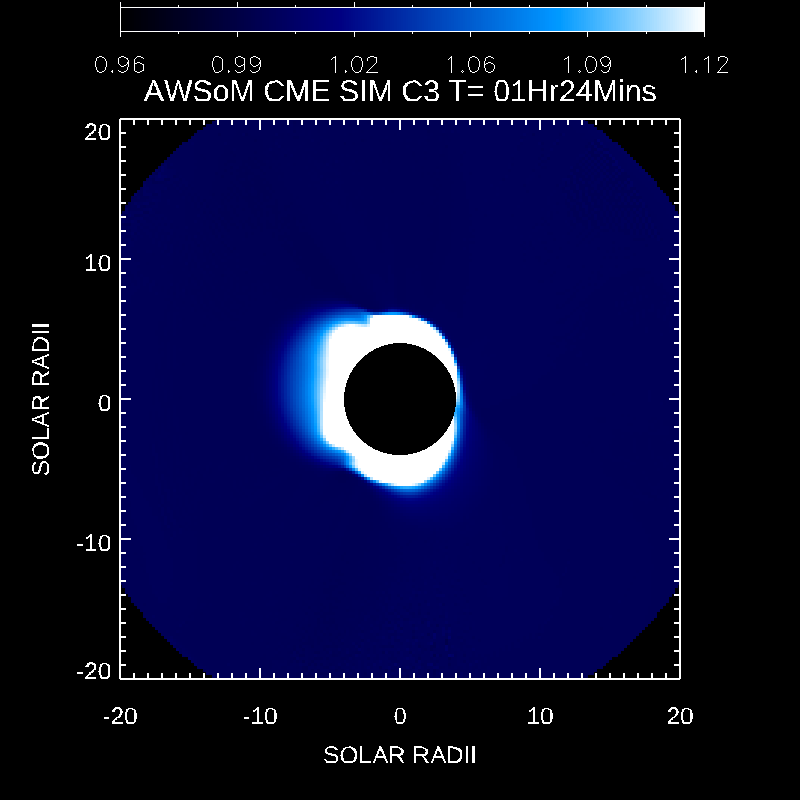}%
        \includegraphics[width=0.25\textwidth]{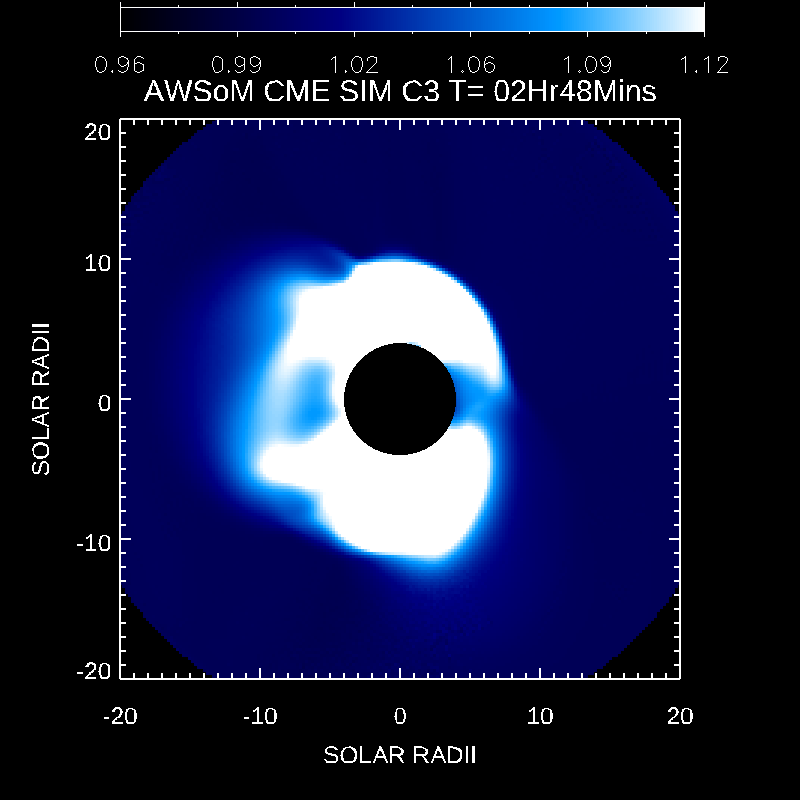}
         \caption{Observed (top row) vs. synthetic (bottom row) LASCO C2/C3 white light images for the CME of the 2013 event. The timestamps for observations and synthetic images are synchronized.
         The color scales are the same for C2 images but different for observed and modeled C3 images.}
         \label{fig:AR11692_WL}
\end{figure}

\begin{figure}[htb!]
     \centering
         \includegraphics[width=0.8\textwidth]{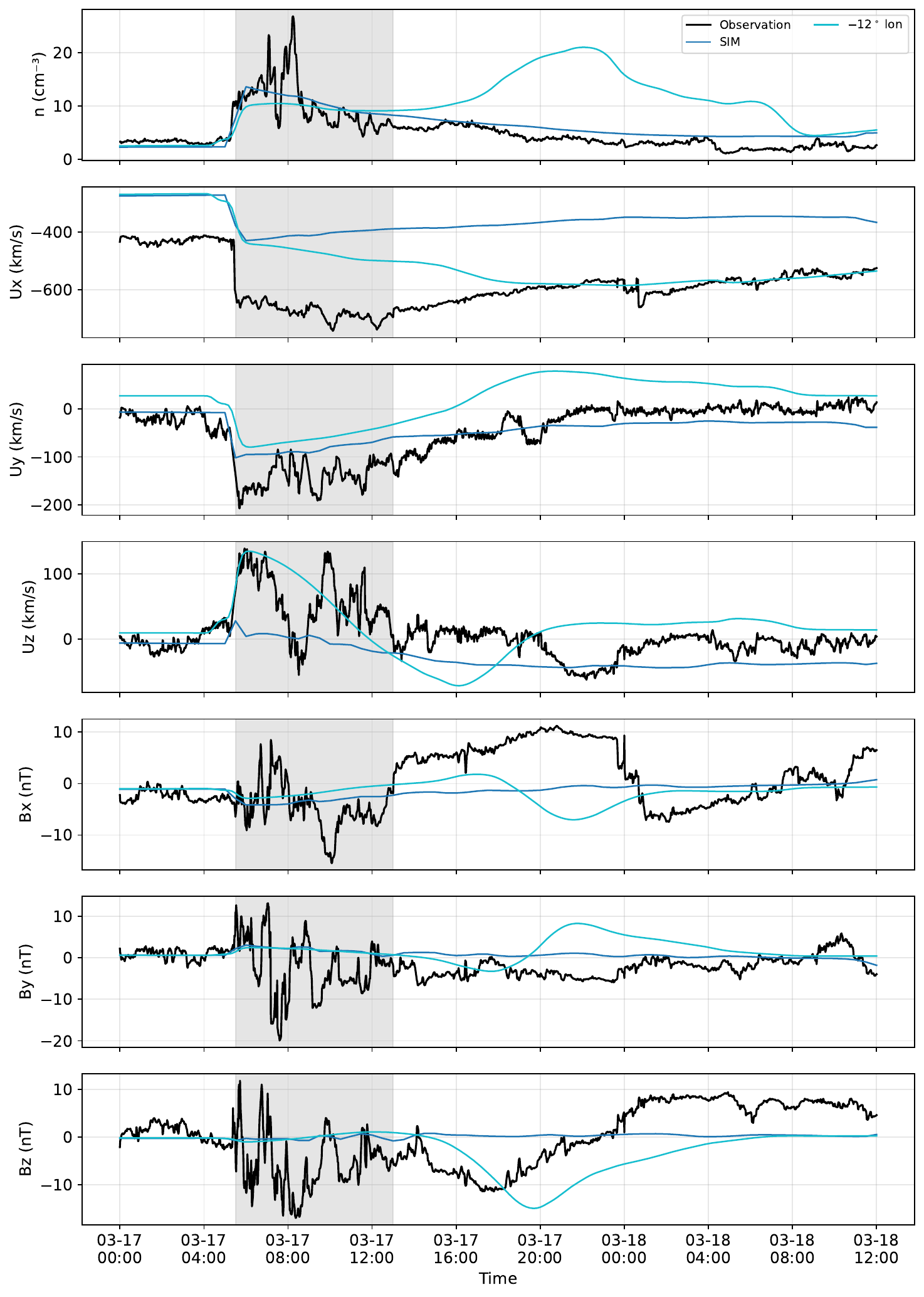}
         \caption{Observed vs. simulated plasma quantities in GSM coordinates at L1 for the 2013 event : $n$ is the number density, $U_x$, $U_y$ and $U_z$ the solar wind velocity components, and $B_x$, $B_y$ and $B_z$ the magnetic field components. The shade represents an educated guess of the sheath crossing. 
         %We include results from two simulations using different values for $c_B$, the factor for field strength reduction. 
         %For the simulation with $c_B=1.25$, w
         In addition to the real trajectory of Earth, we select a hypothetic one $12^\circ$ retrograde in Earth's orbit. The two curves are shifted by 9 and 2 hours respectively to match the shock arrival time. 
         %For the simulation with $c_B=1.0$, we pick a hypothetic trajectory $11^\circ$ retrograde. The two curves are shifted by 14 and 17.5 hours respectively.
         }
         \label{fig:AR11692_1AU}
\end{figure}

Figure \ref{fig:AR11692_WL} compares synthetic white light images from the simulation using $c_B=1.25$ with observations.
The timestamps for the four pairs of comparisons are synchronized.
Again, we find a close match in the silhouettes of the synthetics and the observations, resulting in a close linear speed between the two.
The FoV with weaker enhancement also appears saturated in the simulation as a result of the higher number density.
Figure \ref{fig:AR11692_1AU} plots the plasma states from simulations vs. in-situ measurements at L1.
For the simulation with $c_B=1.25$, we find better agreement in the magnetic field along a hypothetical trajectory $12^\circ$ retrograde in Earth's orbit.
The CME arrival time along Earth's real trajectory is 9 hours late, while that along the hypothetical trajectory is only 2 hours late.
Their curves are shifted accordingly to match the shock arrival time with the observation.
Although the background solar wind is slower than that observed, we find a matching value in the jump in $U_x$ and $U_z$ across the shock front.
The number density agrees well before and after the shock, suggesting that the shock strength in the simulation is correct.
As usual, we identify the sheath region and mark it in light shade.
The sheath crossing time is around seven hours.
%A $11^\circ$ retrograde shift in location finds a better magnetic field and a shock arrival time 17.5 hours early.
After the sheath region, the simulation sees a significantly stronger magnetic field along the hypothetical trajectory.
The profile of $B_z$ closely matches the observation in magnitude and sign.
Other notable findings from the simulation are an over-estimated number density, a matching recovery phase speed, and rotating $B_x$ and $B_y$ in the IMFR.

%Along the hypothetic trajectory, we find excellent speed in the recovery phase.
%The sheath crossing lasts about seven hours and we find matching signs and magnitude in $B_z$ after that.
%Along the hypothetic trajectory for $c_B=1.0$, we find matching signs and larger magnitudes for all three components of the magnetic field in the initial hours.
%Since the shock is much stronger than observations, this simulation also records much stronger number density, speed, and ion temperature post-shock.
In addition, we perform another simulation for this event with $c_B=1.0$ to investigate the impact of $c_B$ on the results. 
It produces a MFR with the same orientation of its axis and a stronger shock that arrives at L1 14 hours early.
We find the location with the strongest $B_z$ to be $11^\circ$ retrograde in Earth's orbit, where the shock arrives 17.5 hours early.
This illustrates how $c_B$ changes the free magnetic energy and the strength of the CME, and thus how $c_B$ acts as a parameter that we can tune to match simulations to observations. 

\subsection{Discussion}

As a first-time endeavor to initiate CME simulations in the SWMF only with a vector magnetogram, we have achieved successful results.
Simulations for all three events show the development of a MFR and later an IMFR, either with controlled photospheric driving or with spontaneous eruption.
We consider NLFFF eruptions with controlled driving to be an ideal paradigm of our methods, but the spontaneous eruption is not very different from the eruption of a Gibson-Low MFR, which has been our standard approach to generating CMEs.
The highlight of the comparison with observables is the highly-matching silhouettes of synthetic LASCO C2/C3 images.
Furthermore, for the three geo-effective events, we are able to identify hypothetical trajectories near the real location of Earth where the crossing of the IMFR produces magnetic signatures that often match observations in terms of magnitudes and signs.
The errors in the CME arrival times are usually small and can be as low as 1.5 hours. 
This agreement is, of course, a result of tuning the magnetic field strength in the erupting active region with the $c_B$ parameter. 
On the other hand, we found that the same $c_B=1.25$ worked well for all three events, which is quite promising.

We notice that CMEs whose MFR axes are not close to being perpendicular to the ecliptic plane produce less accurate results.
In the case of the 2015 event, a previous study finds that the CME is likely to be deflected about $12^\circ$ towards Earth during its transit \citep{wang2015}.
In the case of the 2013 event, the $11^\circ-12^\circ$ displacement needed for Earth to intercept the main body of the MFR may be attributed to a few possible reasons.
First, as real pre-eruption filament channels can build up the energy for a CME over days to weeks, the discretionary selection of one vector magnetogram before the eruption may not correspond to a nearly force-free state of the magnetic field at the same time.
Second, while the $0.175^\circ$ ($0.088^\circ$ effective) resolution is among the highest in simulations using the AWSoM model, it is still insufficient to resolve specific field topologies.
Together, these reasons may result in a solution with a high $\sigma_J$, which may contain an artificial Lorentz force that introduces an inaccurate deflection to the CME.
%Third, real pre-eruption filament channels can build up the energy for a CME over days to weeks, so the use of the vector magnetogram observed at a single time may not be sufficient.
%First, using the vector magnetogram observed at only one instant may not adequately capture all physics in the relaxation for a NLFFF solution.
Finding the exact reasons behind these deflections is beyond the scope of this paper but motivates improvements to our method in the future.

Despite the success of this novel method, our simulations are driving the existing hardware and software to their limits.
The greatest hardware-related challenge is the high computational costs. 
Each magneto-frictional run requires more than 40,000 core-hours on Intel Xeon Platinum 8280 ``Cascade Lake'' CPUs.
The cost for a quasi-steady state run before eruption is relatively low, at 10,000 core-hours.
The cost for the time accurate run is the highest: each run takes around 350,000 core-hours to complete.
This makes the total computational cost for each simulation 400,000 core-hours.
The entire project has cost over ten million core-hours over more than two years, including experimentation and failed simulations.
The high computational cost is mainly due to the fine grid cells near the inner boundary of the SC domain, which are needed to resolve the magnetic field structures in the erupting active region. 
%AWSoM requires to resolve sharp gradients in the transition region \citep{vanderHolst:2014awsom}.
The initial hour of simulation takes a disproportionately long time -- we estimate it to be over one third of the core-hours for the entire simulation, even when we turn on local time stepping and apply AMR efficiently.
A potential solution is to apply the so-called ``Boris Correction'' \citep{Gombosi:2002}, which relaxes the MHD equations to allow for larger time steps.
This option is routinely used in magnetospheric simulations, but it is not yet supported in the SC or IH domains of the SWMF.
Meanwhile, improving our software infrastructure is beneficial for enhancing the accuracy of our proposed methods, such as including the preprocessing method by \citet{Wiegelmann:2006} and adding support for time-dependent boundary conditions in the relaxation of NLFFF solutions.
Another approach to reduce the systematic error is to improve the resolution of the potential background magnetic field ($\mathbf B_P$) in the AR.
BATS-R-US normally operates on CPUs with Message Passing Interface (MPI) in Single Instruction, Multiple Data (SIMD) mode, which poses a limit on the resolution of lookup tables such as $\mathbf B_P$.
A $\mathbf B_P$ with $1^\circ$ resolution in both longitude and latitude and 100 logarithmically varying radial grid points ($100\times360\times180$ grid points in total) is our common choice.
However, this $\mathbf B_P$ induces a non-negligible error in its curl when interpolated in high-resolution regions, such as the AR.
This leaves $\mathbf B_0=\mathbf B_P$ with non-zero $\mathbf J_0$ when solving for a NLFFF solution.
It is uniformly neglected as the background in BATS-R-US, but it can give rise to errors in calculating the Lorentz force under the $\mathbf J_0=0$ assumption, i.e., $\mathbf J \times \mathbf B \approx \mathbf J_1\times(\mathbf{B_0+\mathbf{B_1}})$, which then affects the NLFFF solutions in magneto-frictional runs.
Here, $\mathbf J_0$ and $\mathbf J_1$ are the current in $\mathbf B_0$ and $\mathbf B_1$, respectively.
%\textcolor{red}{Doesn't the local B0 table have much higher resolution?}
Finally, we acknowledge that state-of-the-art methods in NLFFF reconstruction can be implemented into the SWMF in the future.
For example, the global NLFFF relaxation method by \citet{Jiang2012} may eliminate the residual Lorentz force near where our NLFFF merges with the background potential field and eliminate spontaneous shock formation in those regions.
The TMFM \citep{Daei2023, Guo2024} can be tested on ARs where the NLFFF solution we find does not have high quality in terms of $\sigma_J$, such as AR 11692, where the 2013 event originates.

\section{Conclusion}
\label{section_conclusion}
For the first time, we run Sun-to-Earth CME simulations in the SWMF using only a vector magnetogram.
We select three geo-effective space weather events to test this novel method.
For each event, we implement an established magneto-frictional method that reconstructs an approximate NLFFF solution in the AR from a vector magnetogram observed prior to the CME.
The AR is artificially enlarged by a factor of two and has an effective angular resolution of $0.88^\circ$.
We achieve the same quality in the reconstructed NLFFF in terms of established metrics as comparable magneto-frictional methods \citep{Guo:2016II}.
The NLFFF is then merged with a potential background magnetic field in a quasi-steady state run of the AWSoM model for a pre-eruption solar corona.
We temporarily neglect the current in the NLFFF, as even a high quality NLFFF contains Lorentz force that can trigger an eruption prematurely.
We compare both the field topology and the pre-eruption corona to AIA observations and find reasonably good agreement for two of the three events.
Next, we turn on the current in the NLFFF and the quasi-steady state erupts spontaneously.
We apply volumetric STITCH to strengthen the CME shock as needed.
The eruption produces MFRs that propagate into interplanetary space and reach  1\,au.
We compare synthetic white light images of the simulated CMEs to LASCO C2/C3 observations and find good resemblance in their angular sizes and shapes.
This suggests that the CMEs in our simulations have roughly the correct transit speed.
Indeed, the error in the arrival time at 1\,au ranges from 1.5 to 9 hours, which is on par with other CME modeling methods.
Through a comparison of the simulated plasma at 1\,au with in-situ measurements, we find that the MFR can produce southward $B_z$ which, in two of the three events, occurs at the correct time.
We are also able to match the magnitude of the southward $B_z$ in two of the three events and occasionally the magnitudes and signs of $B_x$ and $B_y$.
The simulation for one event also successfully reproduces the NE-SW orientation of the MFR axis.
 In conclusion, this novel method of CME simulation shows great promise.
We hope that, through further improvements, this work can open up new ground in CME simulation methods. 

%% Please use the acknowledgment and contribution environments. This will 
%% be anonomyized when the "anonymous" style option is used. 
\begin{acknowledgments}
The authors acknowledge support from the National Science Foundation grant PHY-2027555 and support from the CLEAR NASA Space Weather Center of Excellence. 
The authors thank Prof. Lulu Zhao and Prof. Tamas Gombosi for their support of this project.
The authors thank Dr. Nishtha Sachdeva for sharing her knowledge in CME simulation and her various post-processing and script-generating scripts.
The authors thank Prof. Ward B. (Chip) Manchester for his suggestions on improving the paper.
The authors acknowledge the Texas Advanced Computing Center (TACC) at The University of Texas at Austin for providing computational resources that have contributed to the research results reported in this paper.
This work would also not be complete without the prolonged operation of the Frontera Supercomputer, so we thank all relevant TACC personnel for their efforts to keep Frontera operational.
This work utilizes data produced collaboratively between Air Force Research Laboratory (AFRL) \& the National Solar Observatory (NSO). The ADAPT model development is supported by AFRL.
Artificial Intelligence is used to generate scripts that produce some of the figures in this paper and helps with an edit of English.
\end{acknowledgments}

\begin{contribution}
%%This section gives authors the space to recognize author contributions. The text inside this environment is NOT counted towards the total word quanta. At a minimum, manuscripts are expected to include this text:

YFA is responsible for implementing codes changes, running numerical simulations and writing the manuscript.
GT came up with the research concept and supervised YFA as his research advisor. He edited the manuscript.
BPB first implements the magneto-friction \textit{user module} in BATS-R-US.

%% But authors are expected to provide more specific details, e.g. 
%%
%%SC was responsible for writing and submitting the manuscript.
%%WWM came up with the initial research concept and edited the manuscript.
%%OTS obtained the funding and edited the manuscript.
%%EBF provided the formal analysis and validation. He also edited the manuscript.
%%GEH Supervised the undergraduates, wrote the software and administers the project github and Zenodo repositories.
%%
%% Authors can use the Contributor Role Taxonomy (CRediT) at
%% for ideas on how write a good statement tailored to their needs.

\end{contribution}

\bibliography{csem,mybib}{}
\bibliographystyle{aasjournalv7}



\end{document}